\documentclass[traditabstract]{aa} % for the abstract without structuration 
\newcommand\sersic{S\'{e}rsic}
\newcommand\cote{C{\^ o}t{\' e}}
\newcommand\jordan{Jord{\'a}n}
\newcommand\etal{{et~al.}} 
\newcommand\esogxy{ ESO\,306-017}
\usepackage{graphicx}
\usepackage{txfonts}
\begin{document}
   \title{Globular cluster systems in fossil groups: NGC\,6482, NGC\,1132 and ESO\,306-017
\thanks{Based on observations made with the NASA/ESA Hubble Space Telescope, obtained at the Space Telescope Science Institute, which is operated by the Association of Universities for Research in Astronomy, Inc., under NASA contract NAS 5-26555. These observations are associated with program \# 10558.}}

 %  \subtitle{I. Overviewing}

   \author{K. A. Alamo-Mart\'inez\inst{1,2}
          \and
          M. J. West\inst{1}
          \and
          J. P. Blakeslee\inst{3}
          \and
          R. A. Gonz\'alez-L\'opezlira\inst{2,8}
          \and
          A. Jord\'an\inst{4}
          \and
          M. Gregg\inst{5,6}  
           \and
           P. C\^ot\'e\inst{3}
            \and
          M. J. Drinkwater\inst{7}
           \and
          S. van den Bergh\inst{3}
          }

   \institute{European Southern Observatory, Alonso de C\'ordova 3107, Vitacura, Santiago, Chile\\
          \email{k.alamo@crya.unam.mx}
             % \email{kalamo@eso.org, mwest@eso.org}
         \and
             Centro de Radioastronom\'ia y Astrof\'isica, Universidad Nacional Aut\'onoma de M\'exico, Morelia 58090, M\'exico 
            % \email{k.alamo@crya.unam.mx, r.gonzalez@crya.unam.mx}
              \and
            Dominion Astrophysical Observatory, Herzberg Institute of Astrophysics, National Research Council of Canada, Victoria, BC V9E 2E7, Canada  
            % \email{john.blakeslee@nrc-cnrc.gc.ca; patrick.cote@nrc-cnrc.gc.ca;sidney.vandenbergh@nrc-cnrc.gc.ca}
              \and
          Departamento de Astronom\'ia y Astrof\'isica, Pontificia Universidad Cat\'olica de Chile, 7820436 Macul, Santiago, Chile 
           %\email{ajordan@astro.puc.cl }
                                        \and
            Department of Physics, University of California, Davis, CA 956160, USA 
             %\email{gregg@igpp.ucllnl.org}
                           \and
               Institute for Geophysics and Planetary Physics, Lawrence Livermore National Laboratory, L-413, Livermore, CA 94550, USA  
             %\email{gregg@igpp.ucllnl.org }
             \and
                       School of Mathematics and Physics, The University of Queensland, Brisbane, QLD 4072, Australia 
             %\email{m.drinkwater@uq.edu.au}
             \and
             Argelander-Institut f\"ur Astronomie, Auf dem H\"ugel 71, D-53121 Bonn, Germany
            % \email{ragl@astro.uni-bonn.de}
             }

   \date{}

% \abstract{}{}{}{}{} 
% 5 {} token are mandatory
 
 \abstract
{
We study the globular cluster (GC) systems in three representative fossil group galaxies:
the nearest (NGC\,6482), the prototype (NGC\,1132) and the most massive known to date
(ESO\,306-017).  This is the first systematic study of GC systems in fossil groups.
Using data obtained with the \textit{Hubble Space Telescope} Advanced Camera for Surveys
in the F475W and F850LP filters, we determine the GC color and magnitude distributions,
surface number density profiles, and specific frequencies.
In all three systems, the GC color distribution is bimodal, the GCs are spatially more extended than the starlight, 
and the red population is more concentrated than the blue.
The specific frequencies seem to scale with the optical luminosities of the central
galaxy and span a range similar to that of the normal bright elliptical galaxies
in rich environments.
We also analyze the galaxy surface brightness distributions to look for deviations from
the best-fit S\'ersic profiles; we find evidence of recent dynamical interaction in all
three fossil group galaxies. 
Using X-ray data from the literature, we find that luminosity and metallicity appear 
to correlate with the number of GCs and their mean color, respectively. 
Interestingly, although NGC\,6482 has the lowest mass and luminosity in our sample, 
its GC system has the reddest mean color, and the surrounding X-ray gas has the highest metallicity.
}
%Further studies of GC systems in larger samples of fossil groups are needed to confirm these trends.

   {}


   \maketitle
%
%________________________________________________________________

\section{Introduction}

The most accepted mechanism of galaxy formation is hierarchical assembly, whereby large structures are formed by the merging of smaller systems  (Press \& Schechter 1974; De Lucia et al. 2006); nevertheless, there are some open questions and this scenario is debated (Nair et al. 2011, see also Sales et al. 2012). 
The hierarchical assembly is supported by the frequent observation of galactic interactions, and predicts that all the galaxies within a galaxy group or cluster will eventually merge into a single massive elliptical galaxy; this would be the last step of galaxy formation. Indeed, Ponman et al.\ (1994) found an extreme system consisting of a giant elliptical galaxy surrounded by dwarf companions, all immersed in an extended X-ray halo. This system was interpreted as the end product of the merger of galaxies within a group, and was thus called a fossil group (FG). A formal definition was introduced by Jones et al.\ (2003), according to which FGs are systems with $L_x\ge10^{42}h_{50}^{2}\rm{erg~s}^{-1}$, and an optical counterpart where the difference in magnitude between the first and the second brightest galaxies is ${\Delta}m_R>2~$mag. 
Because of their regular X-ray morphologies and lack of obvious recent merger activity, FG galaxies are usually considered to be ancient and unperturbed systems (Khosroshahi et al. 2007). 
This is supported by numerical simulations that suggest FGs formed at early epochs ($z>1$) and afterwards evolved fairly quiescently (D'Onghia et al.\ 2005; Dariush et al.\ 2007; D{\'{\i}}az-Gim{\'e}nez et al.\ 2011).

At present, there is no consensus on the mass-to-light ($M/L$) ratios of FGs; measurements span
from $\sim100M_\odot/L_\odot$ (Khosroshahi et al.\ 2004) to $\sim1000M_\odot/L_\odot$ (Vikhlinin et
al.\ 1999; Yoshioka et al. 2004; Cypriano et al.\ 2006) in the $R$-band, but there is a tendency
towards higher values (Proctor et al. 2011; Eigenthaler \& Zeilinger 2012).
Such high $M/L$ values, together with the fact that FGs show unusually high $L_X / L_{opt}$ ratios
compared with typical galaxy groups, appear inconsistent with the assumption that FGs are formed by
the merging of the members in ordinary galaxy groups (see Fig.\,2 in Khosroshahi et al.\ 2007).
Compact galaxy groups in the nearby universe tend to have lower masses and X-ray luminosities
than observed in FGs, although there is evidence that high-mass compact groups may have been more
common in the past and could represent the progenitors of today's FGs (Mendes de Oliveira \&
Carrasco 2007).
In addition, the optical luminosities of the dominant giant ellipticals in FGs (Khosroshahi et
al. 2006; Tovmassian 2010) are similar to those of brightest cluster galaxies (BCGs).
For these reasons, it has been suggested that FGs are more similar to galaxy clusters, but simply
lack other early-type galaxies with luminosities comparable to the central galaxy (Mendes de
Oliveira et al.\ 2009).
It remains an open question whether FGs represent the final phase of most galaxy groups, or if they
constitute a distinct class of objects which formed with an anomalous top-heavy luminosity function
(Cypriano et al.\ 2006; Cui et al.\ 2011; see also M{\'e}ndez-Abreu et al.\ 2012).

Globular clusters (GC) systems are a powerful tool to study galaxy assembly (Fall \& Rees 1985;
Harris 1991; Forbes et al.\ 1997; West et al. 2004; Brodie \& Strader 2006). GCs are among the
oldest objects in the universe, and dense enough to survive galactic interactions. Most of them
have ages older than 10\,Gyr (Cohen et al. 1998; Puzia et al. 2005), and therefore probably formed
before or during galaxy assembly.  The correlation of the optical luminosity of a galaxy with the
mean metallicities of the field stars and GCs (van den Bergh 1975; Brodie \& Huchra 1991; 
Forbes et al.\ 1996; C\^ot\'e et al.\ 1998) supports the idea that GC and galaxy formation are
strongly linked.  Nevertheless, there is a shift in metallicity, in the sense that field stars are 
on average more metal-rich than GCs.  
Interestingly, Cohen et al. (1998) noted that the metallicity of the X-ray gas in M87 is similar to
the metallicity of the field stars.  Thus, the GC population could constitute a record of the
initial chemical enrichment of the parent galaxy.  

The GC systems of massive early-type galaxies are generally bimodal in their optical
color distributions (Zepf \& Ashman 1993; Geisler et al.\ 1996; Gebhardt \& Kissler-Patig
1999). 
The two GC color components are commonly referred to as the metal-poor (blue) and
metal-rich (red) subpopulations, and could indicate distinct major episodes of star formation.
In the Milky Way, spectroscopic studies reveal two populations of GCs with
distinct metallicities (Zinn 1985).  However, the situation is less clear for giant ellipticals and other
massive galaxies with more complex formation histories, which may simply have broad metallicity
distributions with a nonlinear dependence of color on metallicity (Yoon et al.\ 2006, 2011; Blakeslee et al.\
2012; Chies-Santos et al. 2012). 
Another important property of GC systems is their luminosity function (GCLF). 
The GCLF is well described by a Gaussian with peak at ${M_V}^0\approx-7.4$, and the dispersion for giant ellipticals
with well-populated GC systems is ${\sigma_V}^{\rm GCLF}\approx1.4$ (Harris 1991).  
The general homogeneity of the GCLF is unexpected {\it a priori}, given that GC
destruction mechanisms should depend on the environment (McLaughlin \& Fall 2008).
However, recent studies show that the GCLF dispersion depends on galaxy luminosity (\jordan\ et al.\ 2007b; Villegas
et al.\ 2010). 

The most basic measure of a GC system is its richness, or the total number of GCs in the system.
Harris \& van~den~Bergh (1981) introduced the concept of {\it specific frequency}, ${S_N}$, as the
number of GCs per unit galaxy luminosity, normalized to a galaxy with absolute $V$ magnitude $M_V =
-15$. They reported values in the range 2$< S_N<$10 for elliptical galaxies. However, they also
already noted that M\,87, the central galaxy in the Virgo cluster,
had an outstandingly large value of $S_N\sim20$ (a more recent value from Peng et al. 2008 is $S_N=12.6\pm0.8$); 
hence, they suggested that environment may play a role in GC formation. 
Later, other high-$S_N$ galaxies were identified, most of them being either BCGs or
second brightest galaxies in clusters; often, these galaxies were classified as type cD, however
not all galaxies classified as cD have high $S_N$ (Jord\'an et~al.\ 2004).  
It is interesting to note that, although BCGs have quite a uniform luminosity, to the point of
being considered standard candles (Postman \& Lauer 1995), they have a wide range in $S_N$. 
Conversely, Blakeslee et al.\ (1997) found a correlation between $S_N$ of the BCG and the velocity
dispersion and X-ray luminosity of the cluster.  They therefore suggested that the number of GCs
scales roughly with cluster mass, a fact that suggests that galaxies with high values of $S_N$, 
like M87, are not anomalously rich in GCs but, rather, underluminous as a consequence of the lower
overall star formation efficiency in more massive and denser systems (Blakeslee 1999; Peng et al.\ 2008).   

\begin{figure*}[ht]
\centering
\includegraphics[width=0.30\textwidth]{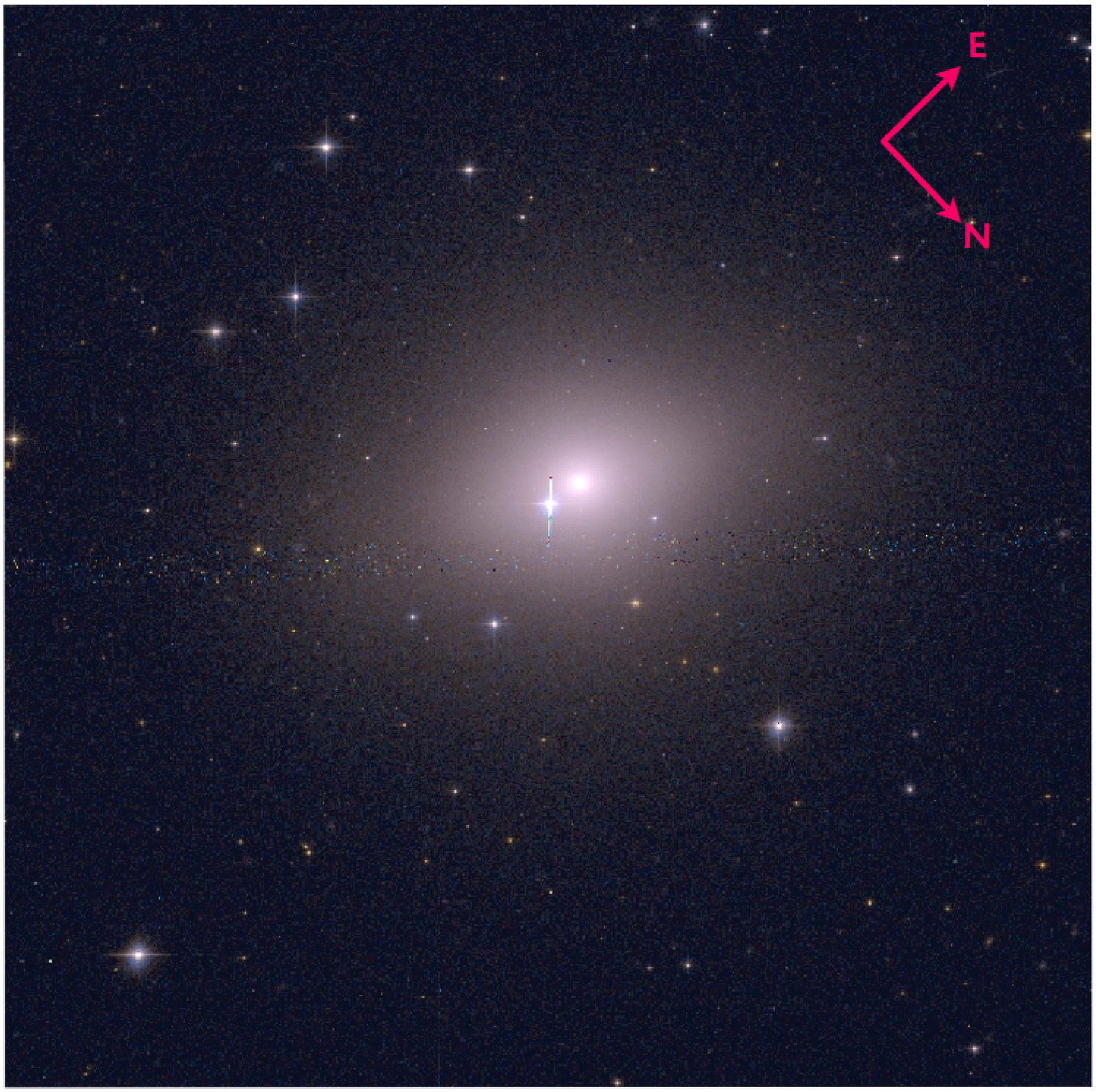}
\includegraphics[width=0.30\textwidth]{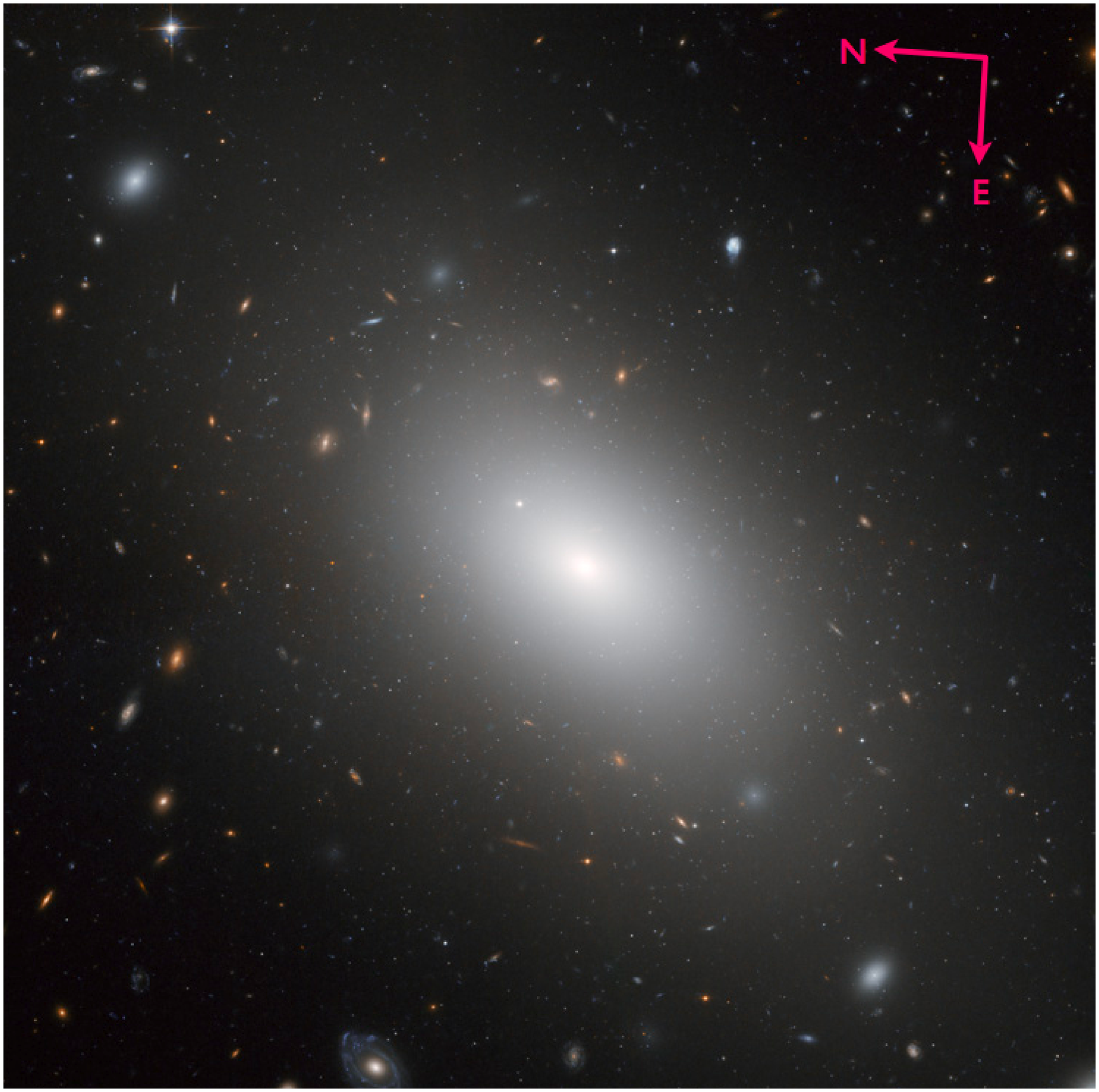}
\includegraphics[width=0.30\textwidth]{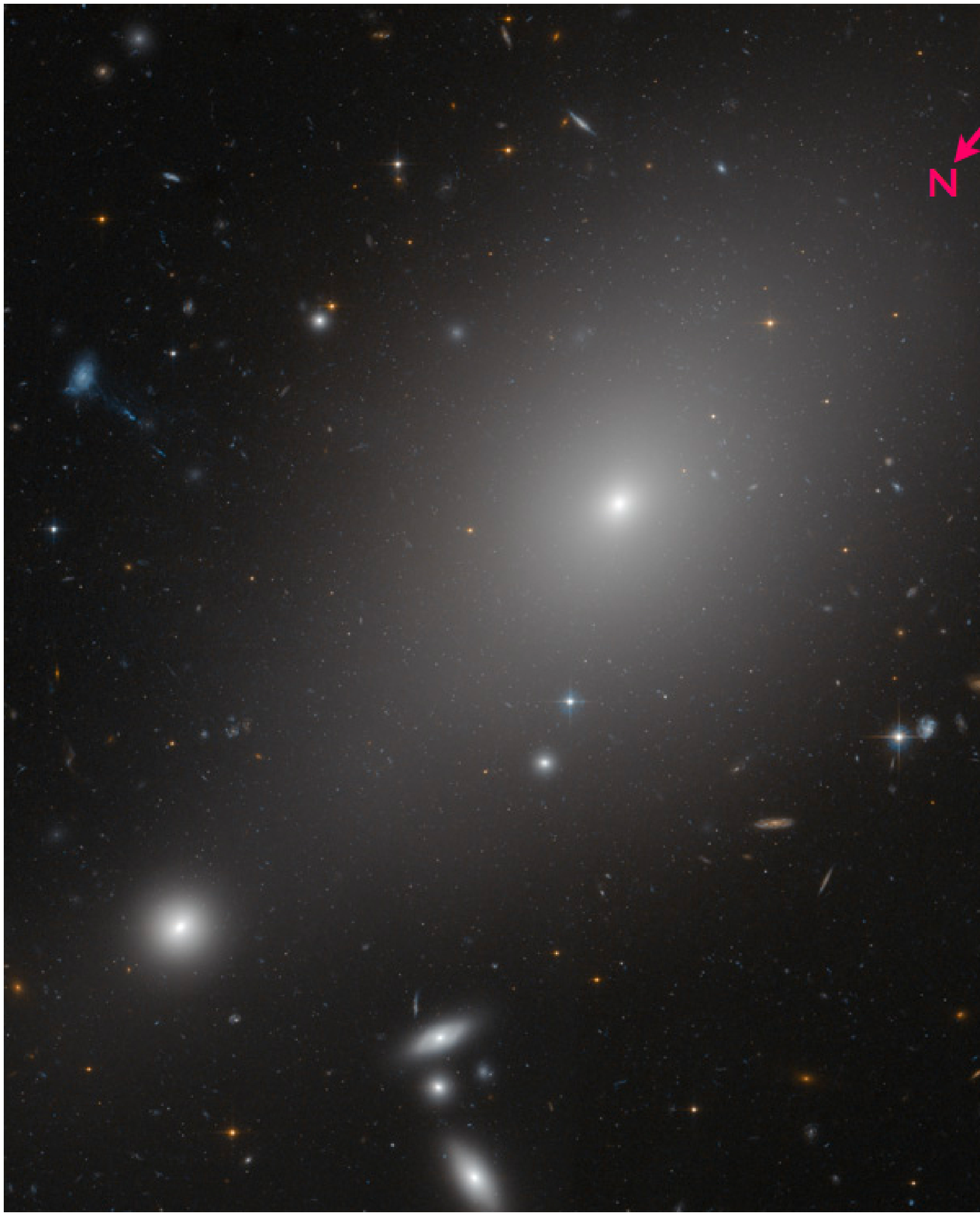}
\caption{ACS color images, FOV$\sim3\farcm3{\,\times\,}3\farcm3$. NGC\,6482 (left), NGC\,1132 (center), and ESO\,306-017 (right) combining the filters F475W and F850LP. {\it Credit:} NASA, ESA, and the Hubble Heritage (STScI/AURA)-ESA/Hubble Collaboration.} 
\label{pretty_pictures}
\end{figure*}

Hence, given that FGs appear to be ancient, highly luminous, but relatively unperturbed
systems with origins that remain poorly understood (identified less than two decades ago),
we study for the first time the GC populations of the dominant elliptical.  We also
compare our measured GC properties with X-ray properties obtained from the literature, and
look for photometric irregularities that could be signs of recent galactic
interactions. In order to have a representative sample we choose: the nearest, NGC\,6482
($z=0.013$), the prototype, NGC\,1132 ($z=0.023$), and the most massive known to date,
ESO\,306-017 ($z=0.036$), shown in Fig.\,\ref{pretty_pictures}. In the following section, we describe the observations, 
data processing, and GC selection criteria; in Sec.~3, we analyze the photometry and present
our measurements; in Sec.~4, we discuss the implications of our results
in the larger context of FG and early-type galaxy formation scenarios.
The final section summarizes our conclusions. 
Throughout, we estimate distances assuming $H_0=71$ km~s$^{-1}$Mpc$^{-1}$.

%%%%%%%%%%%%%%%%%% Observations and Reductions %%%%%%%%%%%%%%%%%%

\section{Observations and Reductions}

\begin{table*}[ht!]
\begin{tabular}{c  c  c  c  c  c  c  c  c  c  c  c  c  p{1in}}
\multicolumn{13}{c}  {Sample data}  \\
\hline \hline
  $Fossil~Group$ & Morph\tablefootmark{a} & $z$ & m-M\tablefootmark{b} & $d$\tablefootmark{b} & B$_{T}$\tablefootmark{a} & A$_{B}$\tablefootmark{a} & M$_{B}$\tablefootmark{a} & ${m^0}_g$ & ${m^0}_z$ & $t_{exp,g}$ & $t_{exp,z}$ & 1 arcsec\\  
   &  &  &  &  (Mpc) & &  &  &  &  &  (s) & (s) & (pc)\\
\hline
 NGC\,6482 & E? & 0.013 & 33.74 & 56.02 & 12.35 & 0.43 & -21.82  & 26.54 & 25.34 & 928 & 1200 & 264 \\
 NGC\,1132 & E & 0.023 & 34.99 & 99.47 & 13.25 & 0.27 & -22.01 &27.79 & 26.59 & 7800\tablefootmark{\ast} & 9630\tablefootmark{\ast} & 461 \\
 ESO\,306-017 & cD3 & 0.036 & 35.96 & 155.46 & 13.33 & 0.14 & -22.77 & 28.76 & 27.56 & 7057 & 8574 & 702 \\
\hline
\end{tabular}
\caption{{\it Col.\,1}: System; {\it Col.\,2}: morphology; {\it Col.\,3}: redshift; {\it Col.\,4}: distance modulus; {\it Col.\,5}: luminosity distance; {\it Col.\,6}: apparent B magnitude; {\it Col.\,7}: extinction in B; {\it Col.\,8}: absolute magnitude in B; {\it Col.\,9} and 10: expected apparent magnitude of the GCLF turnover in $g$ and $z$, respectively; {\it Col.\,11} and 12: exposure times for $g$ and $z$, respectively; {\it Col.\,13}: physical scale corresponding to 1\,arcsec. }
\tablefoottext{a}{Data from NASA/IPAC Extragalactic Database (NED).\\}
\tablefoottext{b}{$H_0=71$ km~s$^{-1}$Mpc$^{-1}$. \\}
\tablefoottext{\ast}{Due to a Hubble Observing Problem Report, the observation was repeated, obtaining at the end an image with longer exposure time than requested.}    
\label{sample_basicdata}
\end{table*}

The observations were taken with the Advanced Camera for Surveys on board the {\it Hubble Space Telescope} ({\sl HST}), using the Wide Field Channel with the F475W ($\approx$ the $g$ passband of the Sloan Digital Sky Survey, SDSS) and F850LP ($\approx$SDSS~$z$) filters (GO proposal 10558).\\ 

The broadband $g$ and $z$ filters offer relatively high throughput and $g{-}z$ color is very sensitive to age and metallicity variations (C\^{o}t\'e et al.\ 2004). The integration time for each galaxy was chosen to reach the expected turnover magnitude of the GCLF, ${M^0_g}=-7.2$ and ${M^0_z}=-8.4$ (Jord\'an et al. 2007b), assuming a Gaussian distribution. 
Table\,\ref{sample_basicdata} lists the redshift, distance modulus, exposure time and expected apparent turnover magnitude of the GCLF for each filter, as well as the physical scale corresponding to 1\,arcsec. The ACS pixel scale is 0.05\,arcsec\,pixel$^{-1}$.\\

 For all images the same data reduction procedure was followed. Geometrical correction, image combination, and cosmic ray rejection were done with the data reduction pipeline APSIS (Blakeslee et al.\ 2003). A PSF model was constructed for each image using the DAOPHOT (Stetson 1987)  package within IRAF (Tody 1986). The best fitting function for all the images was Moffat25 (good for undersampled data), with FWHM$\sim$2~pixels and fitting radius=3~pixels. 

\subsection{Galaxy surface brightness distribution}
\label{section_starlightDIST}
 
We are interested in the detection and magnitudes of GCs, which can be affected by drastic changes in the background due to the galaxy light (mainly in the inner region).
Hence, in order to perform the detections over an image with nearly flat background it is necessary to subtract the galaxy light. \\

To model the galaxy light, the brightest objects (foreground stars, background and group galaxies) were masked, and subsequently the surface brightness distribution of the central giant elliptical was modeled with the {\it ellipse} (Jedrzejewski 1987) and {\it bmodel} tasks within the STSDAS\footnote{STSDAS is a product of the Space Telescope Science Institute, which is operated by AURA for NASA.} package in IRAF. 
A second mask was made for objects detected at $> 4.5\sigma$ with SExtractor (Bertin \& Arnouts 1996) in the galaxy-subtracted image. 
A second, final model of the surface brightness distribution of the stellar component (galaxy model) was produced from the newly masked image. \\

Due to the relatively small ACS field-of-view (FOV; $\sim3\farcm3\times3\farcm3$), the giant ellipticals fill the images, and it is not possible to measure an absolute sky value directly. Thus, a surface brightness radial profile of galaxy plus sky was constructed.

Although the value of the sky is not a critical issue for calculating colors of GCs, 
since a local value of the sky is measured for each of them, a reliable estimate of the sky level 
is essential to measure the magnitude of the galaxy, which is needed to obtain $S_N$. Thus, in order to estimate a realistic sky value, we assumed that the starlight follows a S\'ersic profile (S\'ersic 1968), which provides a good description of the outer parts of early-type galaxies (Trujillo et al. 2004; Ferrarese et al. 2006). We fit, to the constructed intensity radial profile, the function:

\begin{equation}
I(r)=I_e\exp\left\{-b_n \left[ {\left( \frac{r}{R_e}\right) }^{1/n}- 1\right] \right\} + I_{sky}
\end{equation} 

\noindent
where $R_e$ is the effective radius that encloses half of the light; $I_e$ is the effective intensity at $R_e$; $n$ is the S\'ersic index, which indicates the slope; $I_{sky}$ is the sky intensity; $b_n\approx1.9992n - 0.3271$ (Graham \& Driver 2005). 
The technique to estimate the best fit model was $\chi^2$ minimization, using the task {\it Optim} inside {\rm R: A Language and Environment for Statistical Computing}\footnote{http://www.R-project.org}. {\it Optim} uses the algorithm of Nelder-Mead (1965), which is based on the concept of a simplex, searching for the minimum by comparing the function values on vertices in parameter space. We tried various starting parameters in order to avoid local minima.\\

The fits were done independently for each filter. 
Each point was weighted by a factor $\epsilon{I}$, where $I$ is the point's intensity and $\epsilon$ is chosen to obtain $\chi^2\sim1$ (Byun et al. 1996; Ferrarese et al. 2006); we used $\epsilon=0.05$, 0.01 and 0.03 for NGC\,6482, NGC\,1132, and ESO\,306-017, respectively. 
We performed two fits, one with all the parameters free, and a second fixing $n=4$. 
For NGC\,1132 and ESO\,306-017, the sky values obtained for the $n=4$ fits were 
higher than the mean background values in lowest regions of the images, which suggests that the sky levels obtained with these fits were overestimated.
Moreover, with $n$ as a free parameter, we recovered higher values ($n>4$) in line with expectations
 for such luminous galaxies; thus, we favor the fits with $n$ free (Graham et al.\ 1996; La Barbera
 et al.\ 2012).

\begin{figure*}[ht!]
\centering
\includegraphics[width=0.38\textwidth,angle=-90]{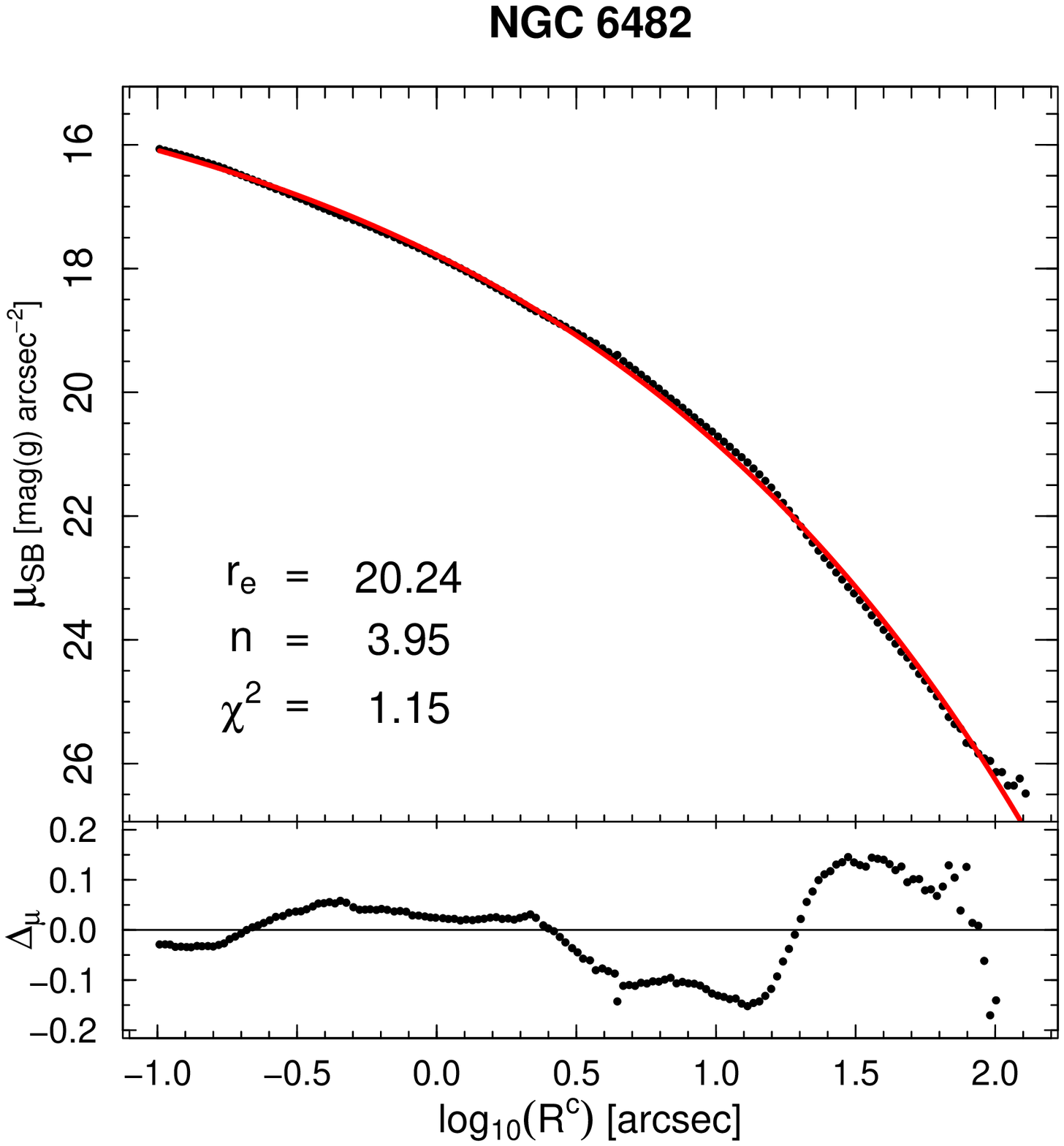}
\includegraphics[width=0.38\textwidth,angle=-90]{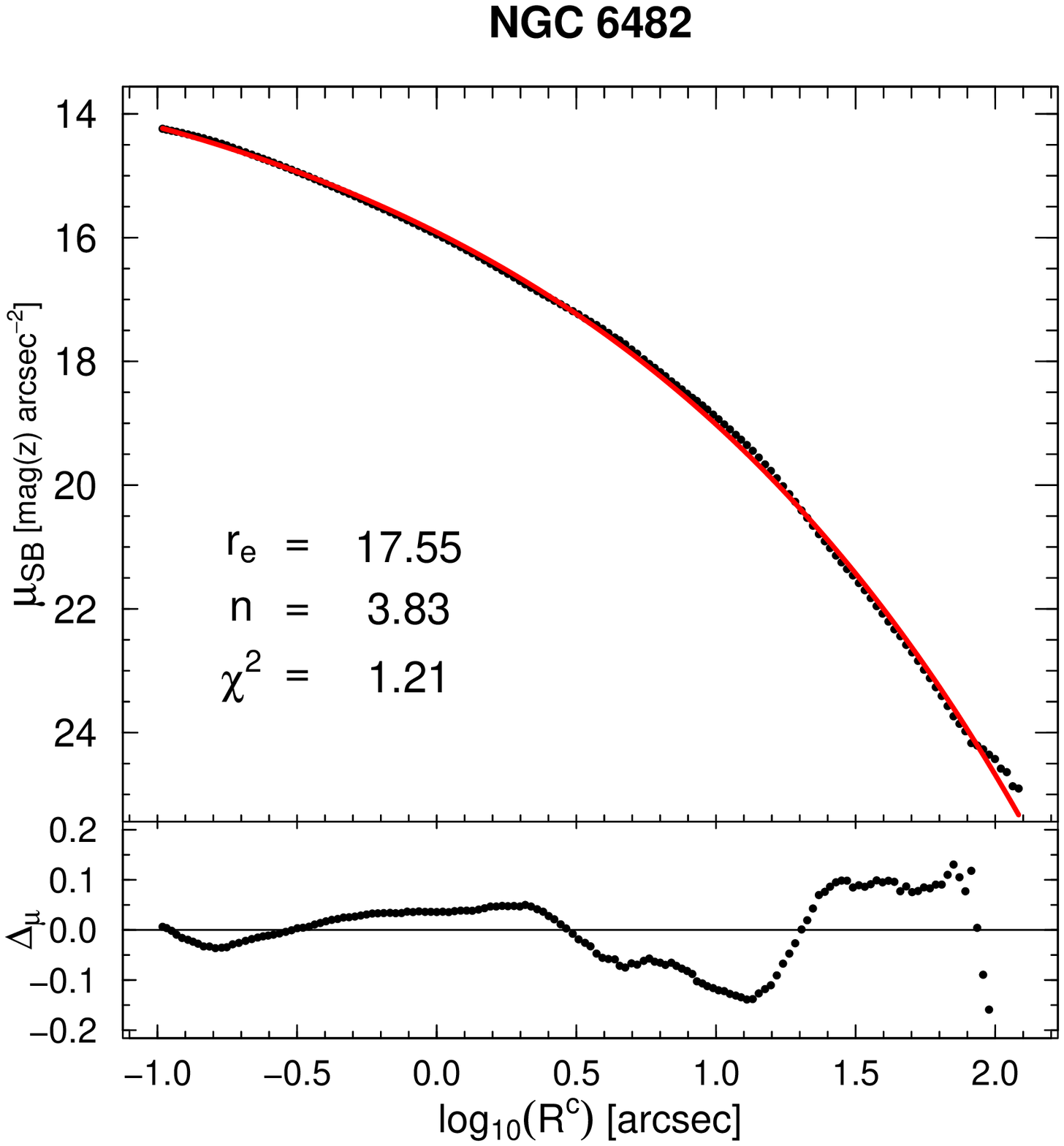}
\includegraphics[width=0.38\textwidth,angle=-90]{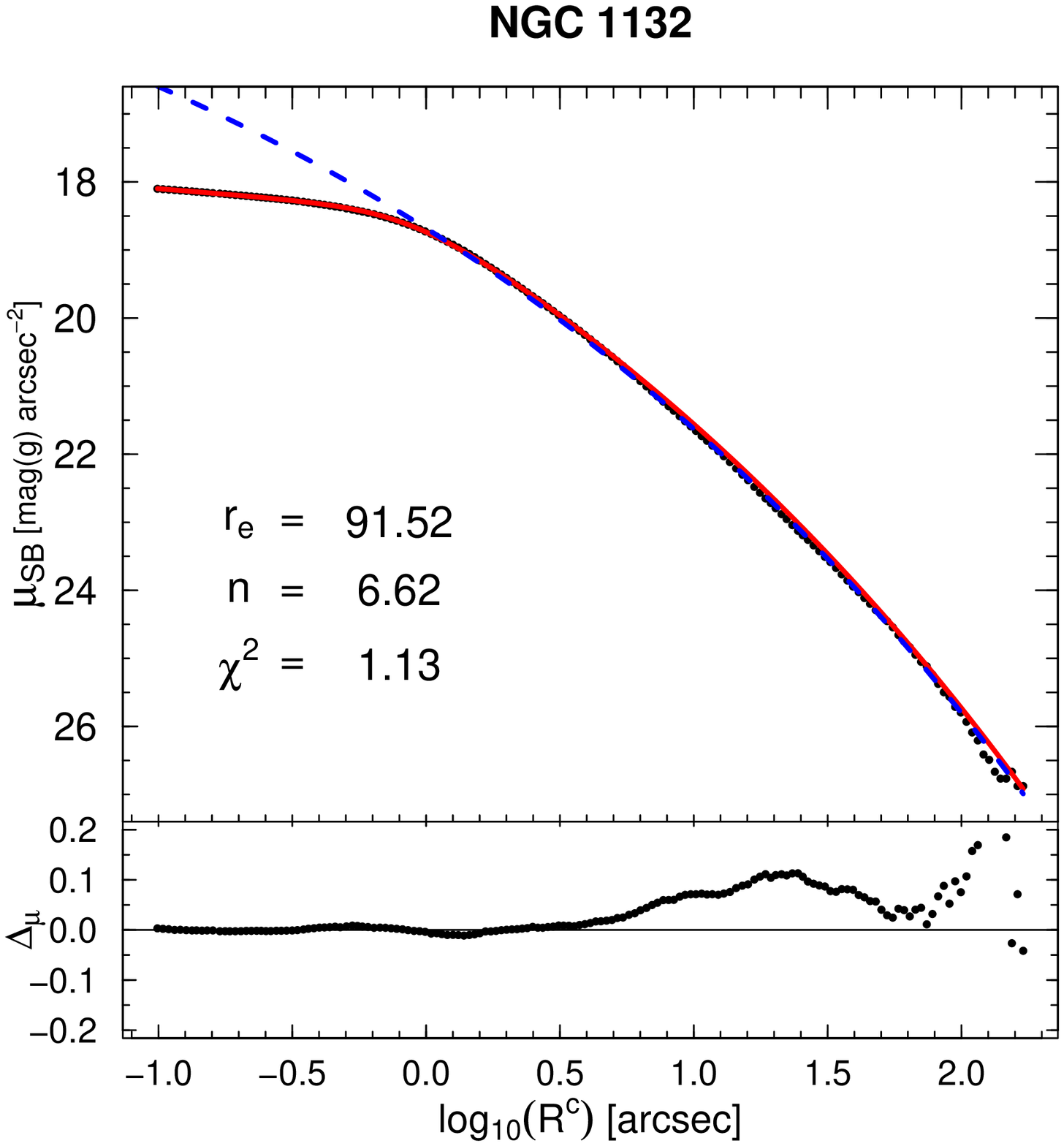}
\includegraphics[width=0.38\textwidth,angle=-90]{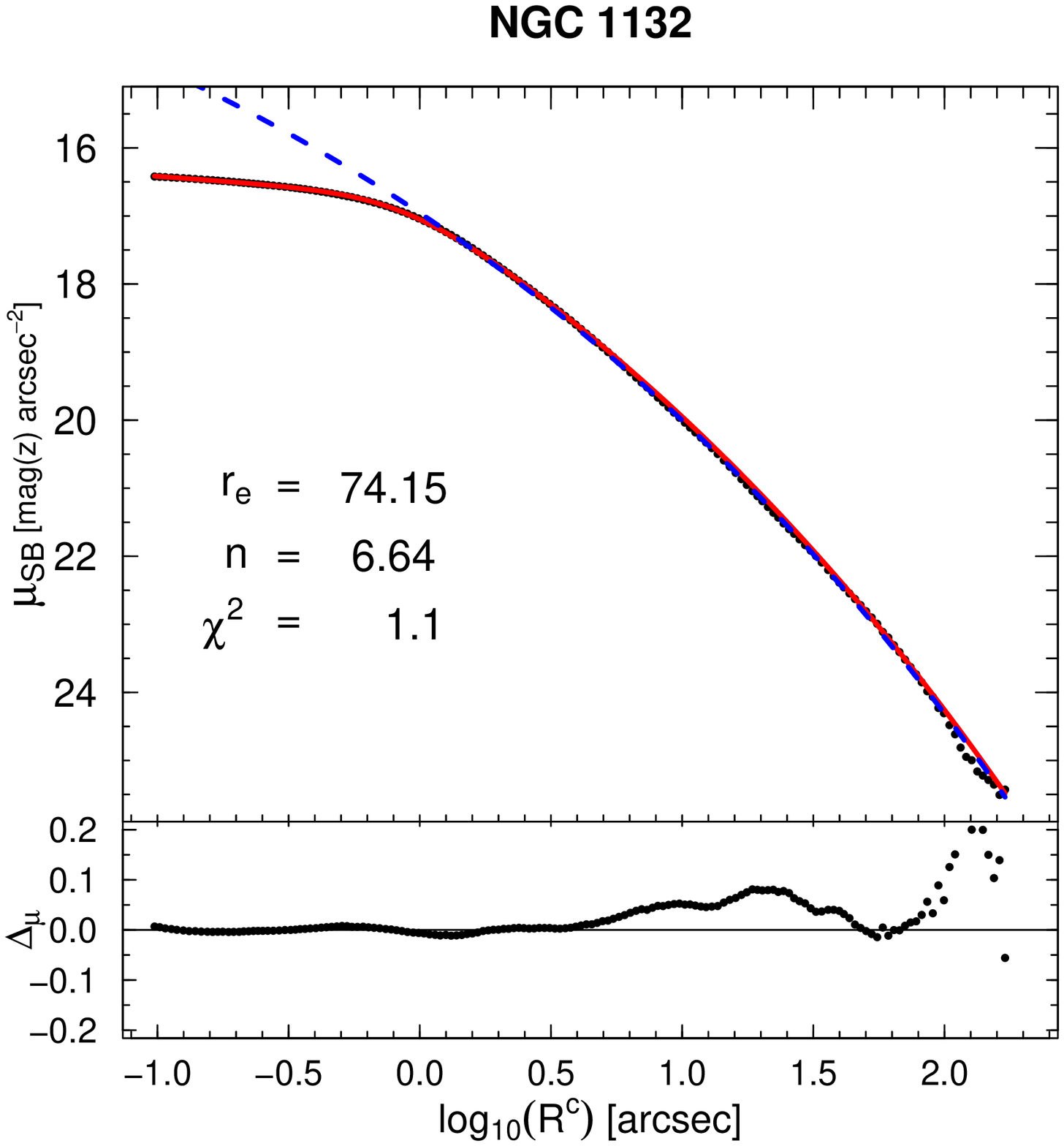}
\includegraphics[width=0.38\textwidth,angle=-90]{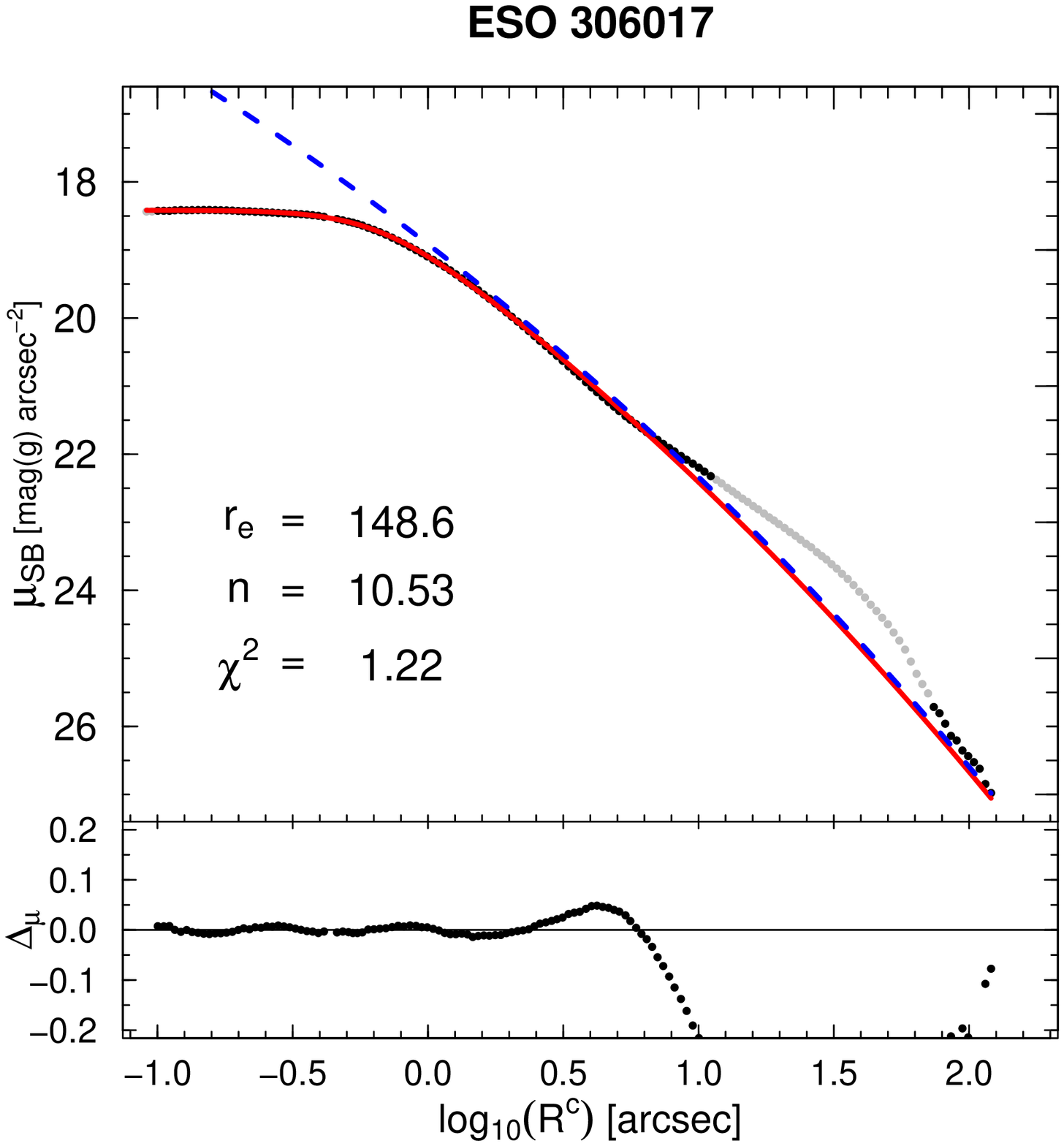}
\includegraphics[width=0.38\textwidth,angle=-90]{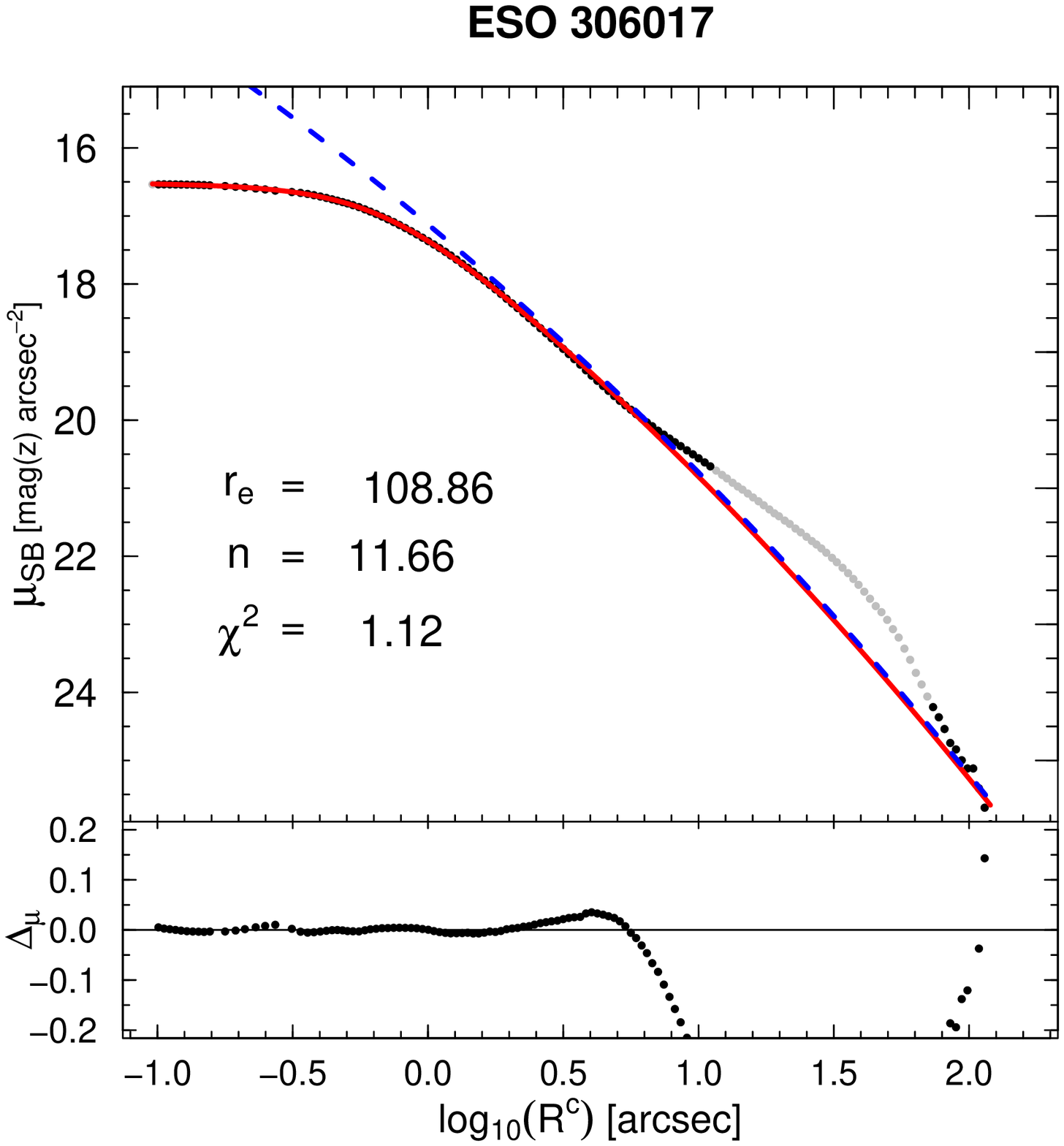}
\caption{Galaxy surface brightness distribution as a function of the 
isophotal equivalent circular radius ${R}^c$. 
The dots are the data after subtraction of the fitted sky value. The solid red lines 
show the best fit profiles: single S\'ersic for NGC\,6482 and core-S\'ersic for NGC\,1132 and ESO\,306-017. For the galaxies where the best fit is core-S\'ersic, the dashed blue line shows the single S\'ersic. {\it Top to Bottom}: NGC\,6482, NGC\,1132 and ESO\,306-017. The left and right panels are the $g$ and $z$ band, respectively. For ESO\,306-017, the gray dots were not included in the fit.}
\label{all_muPROFILES}
\end{figure*}

We next subtracted the estimated sky value and converted the semi-major axis of each isophote 
to the equivalent circular radius (or geometric radius), ${R}^c$, following the relation:
${R}^c=R\sqrt{q}$, where $q$ here is the axis ratio of the isophote.
We then performed a second fit to the sky-subtracted, circularized $\mu_{SB}$ profile 
(Fig.\,\ref{all_muPROFILES}).  This approach allows for direct 
comparisons with literature values (e.g., Ferrarese et al.\ 2006). 

As seen from the fitted profiles, there is less light in the inner regions of NGC\,1132 and ESO\,306-017 than predicted by the S\'ersic profile (blue dashed lines). This is a common feature in giant ellipticals. Studies of the surface brightness, $\mu_{\rm SB}$, profiles of early-type galaxies spanning a range in luminosity have revealed that the most luminous ellipticals often show central deficits, or ``cores,'' with respect to the inward extrapolation of the best-fit S\'ersic models, whereas intermediate-luminosity early-type galaxies are generally well described by S\'ersic models at all radii (Ferrarese et al.\ 2006; C\^ot\'e et al.\ 2007).  At fainter luminosities, early-type dwarf galaxies tend to possess central excesses, or ``nuclei.''

Hence, we add a core component (Trujillo et al.\ 2004, Graham \& Driver 2005; Ferrarese et al.\ 2006) to the function in order to improve the models for these high-luminosity galaxies. The core-S\'ersic function is:  \\
\begin{eqnarray}
I(r)=I_b 2^{-(\gamma/\alpha)} \exp \left[ b\left(2^{1/\alpha}\frac{R_b}{R_e}\right)^{1/n} \right] 
\end{eqnarray} 
\begin{eqnarray*}
\times {\left[ 1+\left(\frac{R_b}{r}\right)^{\alpha}  \right]}^{\gamma/\alpha} \times \exp\left\{ -b \left[  \frac{R^{\alpha} +{R_b}^{\alpha} }{{R_e}^{\alpha}} \right] ^{1/(\alpha{n})}\right\} + I_{sky}
\end{eqnarray*} 

\noindent
where $R_b$ is the break radius (transition point between S\'ersic and inner power-law), $I_b$ is the intensity at $R_b$, $\alpha$ indicates how sharp is the transition, and $\gamma$ is the slope of the inner component. \\

\begin{figure*}[ht]
\centering
\includegraphics[width=0.33\textwidth]{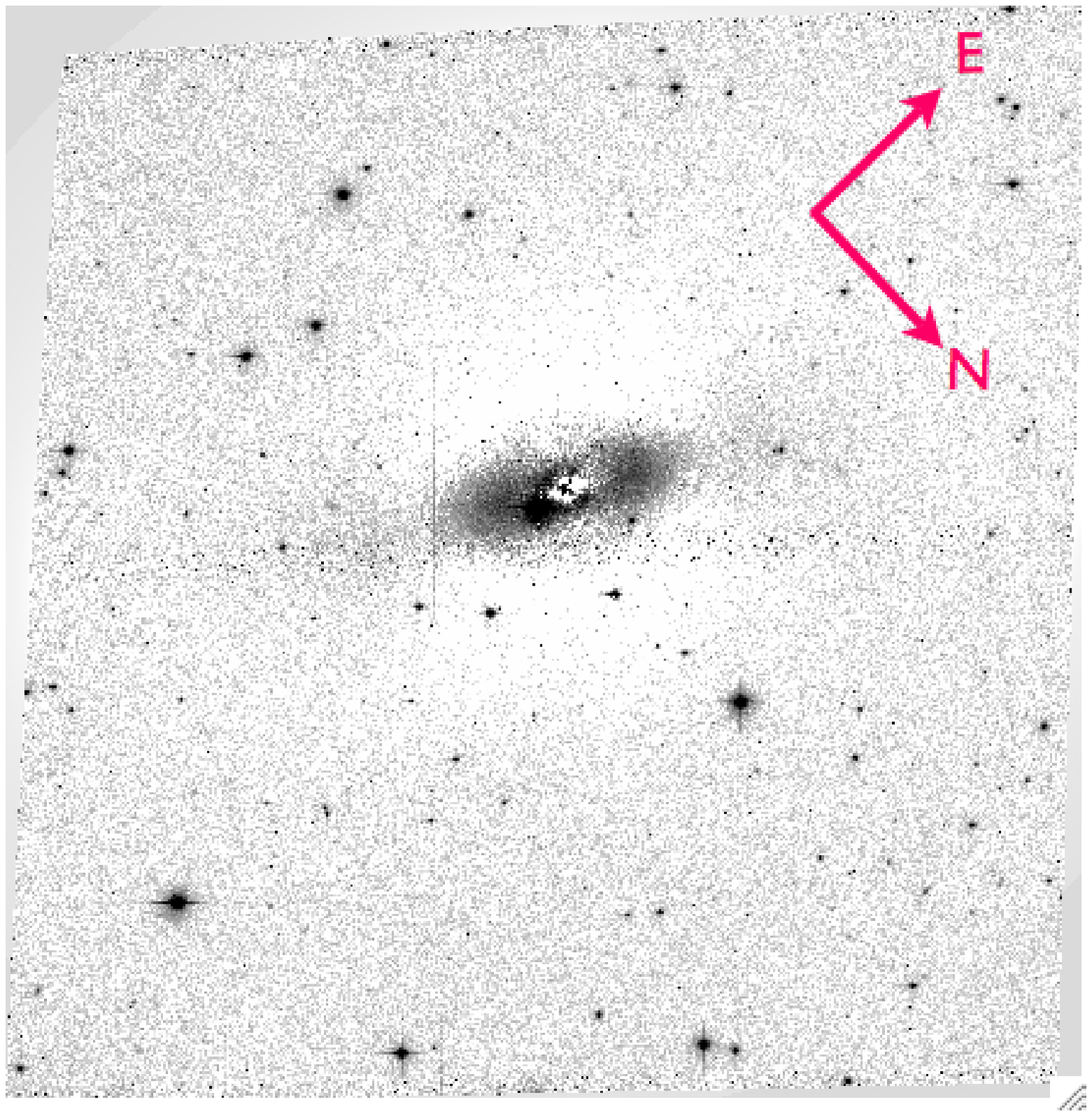}
\includegraphics[width=0.33\textwidth]{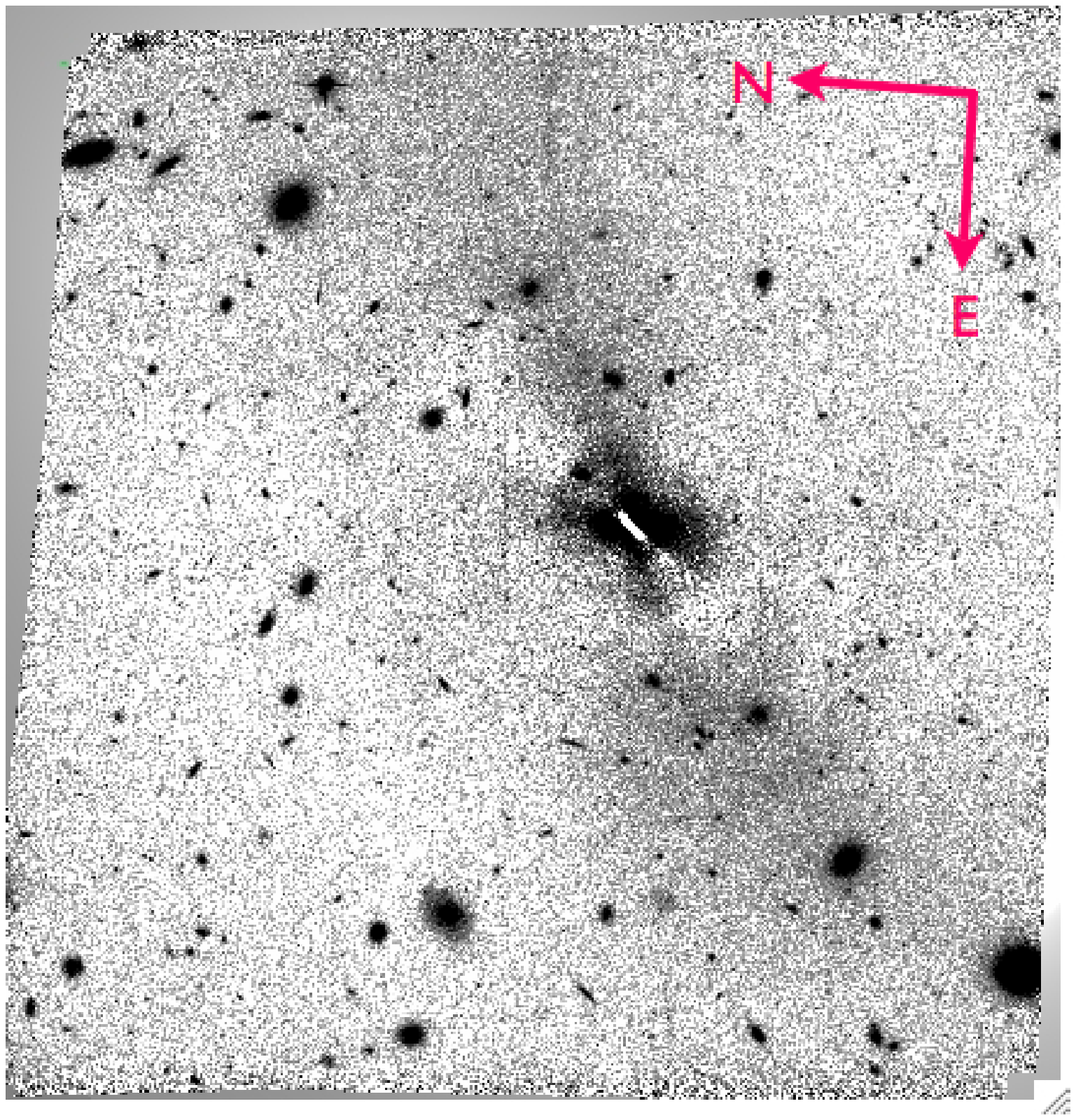}
\includegraphics[width=0.33\textwidth]{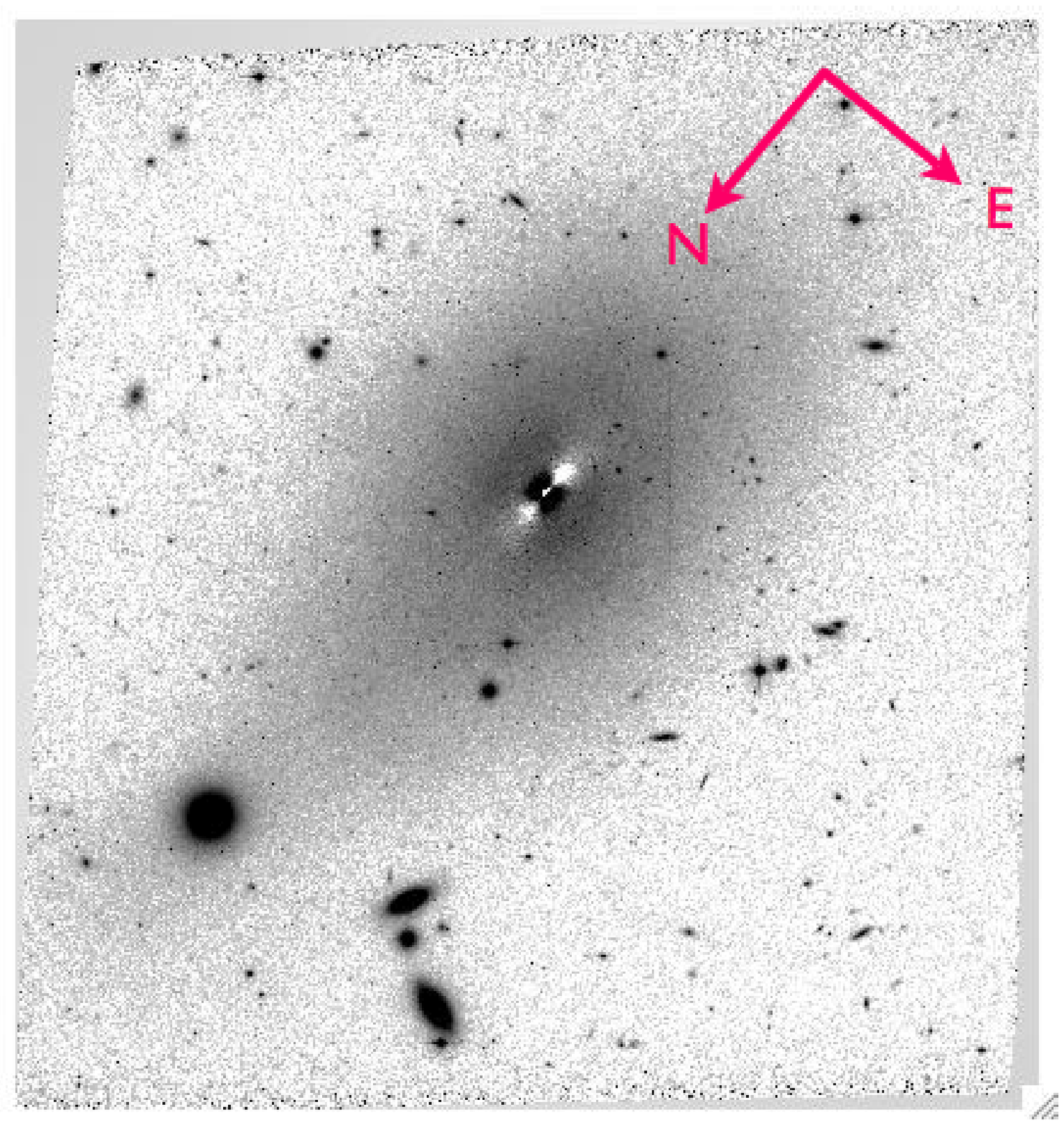}
\caption{Residual images after subtracting the best single S\'ersic model, in $z$ band:
NGC\,6482 (left), NGC\,1132 (center), and ESO\,306-017 (right). 
The galaxies are shown at the observed orientation.}
\label{sersic_residuals}
\end{figure*}

\begin{figure*}[ht]
\centering
\includegraphics[width=0.33\textwidth]{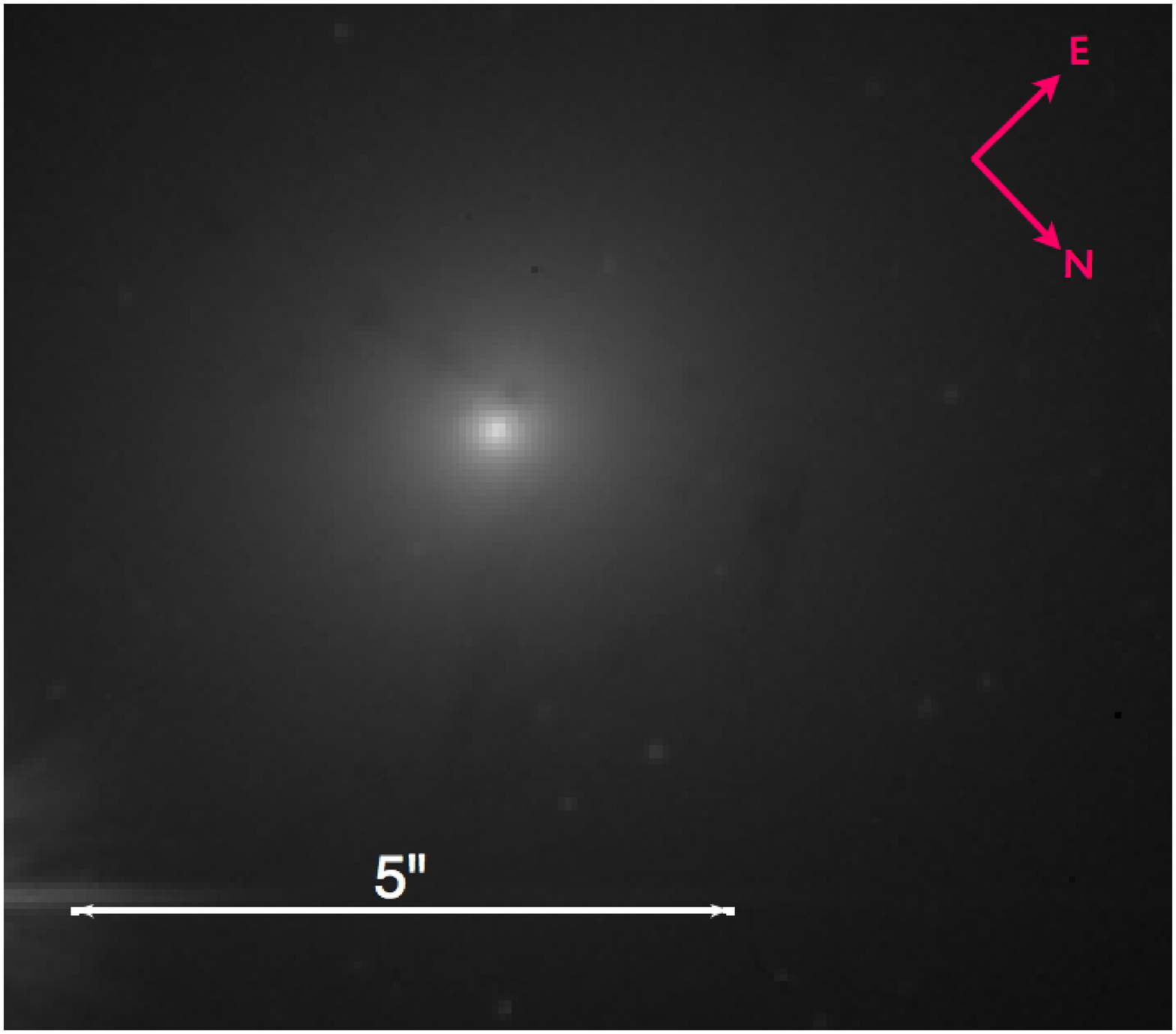}
\includegraphics[width=0.33\textwidth]{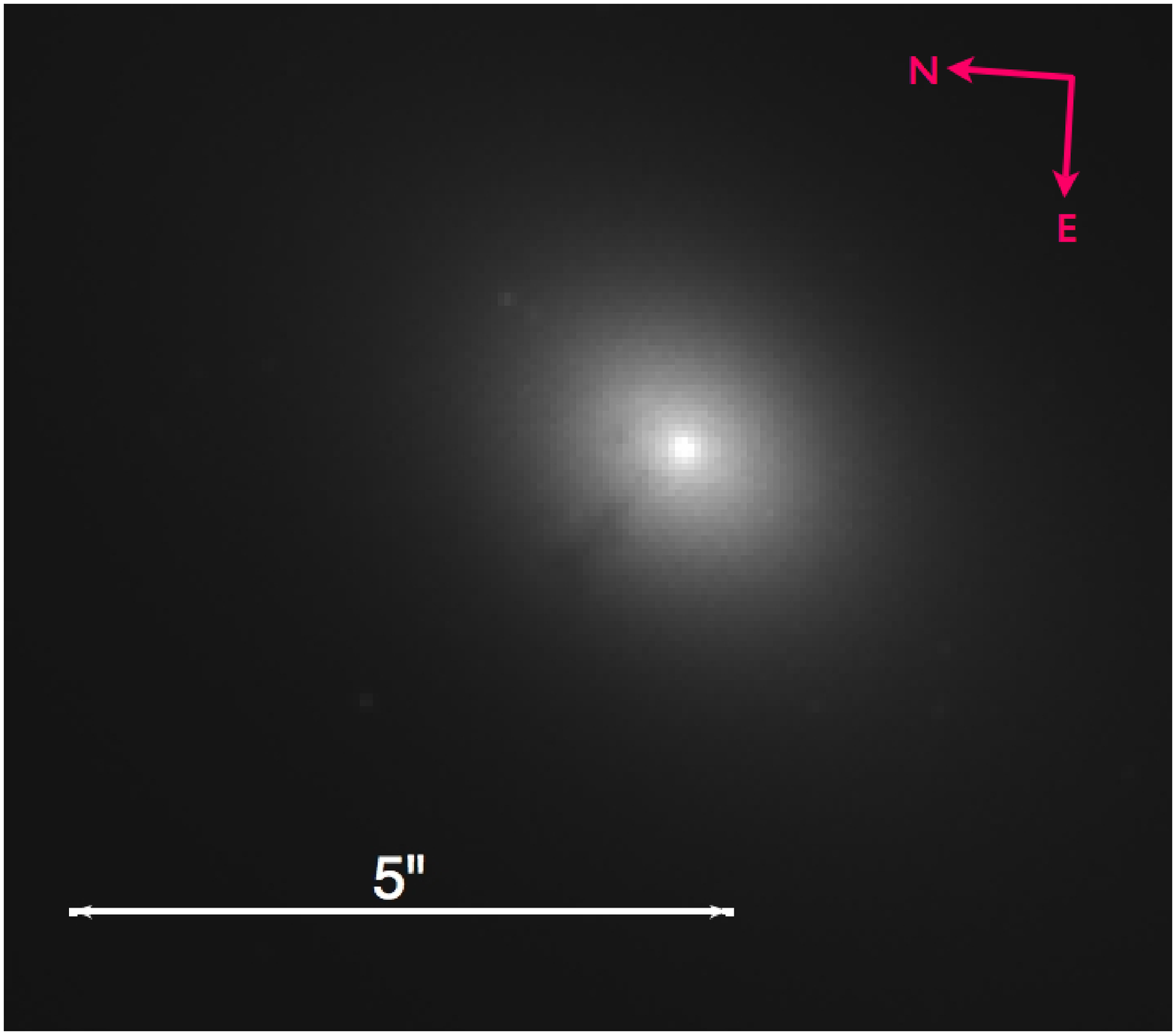}
\includegraphics[width=0.33\textwidth]{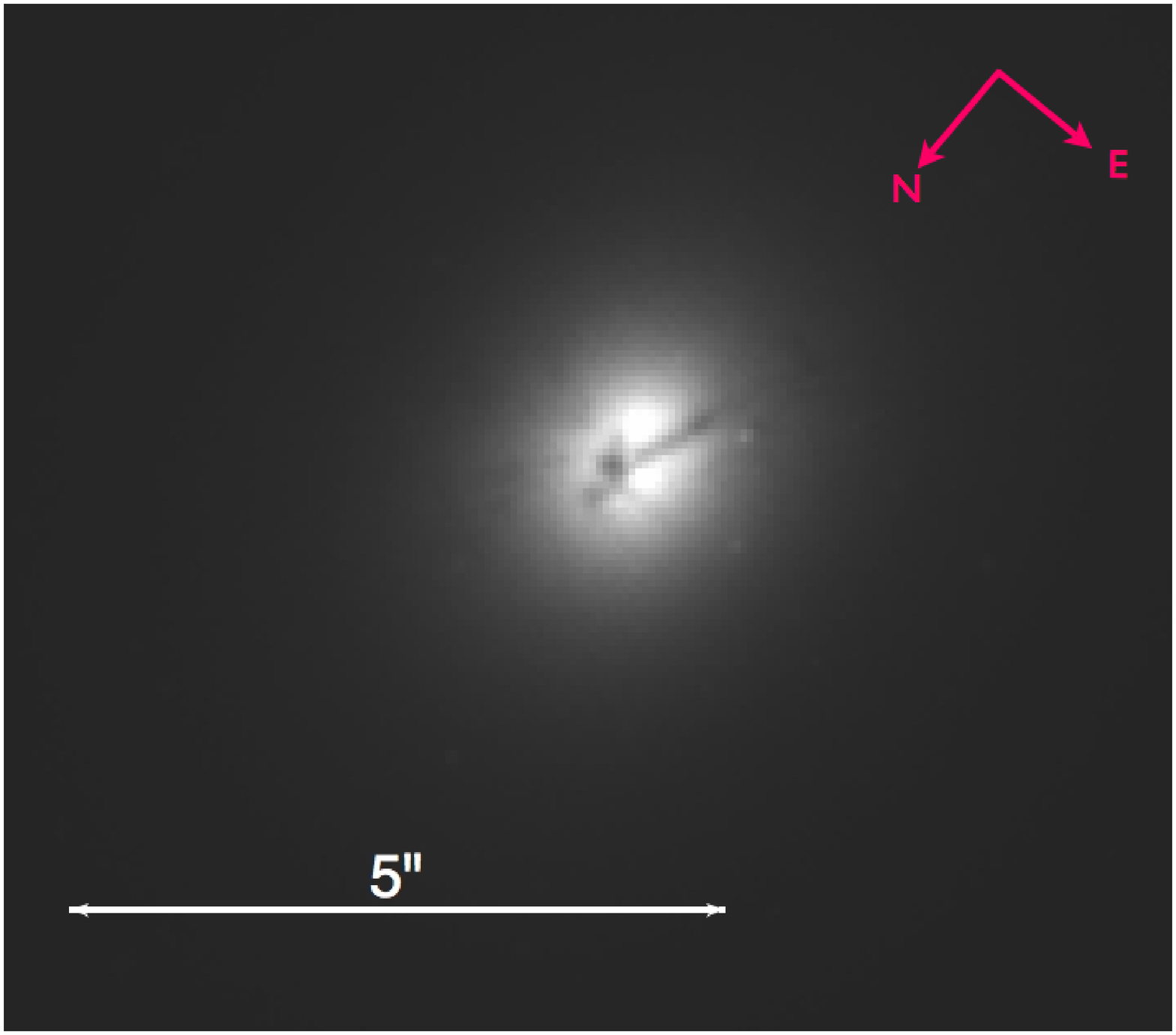}
\caption{Inner regions of $g$-band images:
NGC\,6482 (left), NGC\,1132 (center), and ESO\,306-017 (right). 
Dust is present in all cases.}
\label{zoomcenters}
\end{figure*}

For ESO\,306-017 we can see a feature around $\log_{10}(r)\sim1.7$, that might be stripped material from the second brightest galaxy in the image, resulting in an excess of light. Consequently, we reject the range $12.5^{\prime\prime}<r<90^{\prime\prime}$ from the fit domain in both filters (gray dots in Fig.\,\ref{all_muPROFILES}), whereupon we obtain a good fit. 
Nonetheless, Sun et al.\ (2004) reported a steeper profile over a larger spatial range using ground-based data in the $R$-band, with the profile declining more sharply than $n=4$
at large radii. 
It would be unusual for a galaxy with such high luminosity to have $n<4$.
The inconsistency in $n$ values at different radii in this galaxy reflects the peculiarity of its light
profile, indicating that a single \sersic\ model does not provide an adequate description. 
We therefore encourage the reader to be cautious in interpreting or using the fitted surface
brightness parameters for this galaxy.
For the more regular galaxy NGC\,1132, Schombert \& Smith (2012) reported $n=7.1$ and a circular
effective radius $R_e^c=80.6^{\prime\prime}$, both of which are close to our fitted values.     
Table\,\ref{sersic_parameters} lists our final best-fit \sersic\ parameters and sky values. 

 \begin{table*}
\centering
\begin{tabular}{c  c  c  c  c  c  c  c  c  c  c  c  c  c  c p{1in}}
\multicolumn{15}{c}  {}  \\
\hline \hline
 $Galaxy$  & ${R_{e,g}}^c$ & $n^{g}$ & ${R_b}^{g}$ & $\gamma^{g}$ & $\alpha^{g}$ & ${\chi^2}_{g}$ & ${m_g}^{sky}$
 & ${R_{e,z}}^c$ & $n^{z}$ & ${R_b}^{z}$ & $\gamma^{z}$ & $\alpha^{z}$ & ${\chi^2}_{z}$ & ${m_z}^{sky}$\\
  &(arsec) & & (arsec) &  &  &  & (mag\,${arcsec}^{-2}$) &(arsec) & & (arsec) & & & & (mag\,${arcsec}^{-2}$)\\
 \hline
 NGC6482    &   20.2 & 3.9 & &  & & 1.15 & 23.38 &   17.6 & 3.8 &  &  & & 1.21 & 22.65\\
 NGC1132    & 91.5 & 6.6 & 0.8 & 0.13 & 3.12 & 1.13 & 22.92 & 74.1 & 6.6 & 0.81 & 0.11 & 2.92 & 1.1 & 22.02\\
 ESO306-017 & 148.6 & 10.5 & 0.64 & 0 & 2.97 & 1.22 & 23.22& 108.9 & 11.7 & 0.64 & 0.02 & 2.36 & 1.12 & 22.16\\
\hline
\end{tabular}
\caption{Best galaxy surface brightness model parameters in $g$ and $z$ bands:  
single S\'ersic for NGC\,6484 and core-S\'ersic for NGC\,1132 and ESO\,306-017. }
\label{sersic_parameters}
\end{table*}

Elliptical galaxies are well described by S\'ersic profiles unless they are disturbed due to recent interaction (Ferrarese et al.\ 2006). In order to detect any signature of disturbances, we constructed galaxy models based on the S\'ersic parameters obtained from fitting the bmodel profile. 
The best single S\'ersic parameters were used to construct model images using the program {\rm BUDDA} (de Souza et al. 2004). 
Because the fit we obtained is in one dimension, in order to construct the 2-D model we needed to assign an ellipticity and positional angle (PA), which are taken from the {\it ellipse} fit. These values are nearly constant except in the very inner region (see Fig.\,\ref{xy_all}), probably due to dust or inner structures. Thus, we chose the mean values of ellipticity and PA of the outer region isophotes ($r>30$\,arcsec).  Afterwards, these models were subtracted from the original images. 
For NGC\,1132 we see residuals with an appearance of shells; in order to rule out that these are caused by a bad choice of PA value, we constructed model images changing the PA; we recovered the shell-like structure in all cases.

The existence of dust lanes is evident in the inner regions of all three FGs (Fig.\,\ref{zoomcenters}). Furthermore, the residual images (Fig.\,\ref{sersic_residuals}) show evidence of galactic interaction: apparent shells in NGC\,1132, a tidal tail in ESO\,306-017 and an inner disk in NGC\,6482. It must be noted that these features contradict the claims of previous authors, who did not find signs of merger activity (Jones et al. 2003; Khosroshahi et al. 2004). This point shall be discussed in Sec.\,\ref{section_conclusions}.

 \subsection{Globular Cluster selection}
 
Based on their recession velocities, NGC\,6482, NGC\,1132, and ESO\,306-017 are located at 56, 99, and 155 Mpc, respectively (Table\,\ref{sample_basicdata}); GCs which have typical half-light radii of $\sim$ 3 pc (Jord\'an et~al.\ 2005), will be unresolved and appear as an excess of point sources over the foreground stars and small background galaxies. The number of these contaminants can be estimated using control fields.

The detection and photometry of GC candidates were done with SExtractor, using rms maps that
indicate the weight of each pixel according to read noise, cosmic rays and saturated pixels. The
confidence in the detections is naturally greater for brighter objects, but reaching fainter
magnitudes allows a better sampling of the GCLF. Aiming at a compromise between GCLF completeness
and good rejection of spurious detections, we chose a threshold of 1.5 $\sigma$ per pixel, and a
minimum area for a positive detection MINAREA = 5 pixels; together, these constraints imply a
minimum effective detection threshold of 3.35 $\sigma$. An initial sample was constructed with
objects detected independently in both $g$ and $z$ filters, with matching coordinates within a radius of 2 pixels, ellipticity lower than 0.3 and CLASS\_STAR\footnote{SExtractor parameter that classifies the objects 
according to their fuzziness; point sources have CLASS\_STAR $\sim1$, while for extended objects it
is $\sim0$} greater than 0.7.

We adopt the photometric AB system, for which the zero~points are $zp_{g}=26.081$ and $zp_{z}=24.867$\footnote {http://www.stsci.edu/hst/acs/analysis/zeropoints}. These values are valid for an infinite aperture; we use aperture photometry (with a 3 pixel radius), and apply an appropriate aperture correction (Sirianni et al.\ 2005). Magnitudes are corrected for foreground extinction using the reddening maps of Schlegel et al.\ (1998) for each field and filter. The extinction transformations to $g$ and $z$ are $A_g=3.634\,E(B-V)$ and $A_z=1.485\,E(B-V)$. $E(B-V)$ values for NGC\,6482, NGC\,1132, and ESO\,306-017 are 0.099, 0.065, 0.033\footnote {http://irsa.ipac.caltech.edu/applications/DUST}, respectively.

Objects five magnitudes brighter than the expected turnover magnitude, $m^0$, in both filters ($m_{g,z} < {m^0}_{g,z} - 5$, see Table\,\ref{sample_basicdata}), were rejected from the GC sample as they are undoubtedly contaminants. 
With this cut we reject $\lesssim$0.02\% of the brightest GCs, but we avoid foreground stars. 

Objects with reddening-corrected color $0.5<g{-}z<2.0$ were kept; single stellar populations with age in the range 2-15 Gyr and metallicity $-2.25<[{\rm Fe/H}] < +0.56$ lie in this range (C\^{o}t\'e et al. 2004). Finally, the selected sample of GC candidates was cleaned of objects with an uncertainty in magnitude larger than 0.2 mag in order to have objects with reliable magnitudes.\\

\begin{figure}[h!]
\centering
\resizebox{0.7\hsize}{!}{\includegraphics[angle=-90]{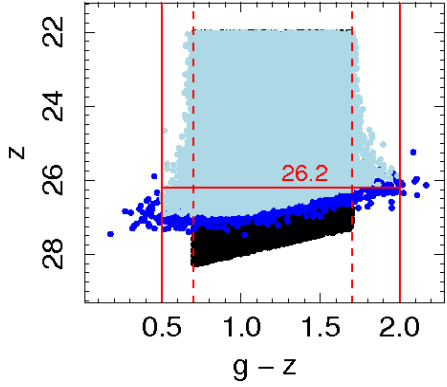}}
\caption{Color-magnitude diagram of added stars. {\it Black points:} input values; {\it dark-blue points:} all recovered objects; {\it light-blue points:} applying same constraints as for selection of GC sample for ESO\,306-017.}
\label{ESO306017_CMDaddstar}
\end{figure}

\begin{figure}[!h]
\resizebox{1.0\hsize}{!}{\includegraphics[angle=-90]{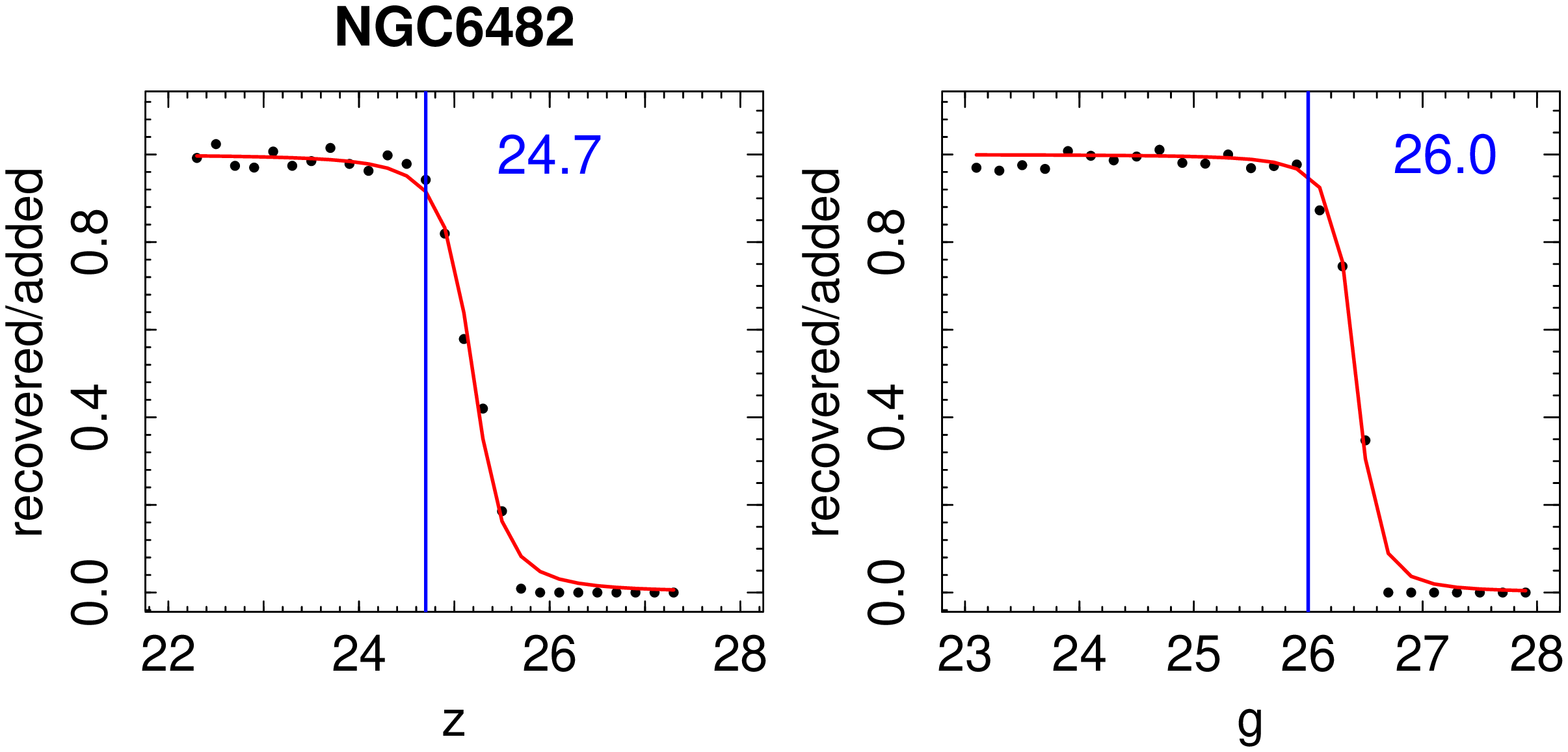}}
\resizebox{1.0\hsize}{!}{\includegraphics[angle=-90]{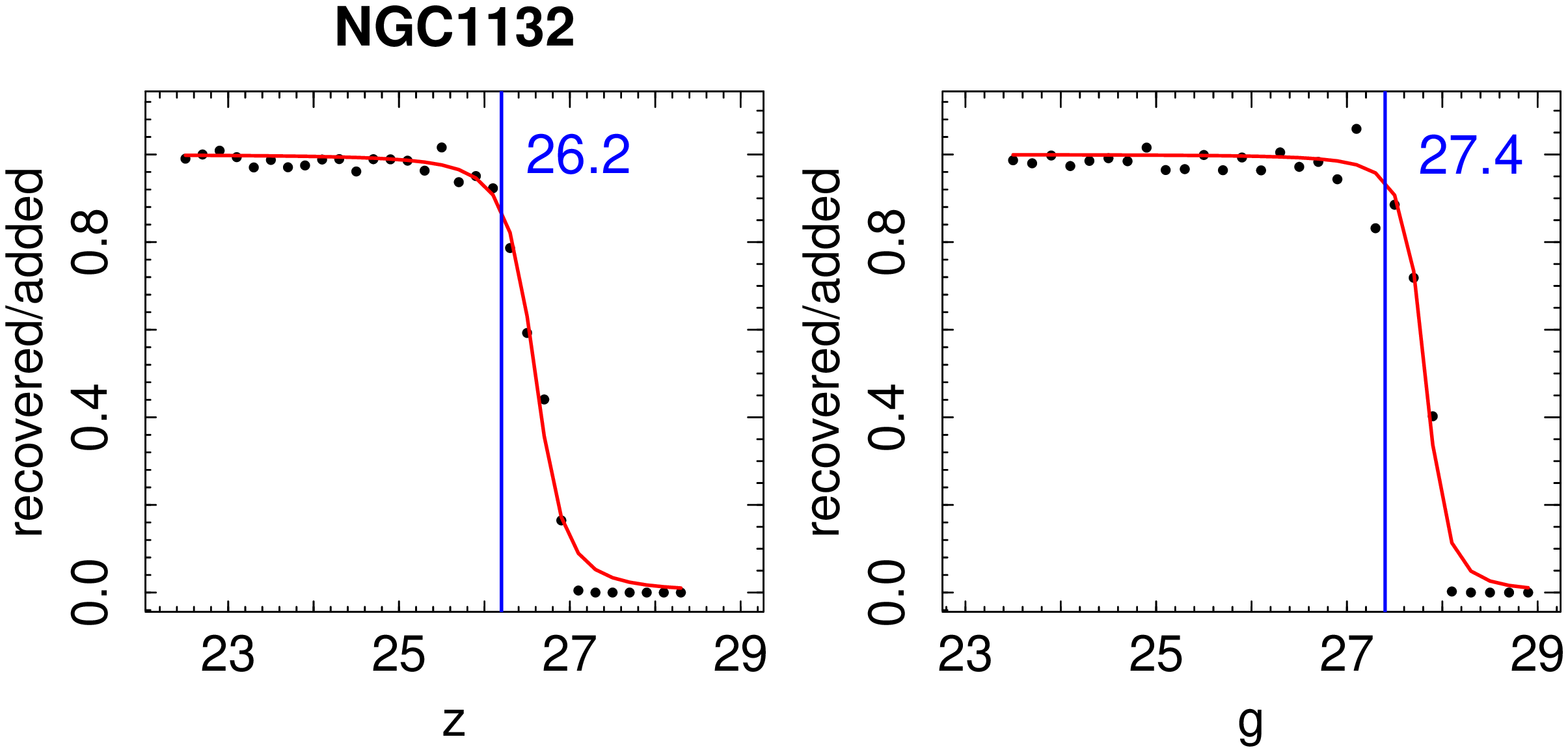}}
\resizebox{1.0\hsize}{!}{\includegraphics[angle=-90]{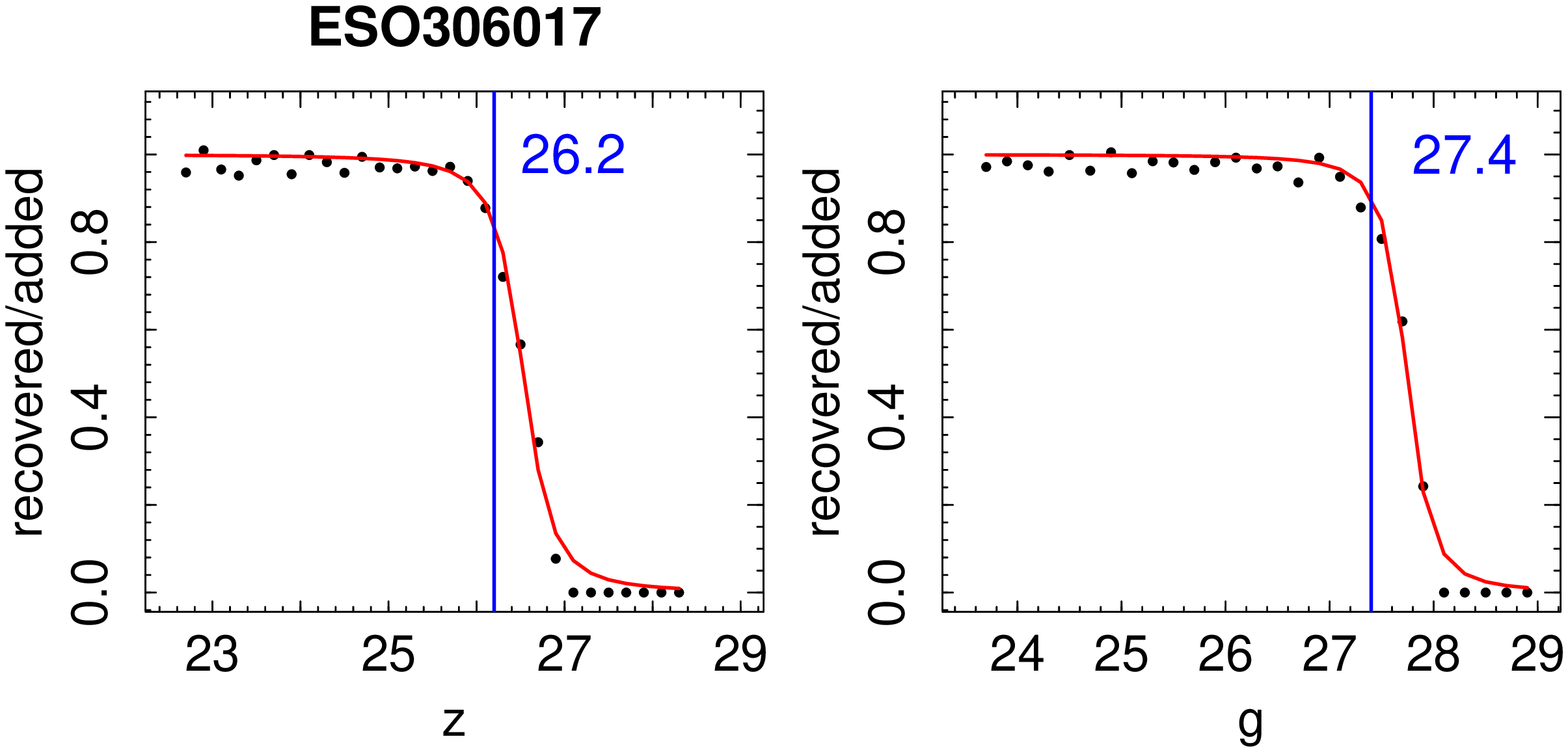}}
\caption{Completeness tests. 
The dots are the fraction of recovered artificial stars as a function of
  magnitude. The solid red line in each panel shows the fitted Pritchet function. The blue line and number indicate where the data are nearly 100\% complete. }
\label{comptest}
\end{figure}

To determine the GC detection completeness as a function of magnitude, $\sim25,000$ artificial stars were constructed based on the PSF of each image, with magnitudes in the interval $22<m_{star}<29$, and color $0.7<g_{star}-z_{star}<1.7$, where most GCs lie. This color range is slightly narrower than the color criteria for GCs because we needed to take into account the dispersion in the magnitude measurements (Fig.\,\ref{ESO306017_CMDaddstar}). 
Only $\sim100$ stars were added randomly at a time, 
in order to avoid crowding and overlapping with original objects. Afterwards, each image was analyzed exactly the same way as the original data. Pritchet curve (Fleming et al. 1995) fitted to the fraction of recovered stars versus magnitude provides our {\it completeness function} (Fig.~\ref{comptest}). 
Based on the completeness test we are confident that the detection is $\gtrsim\,$90\% complete at $z=24.7$, 26.2 and 26.2 for NGC\,6482, NGC\,1132, and ESO\,306-017, respectively.  These values are not as deep as the expected turnover magnitudes (see Table\,\ref{sample_basicdata}), but we prefer to be cautious on the reliability rather than try to push the magnitude limit.

\begin{figure}[ht]
\centering
\includegraphics[width=0.9\textwidth,angle=-90]{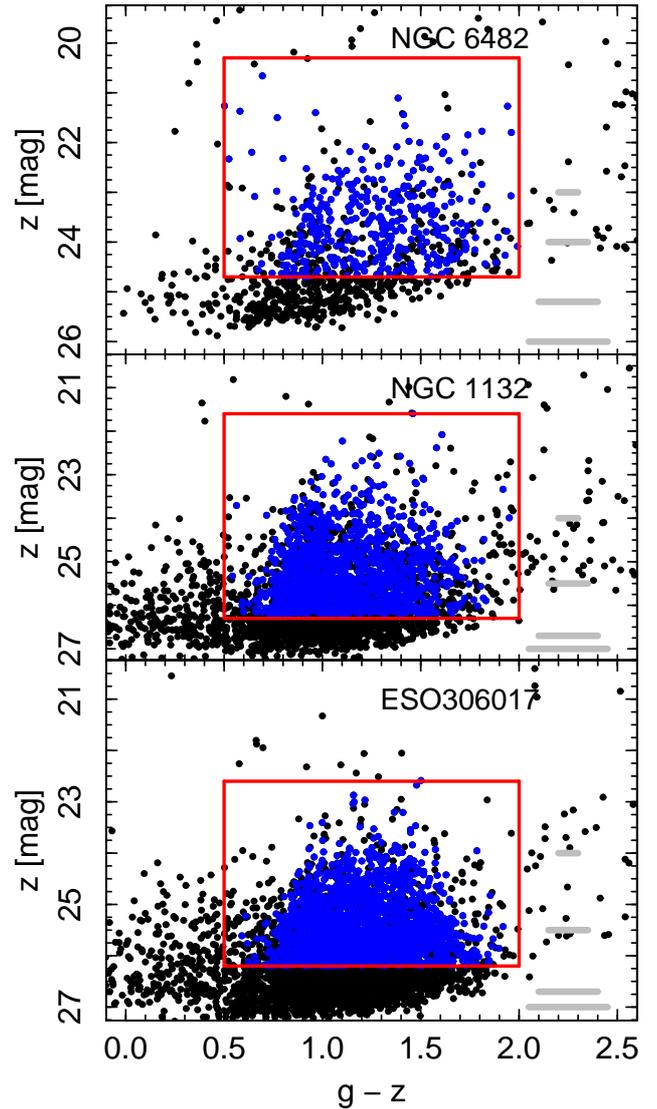}
\caption{Color-magnitude diagram of objects in the FG fields. The black dots are all the
    detected objects in the FOV. The thick gray lines show the error bars in color at different
    magnitudes. The blue dots framed within the red lines are the selected sample of GC
    candidates with color range $0.5< g{-}z < 2.0$. The bright magnitude limit is at
    $m_z = {m_z}^0 - 5$; beyond the faint magnitude limit, there is a dependence of the
    detection limit on color, and redder objects are missed. {\it Top to Bottom}: NGC\,6482, NGC\,1132 and ESO\,306-017. }
\label{cmd_FGall}
\end{figure}

The upper limit in luminosity might still include some ultra-compact dwarfs (UCDs), but we assume that those are just the brightest GCs in the distribution. Their nature is not an issue for this paper (see Madrid 2011). 
The number of detected GCs by direct photometry, without any spatial, luminosity or
contamination correction, are 369, 1410, and 1918 for NGC\,6482, NGC\,1132, and ESO\,306-017,
respectively. In Fig.~\ref{cmd_FGall} we show the color-magnitude diagram of all objects detected
in both filters ({\it black points}), indicating the GC selected sample ({\it blue points}); there
is a clear excess of point sources within the color range expected for GCs in all these fields. 

As mentioned before, some contamination is expected due to foreground stars and unresolved
background galaxies. To estimate the number of these contaminants, $N_{cont}$, the above detection
procedure and selection criteria were applied to three ``blank'' background fields obtained from
the {\sl HST} archive that were chosen to have Galactic latitudes and exposure times in $g$ and $z$
similar to our two deeper fields (NGC\,1132 and \esogxy), but without any large galaxies present in
the FOV.

In order to simulate the same detection conditions for the empty fields, we applied the rms maps of
the FG fields to the background fields. The number of contaminants is taken to be the average of
the three fields. We find $N_{cont}=7$, 6 and 7 for NGC\,6482, NGC\,1132, and
ESO\,306-017, respectively; these values correspond to $\sim$1\% or less of the number of detected
GC candidates, which is negligible. We note that NGC\,6482 has Galactic latitude
$b\approx23^\circ$, about $10^\circ$ less than that of the lowest latitude control field.  Thus, its
contamination may be somewhat higher, but not more than the few per~cent level, as supported by
the steep spatial distribution found below for the GC candidates in this galaxy.

\section{Analysis and Results}
 
\subsection{Globular Cluster Color Distribution}
\label{section_GCcolor}
In order to detect and quantify the existence of colour bimodality, the data were binned in $(g-z)$ using an optimum bin size, $B_{\rm opt}$ (Izenman 1991; Peng et al.\ 2006), that depends on the sample size, $n$, and inter-quartile range, $IQR$ which is a measure of the dispersion of the distribution:

\begin{equation}
B_{\rm opt} = 2 (IQR)n^{-1/3}.
\label{binsize}
\end{equation}

Afterwards, we used the algorithm Gaussian Mixture Modeling (GMM), which identifies Gaussian distributions with different parameters inside a dataset through the expectation maximization (EM) algorithm (Muratov \& Gnedin 2010). GMM also performs independent tests of bimodality (measure of means separation and kurtosis), estimates the error of each parameter with bootstrapping, and estimates the confidence level at which a unimodal distribution can be rejected.

\begin{figure}[ht]
\centering
\includegraphics[width=0.35\textwidth,angle=-90]{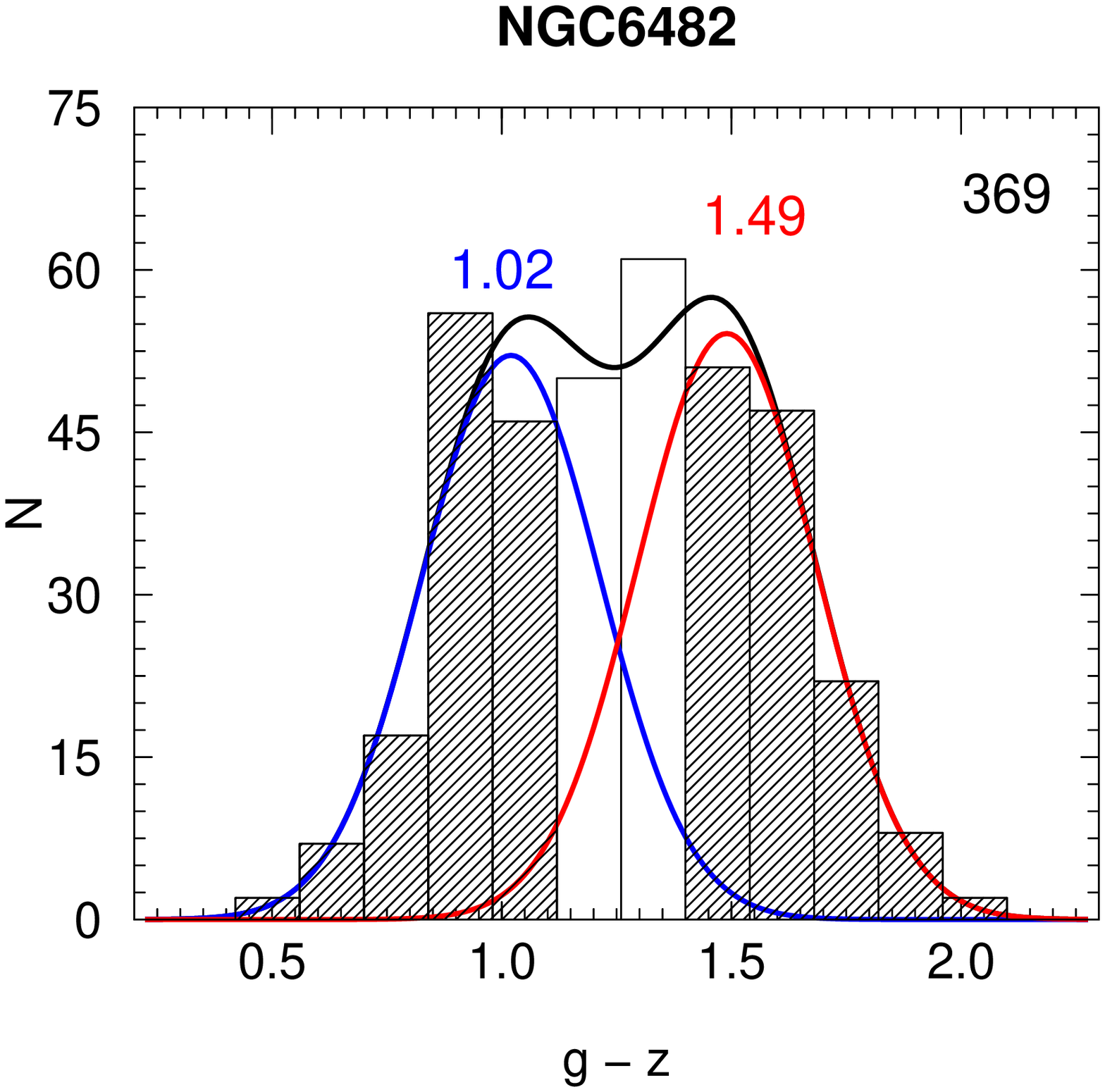}
\includegraphics[width=0.35\textwidth,angle=-90]{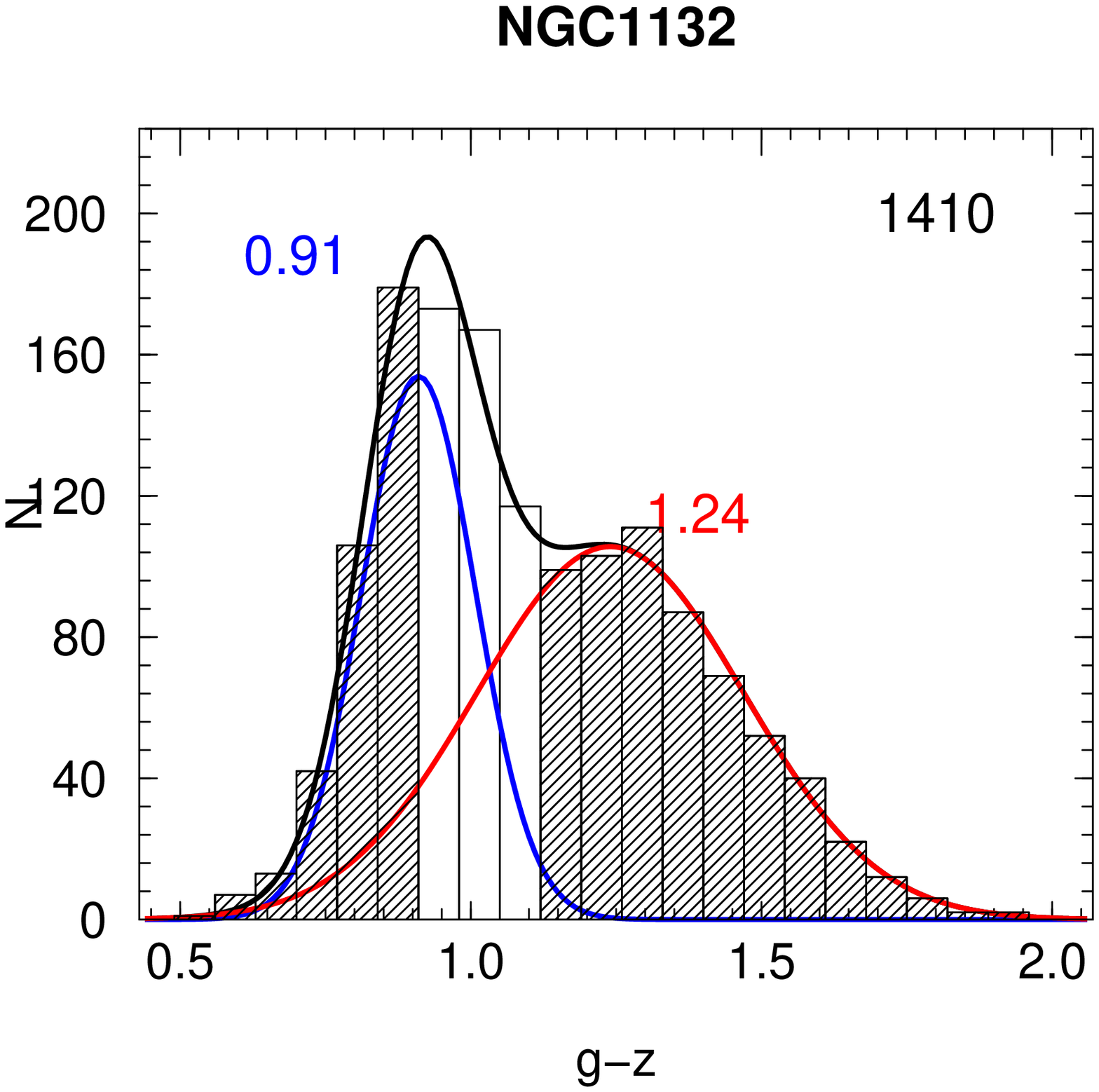}
\includegraphics[width=0.35\textwidth,angle=-90]{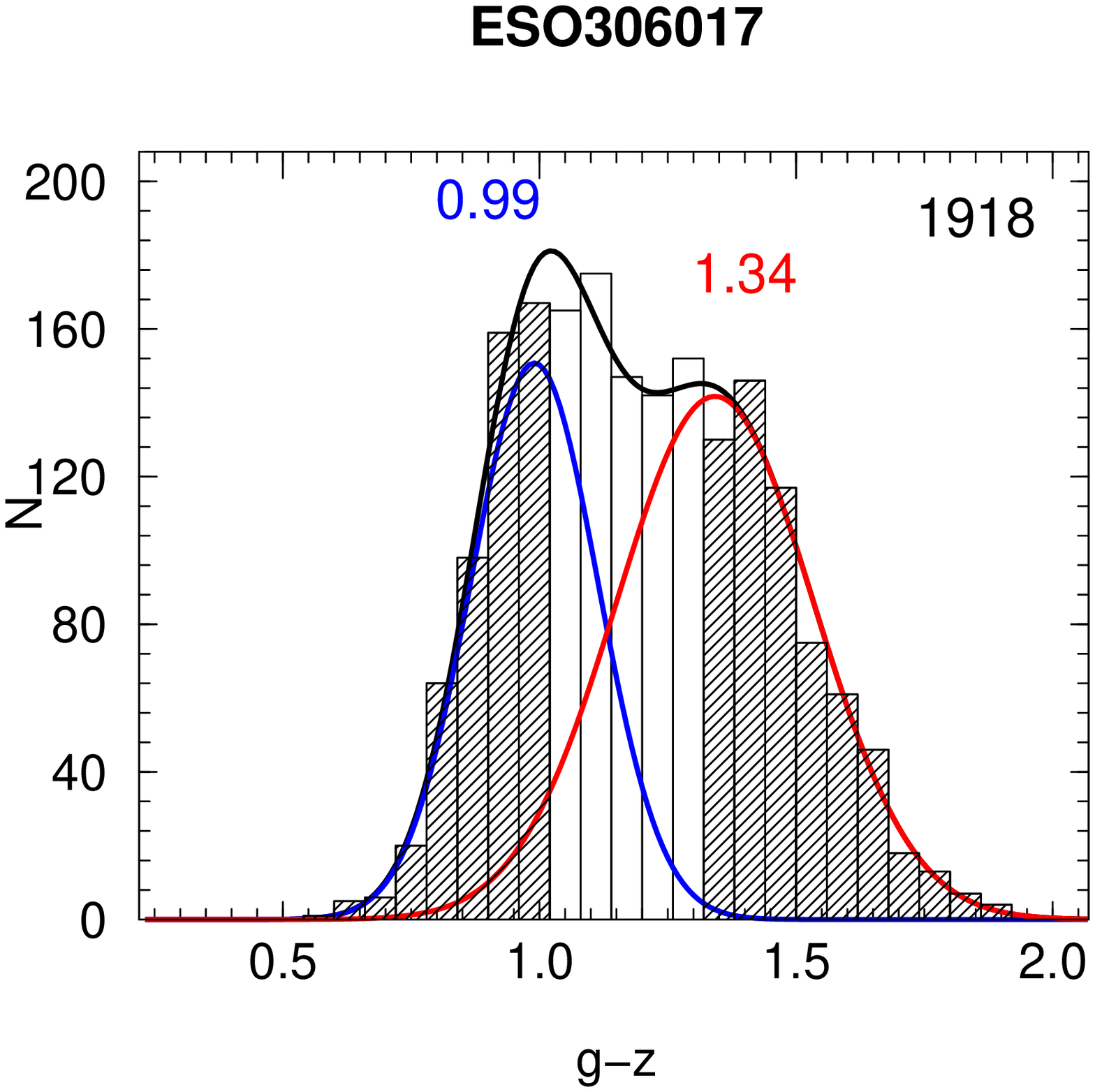}
 \caption{Color histograms with the best GMM two-Gaussian fit. The individual Gaussians with their mean value are indicated with 
a blue or red number and solid curve; their sum is the black solid line.
The hatched areas indicate the objects we include in the blue and red populations. 
The number of GCs is shown in the upper right of each panel.}
  \label{mclust_fits}
\end{figure}

Figure\,\ref{mclust_fits} shows the ($g - z$) color distributions with the best fit Gaussians obtained with GMM. For NGC\,6482 the best model is homoscedastic (same variance), while for NGC\,1132 and ESO\,306-017, a heteroscedastic (different variance) model provides a significantly better fit.
The means, $\mu_{color}$, and dispersions, $\sigma_{color}$, of the best fitting Gaussians are listed in Table~\ref{kmm_parameters}; they are consistent with parameters in the literature (Peng et al.\ 2006). In Table~\ref{kmm_parameters}, Cols. 9 and 10 
we list the probability of obtaining the recovered values from a unimodal Gaussian distribution. 
Since these probabilities are small we are confident that the color distributions are bimodal relative to a null hypothesis of Gaussian unimodality.\\

\begin{table*}
\centering
\begin{tabular}{c  c  c  c  c  c  c  c  c  c  c  p{1in}}
\multicolumn{11}{c}  {}  \\
\hline \hline
  $System$ & $\mu_{\rm blue}$ & $N_{\rm blue}$ & $\sigma_{\rm blue}$& $\mu_{\rm red}$ & $N_{\rm red}$ & $\sigma_{\rm red}$ & $ \overline{g-z} $ & p-value & D & k  \\
  \hline
 NGC\,6482 & 1.02 $\pm$ 0.02 & 181 $\pm$ 16 & 0.19 $\pm$ 0.02 & 1.49 $\pm$ 0.02 & 188 $\pm$ 16 & 0.19 $\pm$ 0.02 & 1.26 & 0.001 & 2.42 (0.001) & -0.7 \\
 NGC\,1132 & 0.91 $\pm$ 0.01 & 540 $\pm$ 87 & 0.10 $\pm$ 0.01 & 1.23 $\pm$ 0.03 & 870 $\pm$ 87 & 0.23 $\pm$ 0.02 & 1.11 & 0.01 & 1.88 (0.13) & -0.41 \\
 ESO\,306-017 & 0.99 $\pm$ 0.02 & 781 $\pm$ 124 & 0.12 $\pm$ 0.01 & 1.34 $\pm$ 0.03 & 1137 $\pm$ 124 & 0.19 $\pm$ 0.01 & 1.20 & 0.01 & 2.17 (0.08) & -0.59\\
\hline
\end{tabular}
\caption{Parameters of the best GMM fit. {\it Col.\,1}: System name; {\it Cols.\,2}, 3 and 4: mean value, size and dispersion of blue population; {\it Cols.\,5}, 6 and 7: mean value, size and dispersion of red population;
 {\it Col.\,8}: mean color of the GC population; 
 {\it Col.\,9}: the probability of recovering the same $\chi^2$ value from a unimodal distribution;
 {\it Col.\,10}: separation of the means, D=$\mid\mu_{\rm blue} -\mu_{\rm red}\mid / [(\sigma_{\rm blue}^2 + \sigma_{\rm red}^2)/2]^{1/2}$, in parenthesis the probability of recovering the same value from a unimodal distribution; 
 {\it Col.\,11}: kurtosis of the color distribution.}
\label{kmm_parameters}
\end{table*}

\subsection{Globular Cluster Spatial Distribution}
\label{gc_spatialDistribution_section}

Harris (1991) noted that in general the GC systems are more extended than the starlight in their parent galaxies. Furthermore, it is known that the blue and red GCs have different spatial distribution, where the blue GCs have a more extended distribution than the red GCs (Geisler et al. 1996; Rhode \& Zepf 2004). 
To see how the distributions compare in these FGs, 
we first need to select the blue and red GC subsamples.\\

\begin{figure}[ht]
\centering
\includegraphics[width=0.27\textwidth,angle=-90]{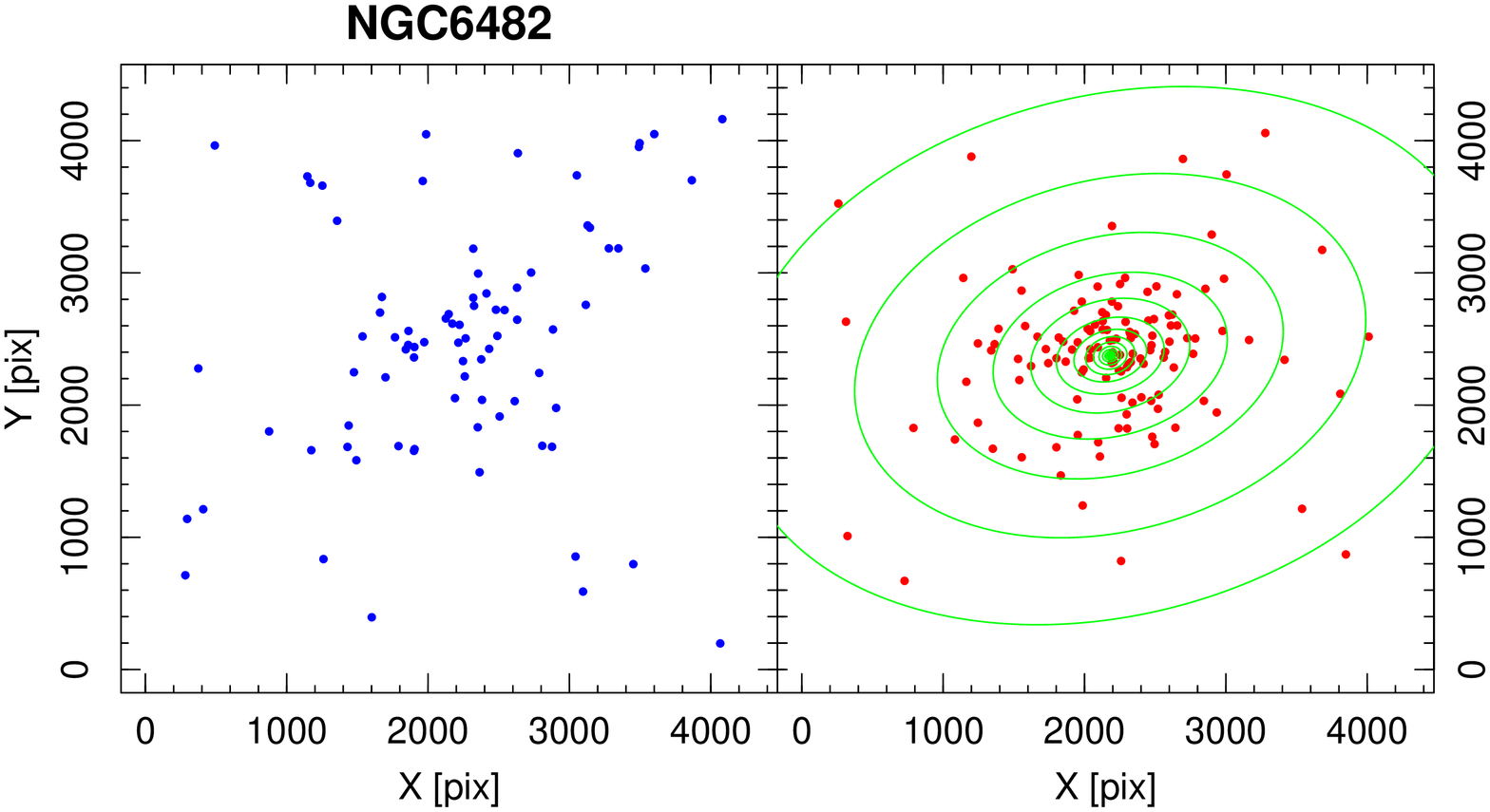}
\includegraphics[width=0.27\textwidth,angle=-90]{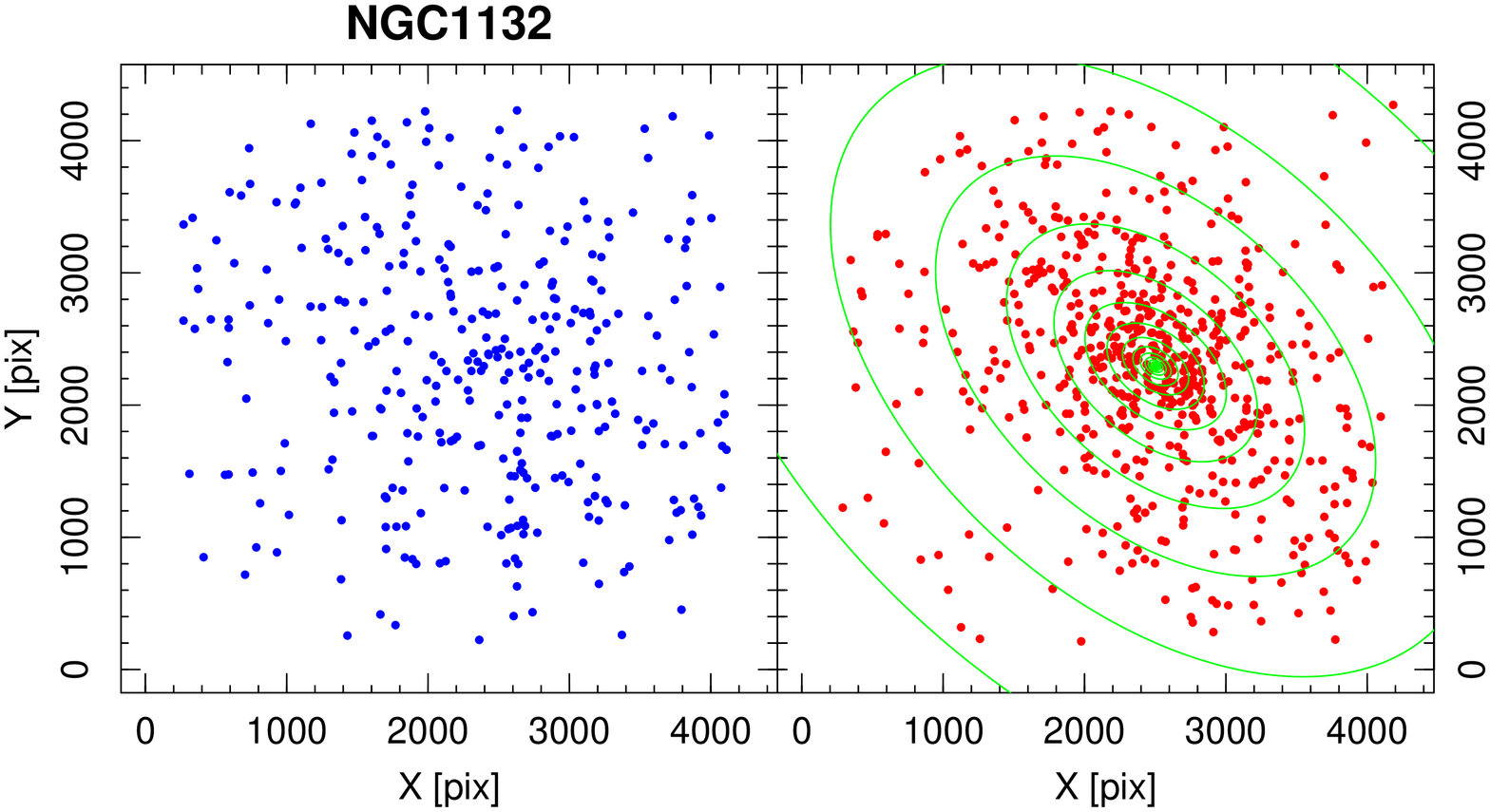}
\includegraphics[width=0.27\textwidth,angle=-90]{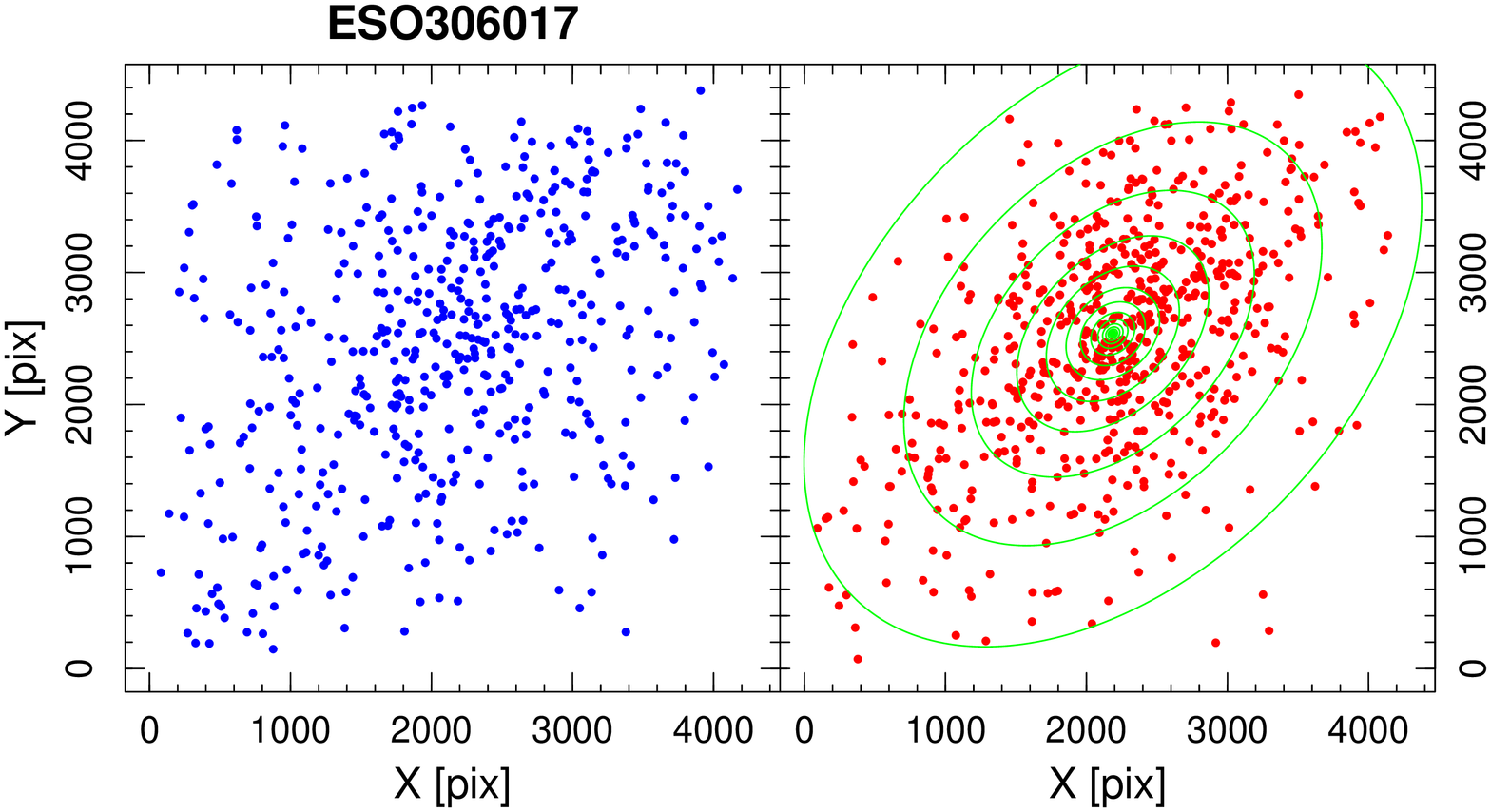}
\caption{Spatial distribution of blue and red populations selected from the color histograms (hatched regions in Fig\,\ref{mclust_fits}). Along with the red population, green ellipses show the starlight distribution.}
\label{xy_all}
\end{figure}

In Fig.\,\ref{mclust_fits}, we see that the two Gaussian components of the color distributions overlap; in order 
to avoid cross-contamination between the two populations, we reject objects in the overlapped regions, selecting blue and red population as indicated by the hatched regions. 
Because the mixed fraction are different for the three FGs color distributions, the cuts are different for each one: $\mu_{\rm blue} + 0.1$ for NGC\,6482 and $\mu_{\rm blue} $ for NGC\,1132 and ESO\,306-017; $\mu_{\rm red} - 0.1$ for NGC\,6482 and NGC\,1132 and $\mu_{\rm red}$ for ESO\,306-017 (see Fig.\,\ref{mclust_fits}). Figure\,\ref{xy_all} shows the spatial distribution of blue and red GCs thus defined.\\

\begin{figure}[ht]
\centering
\includegraphics[width=0.37\textwidth,angle=-90]{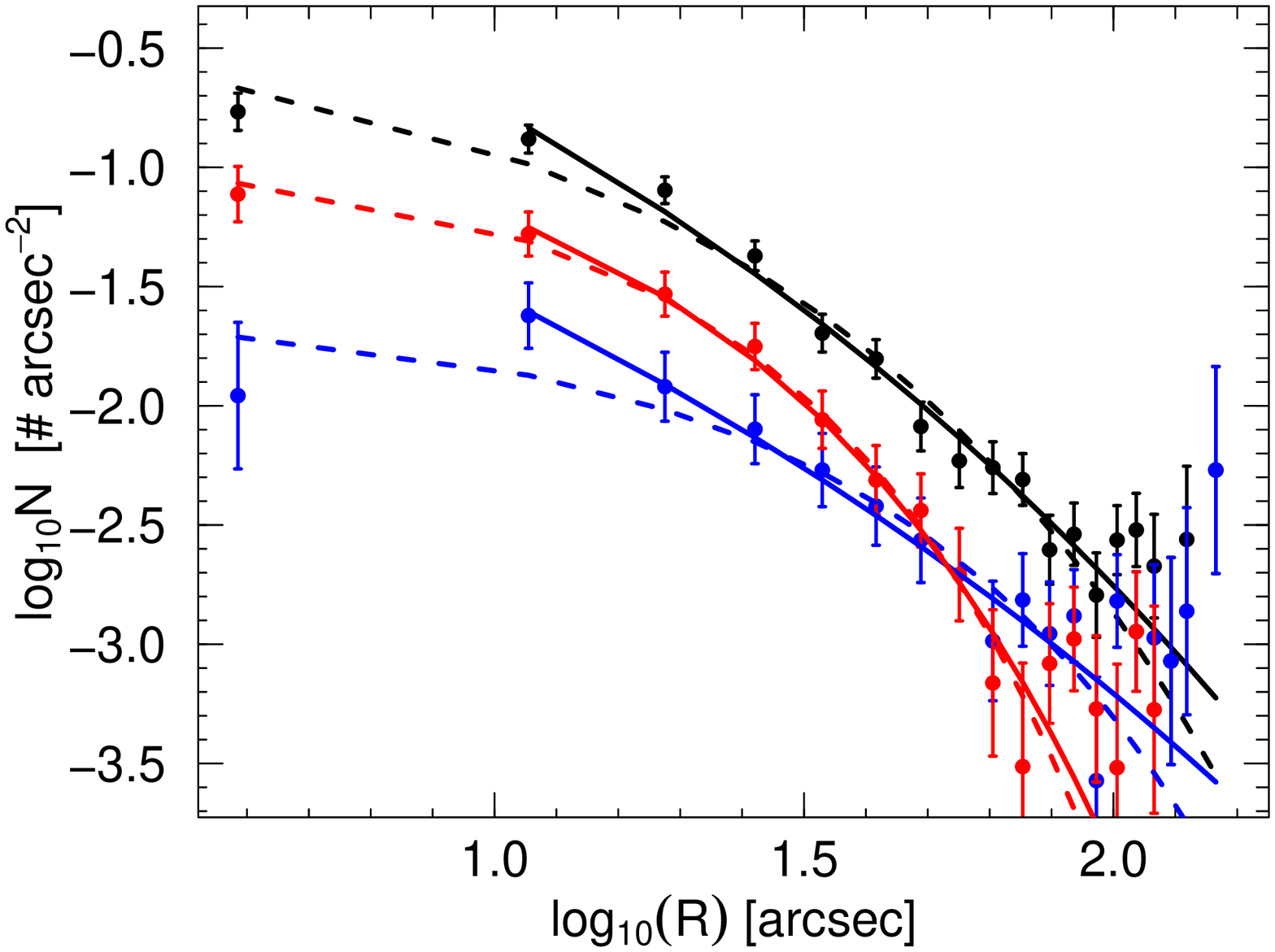}
\includegraphics[width=0.37\textwidth,angle=-90]{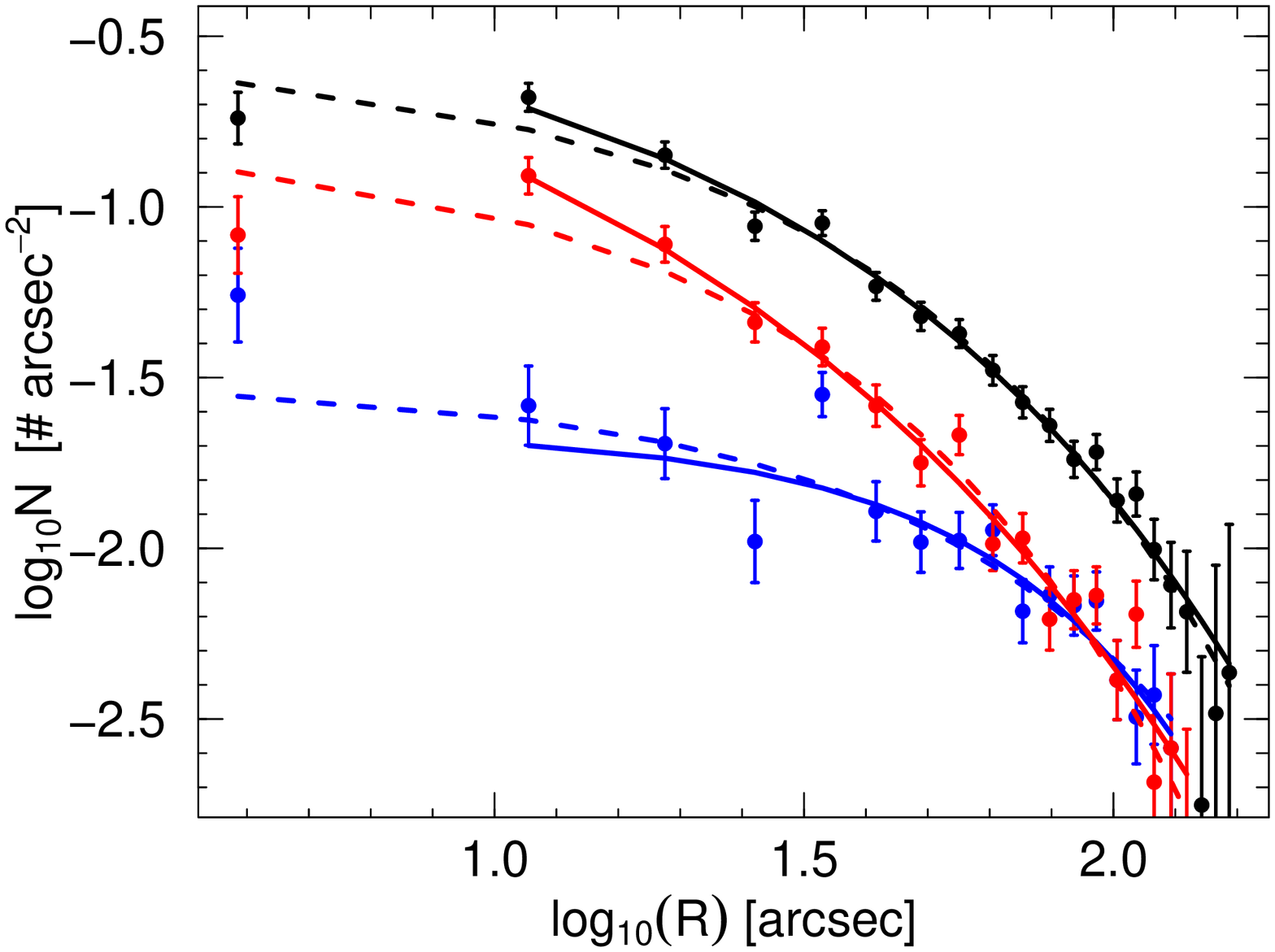}
\includegraphics[width=0.37\textwidth,angle=-90]{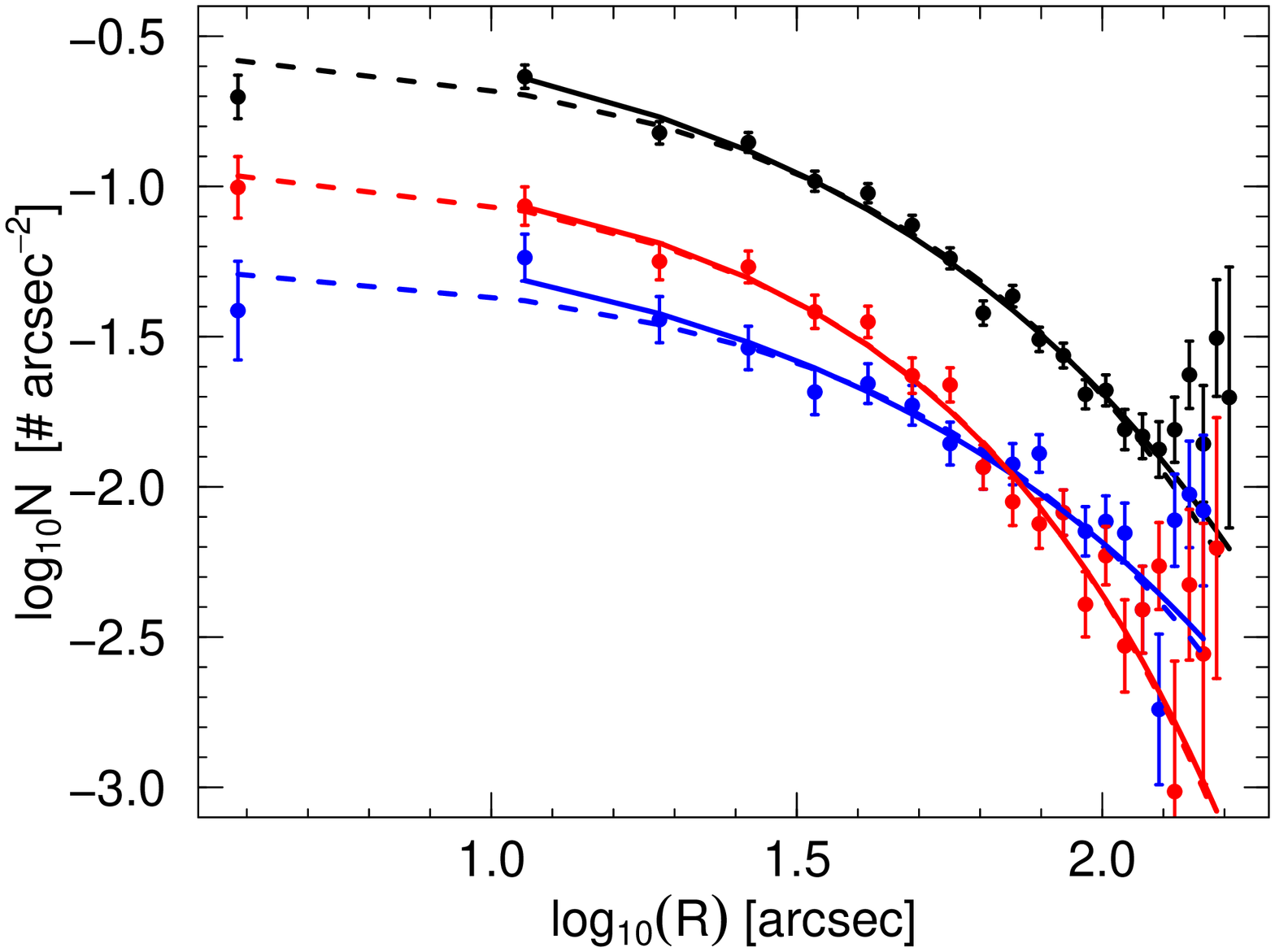}
\caption{Surface number density of GCs. The dots are the data and the lines show the best S\'ersic fits including all the points (dashed lines) and excluding the innermost point (solid lines) for all the GCs detected (black), only blue GCs (blue), and only red GCs (red).
{\it Top to Bottom}: NGC\,6482, NGC\,1132 and ESO\,306-017.}
\label{spatialdist_sersic_all}
\end{figure}

We measured the surface number density $N$ of the blue and red populations by counting the number of GCs inside circular annuli with a width of $7\farcs5=$150\,pixels. To compare the GC spatial distribution with the starlight profiles, we fitted S\'ersic functions to the radial $N$ profiles. The fitted function is:

\begin{equation}
N(r)= N_0\exp\left\{-b_n \left[ {\left( \frac{r}{{R_e}^{\rm GC}}\right) }^{1/n_{\rm GC}}- 1\right] \right\} 
\end{equation} 

\noindent
where ${R_e}^{\rm GC}$ is the effective radius, that contains half of the GC population; $N_0$ is the surface number density at ${R_e}^{\rm GC}$, $n_{\rm GC}$ is the S\'ersic index. 
We performed a second fit excluding the innermost point since 
the GCs closest to the galaxy center are more strongly affected by dynamical friction. 
In Fig.\,\ref{spatialdist_sersic_all}, we show the density profiles of all, blue, and red GCs, together with the best S\'ersic fits. Table\,\ref{GCsersic_table} shows the fitted parameters. 
As expected, the difference between both fits is larger for the closest system, NGC\,6482 because we are looking at smaller physical scales than for NGC\,1132 and ESO\,306-017.

\begin{figure}[ht]
\centering
\includegraphics[width=0.37\textwidth,angle=-90]{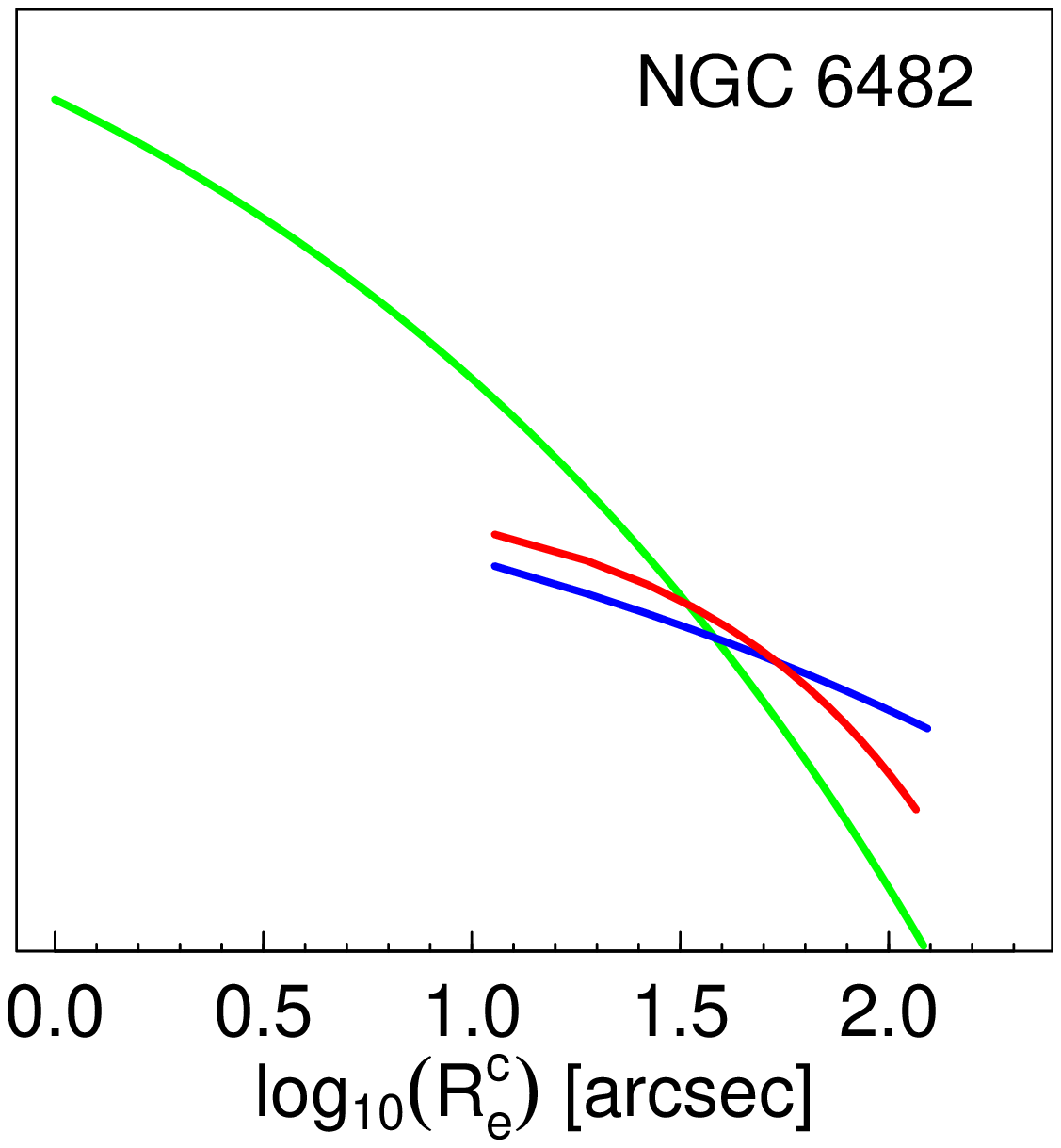}
\includegraphics[width=0.37\textwidth,angle=-90]{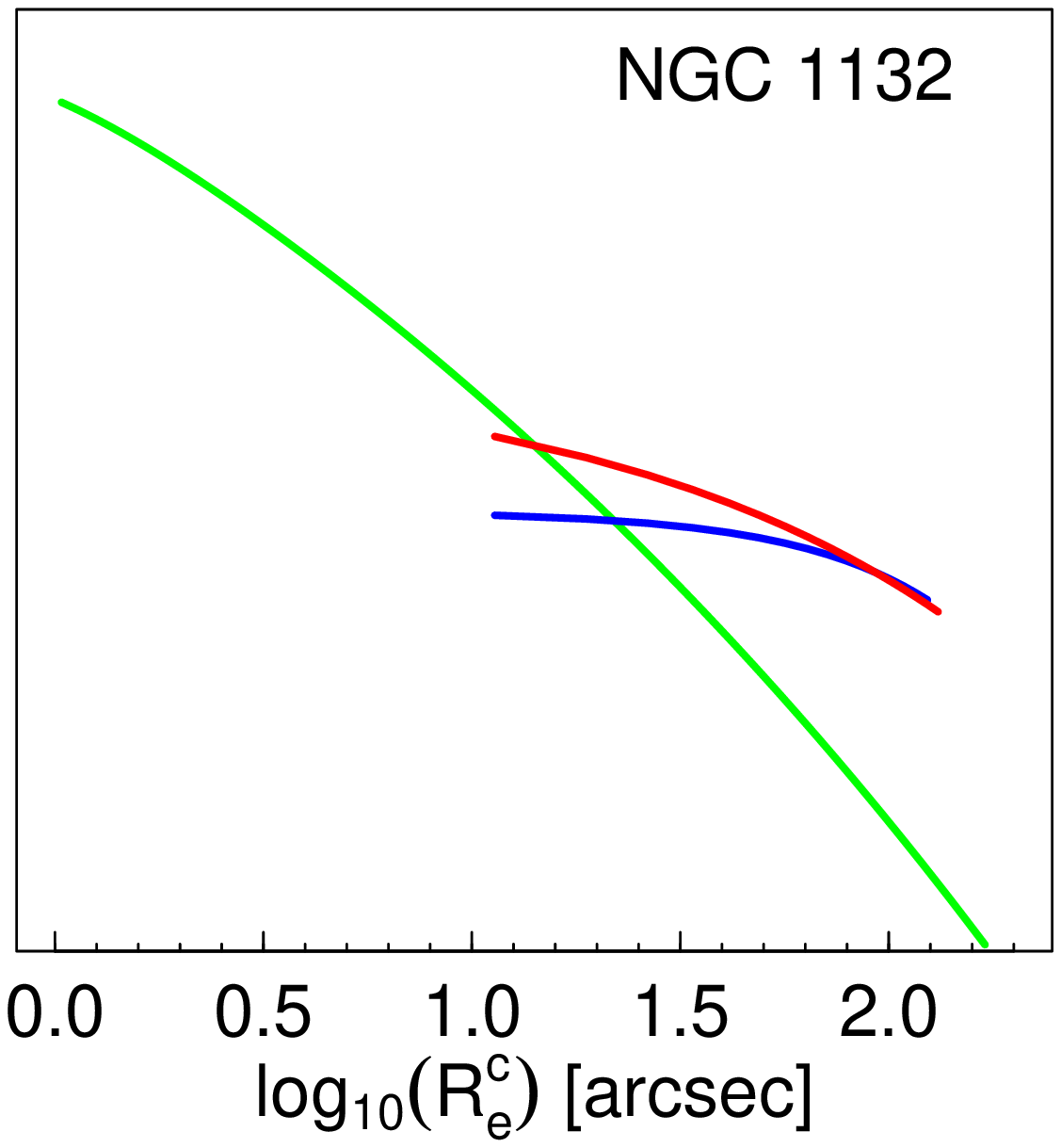}
\includegraphics[width=0.37\textwidth,angle=-90]{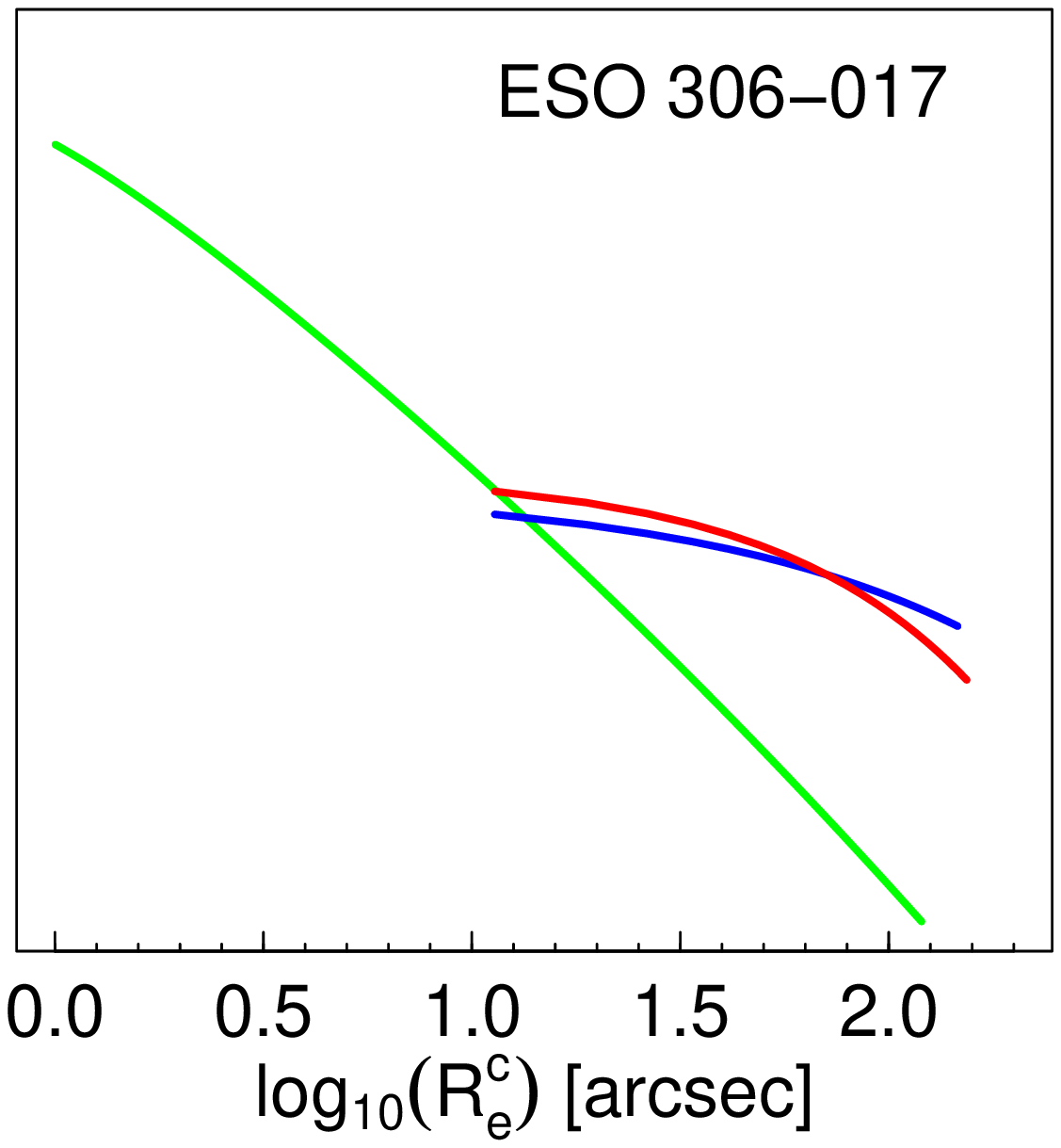}
\caption{
Comparison of S\'ersic profiles of the galaxy in the $z$ band (green line), and of
  the surface number density of blue (blue line) and red (red line) GC populations,
  excluding the inner point in the fit.  The relative scale of the galaxy and GC
  profiles is arbitrary. 
{\it Top to bottom:} NGC\,6482, NGC\,1132 and ESO\,306017.}
\label{sersic_light_gc}
\end{figure}

To compare the S\'ersic profiles of the galaxy light (Fig.\,\ref{spatialdist_sersic_all}) with the GC number density (Fig.\,\ref{all_muPROFILES}), in Fig.\,\ref{sersic_light_gc}, we plot them together using an arbitrary relative scale. 
From this plot we can see that the GC distribution in each galaxy is more extended than its starlight. Nevertheless, we must remember that the comparison of $R_e$ is accurate only if the profiles have similar n.

We find that ${R_e}^{\rm blue}$ is larger than ${R_e}^{\rm red}$, and Figs.\,\ref{xy_all} and \ref{spatialdist_sersic_all} show that the blue populations are more extended than the red ones for these three FGs. 
Also, the distribution of the red populations are more elongated and seem to trace the shape of the galaxy. 

\begin{table*}[ht!]
\centering
\begin{tabular}{c  c  c  c  c  c  c  c  c  c p{1in}}
\multicolumn{10}{c}  {GC surface density S\'ersic parameters} \\
\hline \hline
System & ${R_e}^{\rm all}$ & $n^{\rm all}$  & ${\chi^2}_{\rm all}$ & ${R_e}^{\rm blue}$ & $n^{\rm blue}$  & ${\chi^2}_{\rm blue}$ & ${R_e}^{\rm red}$ & $n^{\rm red}$ & ${\chi^2}_{\rm red}$\\
 & $(arcsec)$ &  & & (arcsec) &  & & (arcsec) & & \\
\hline
\multicolumn{10}{c}  {Including all the points in the fit} \\
\hline
NGC\,6482
 & 32.03 & 1.60 & 1.75
 & 45.73 & 1.17 & 1.04
 & 22.66 & 1.01 & 0.95\\
NGC\,1132 
& 65.68 & 1.27 & 1.36
& 99.06 & 1.09 & 1.85
& 50.70 & 1.21 & 2.10\\
ESO\,306-017
 & 70.29 & 1.19 & 1.96
 & 88.65 & 1.16 & 1.37 
 & 50.57 & 1.04 & 1.24\\
\hline
\multicolumn{10}{c}  {Excluding the innermost point in the fit} \\
\hline
NGC\,6482
 & 27.91 & 3.66 & 1.04
 & 60.64 & 4.40 & 0.75
 & 21.00 & 1.29 & 0.97 \\
NGC\,1132 
 & 70.36 & 1.59 & 1.11
 & 82.73 & 0.77 & 1.74
 & 54.11 & 2.01 & 1.33 \\
ESO\,306-017
 & 77.06 & 1.49 & 1.65
 & 106.38 & 1.56 & 1.36
 & 50.77 & 1.09 & 1.31 \\
\hline
\end{tabular}
\caption{Parameters of the best S\'ersic fit for all GCs, red, and blue populations.}
\label{GCsersic_table}
\end{table*}

\subsection{Globular Cluster Luminosity Function}
  \label{sec:gclf}

\begin{figure*}[ht]
\centering
\includegraphics[width=0.4\textwidth,angle=-90]{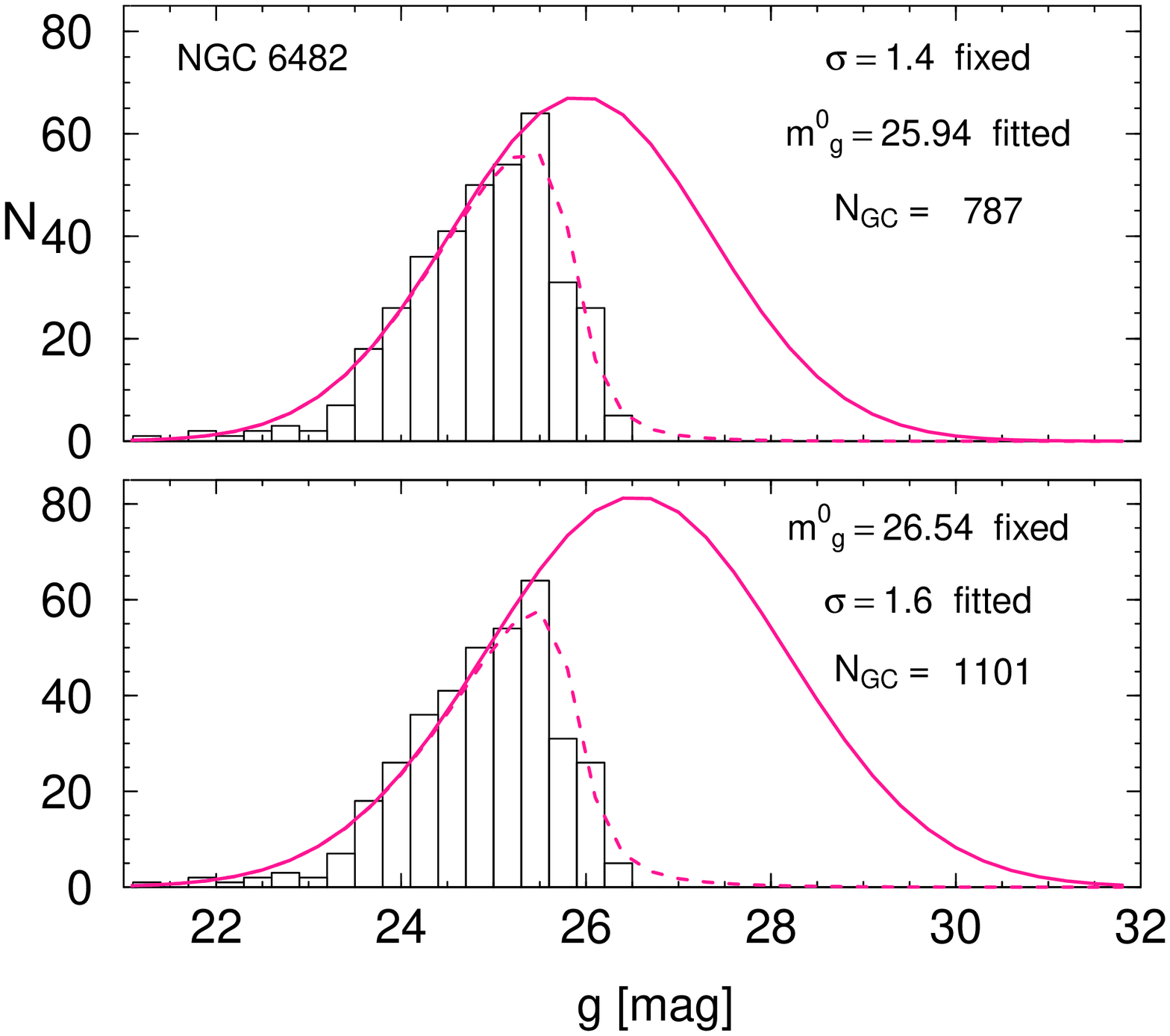}
\includegraphics[width=0.4\textwidth,angle=-90]{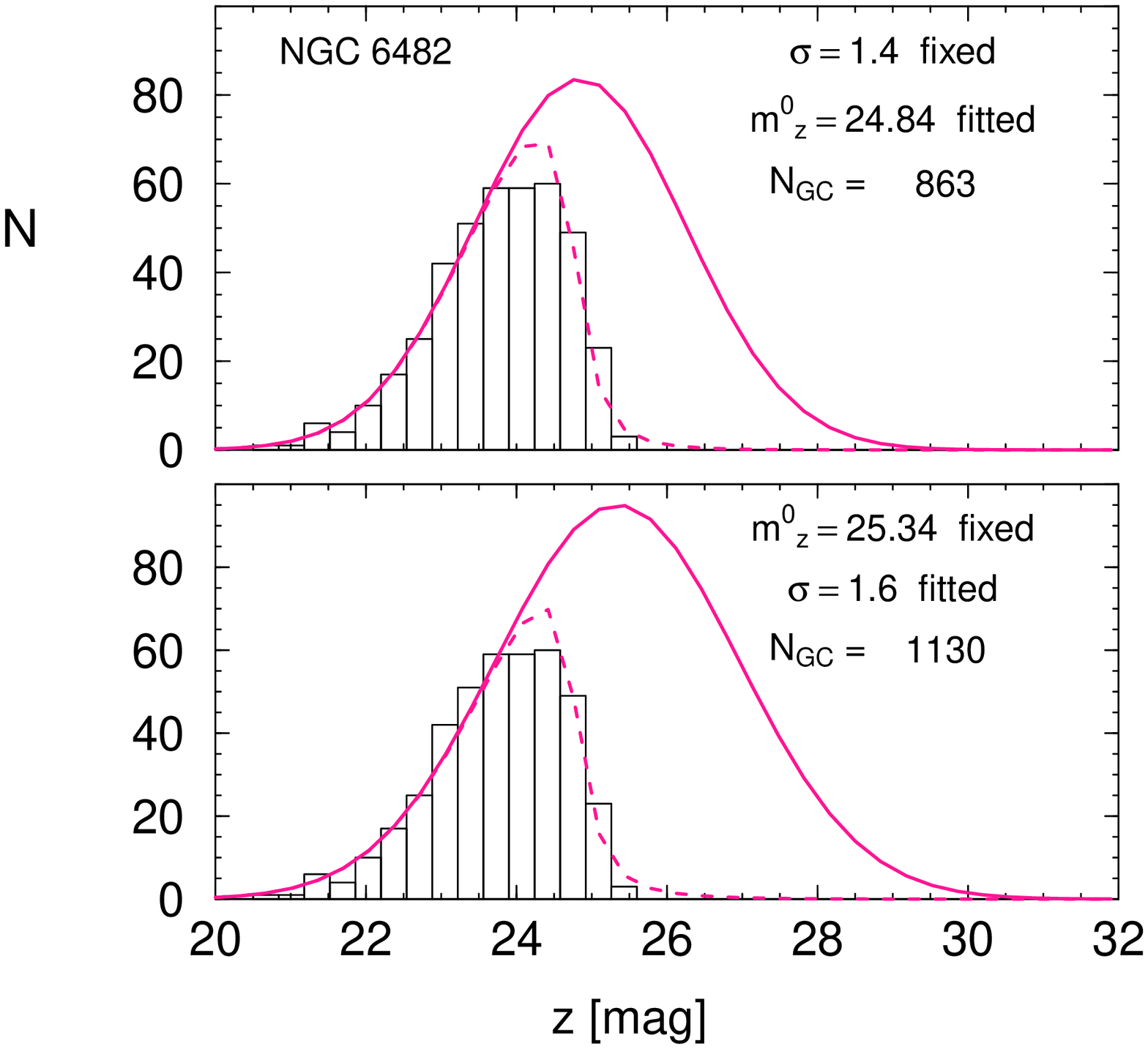}
\includegraphics[width=0.4\textwidth,angle=-90]{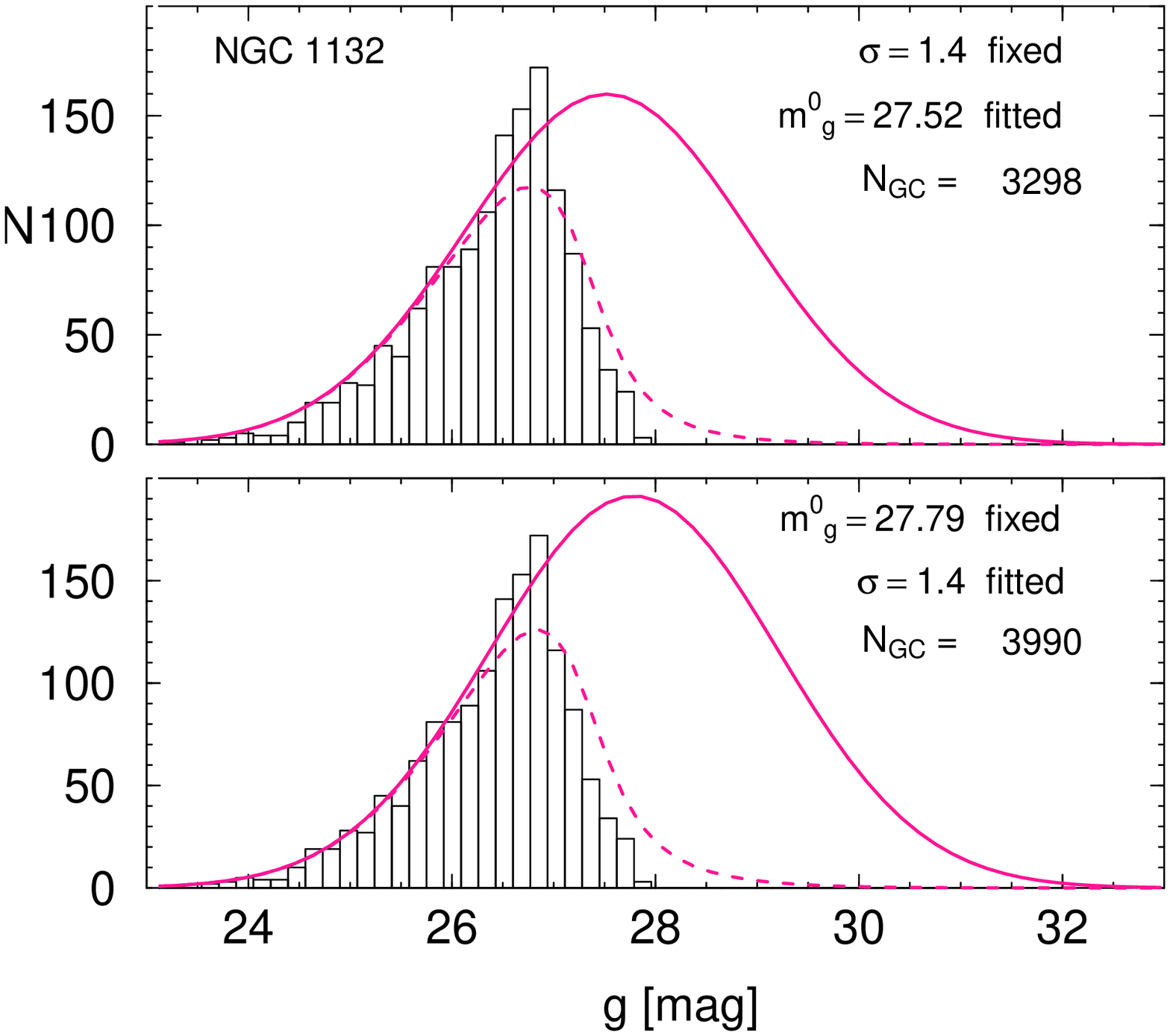}
\includegraphics[width=0.4\textwidth,angle=-90]{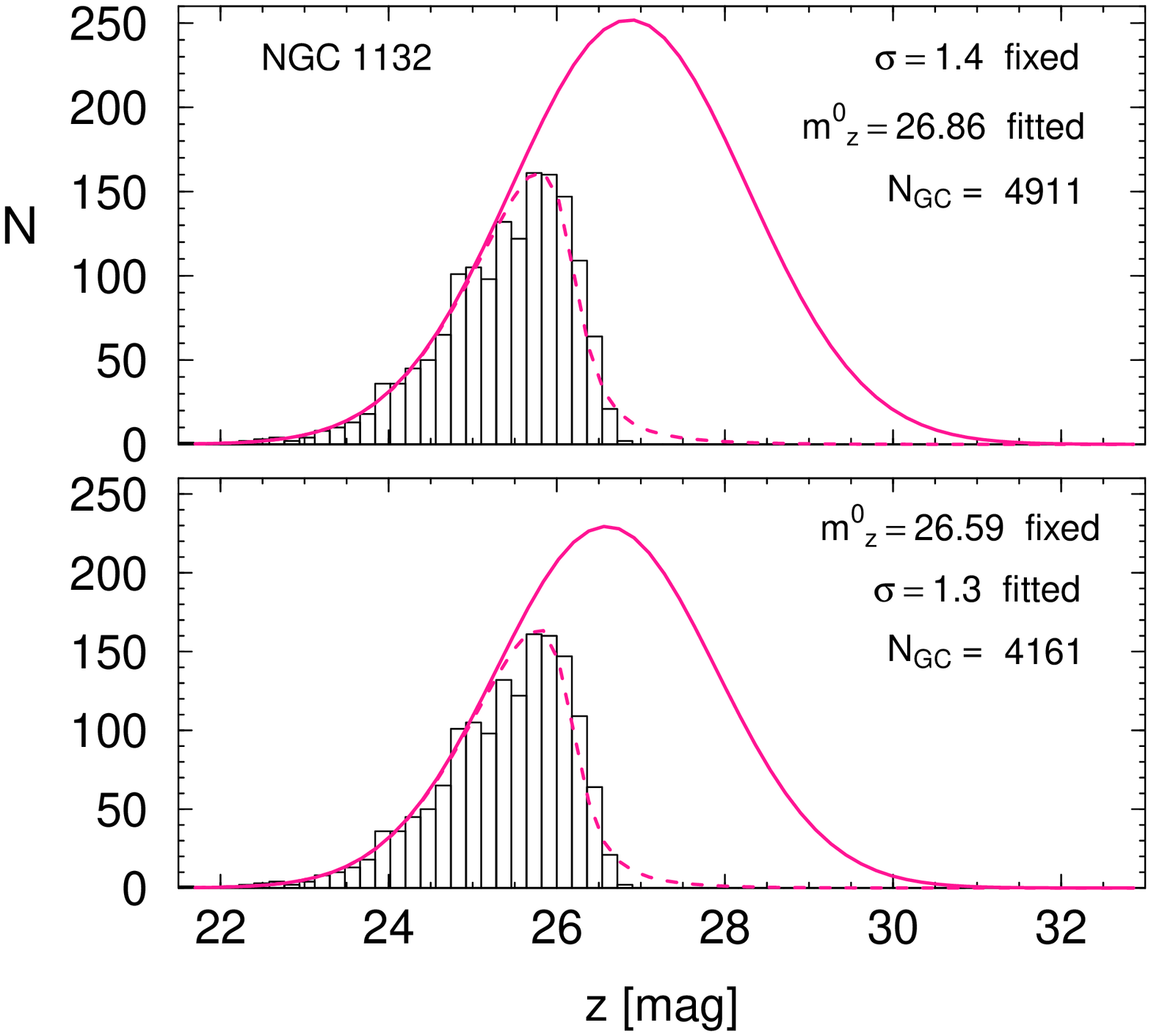}
\includegraphics[width=0.4\textwidth,angle=-90]{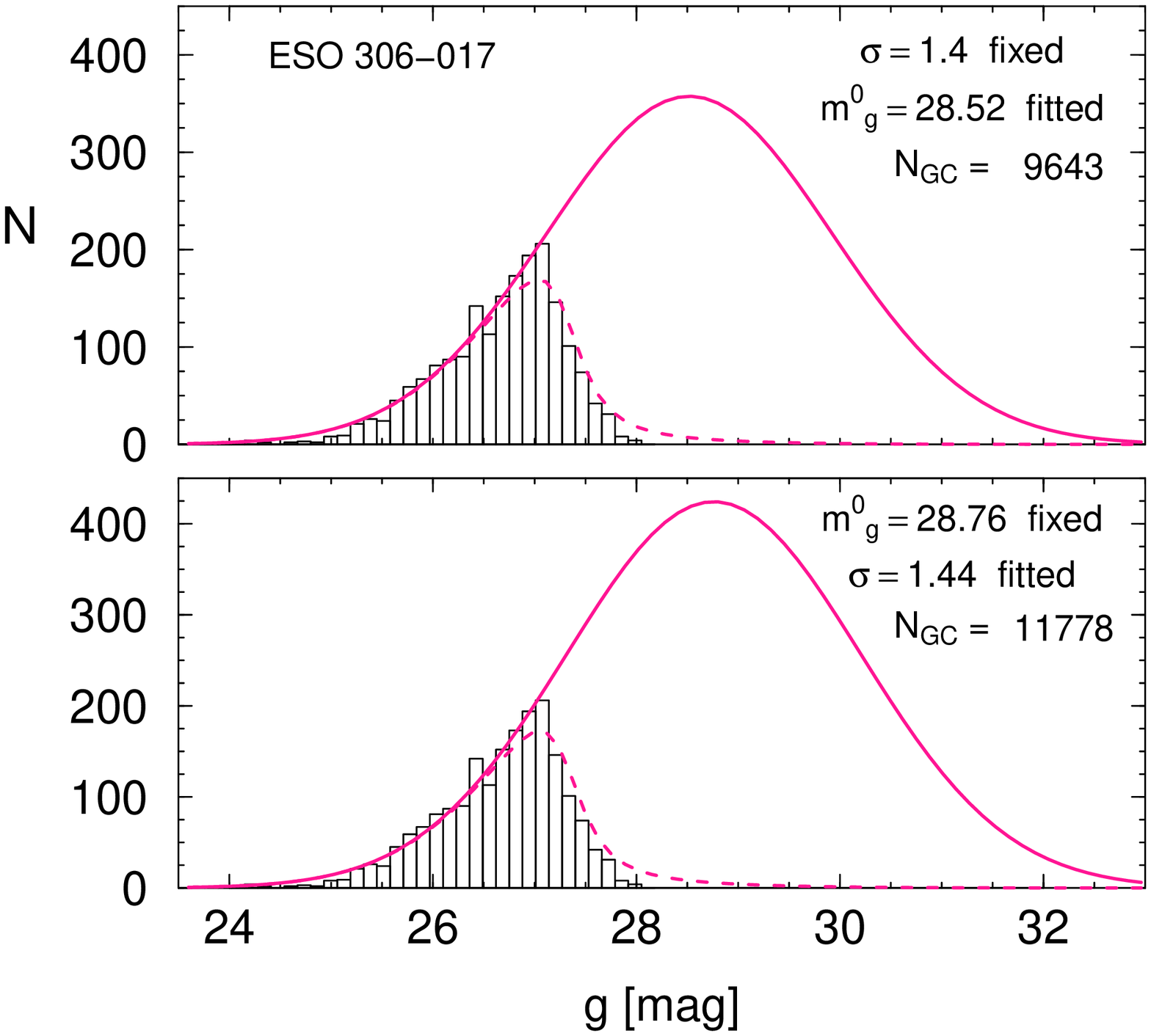}
\includegraphics[width=0.4\textwidth,angle=-90]{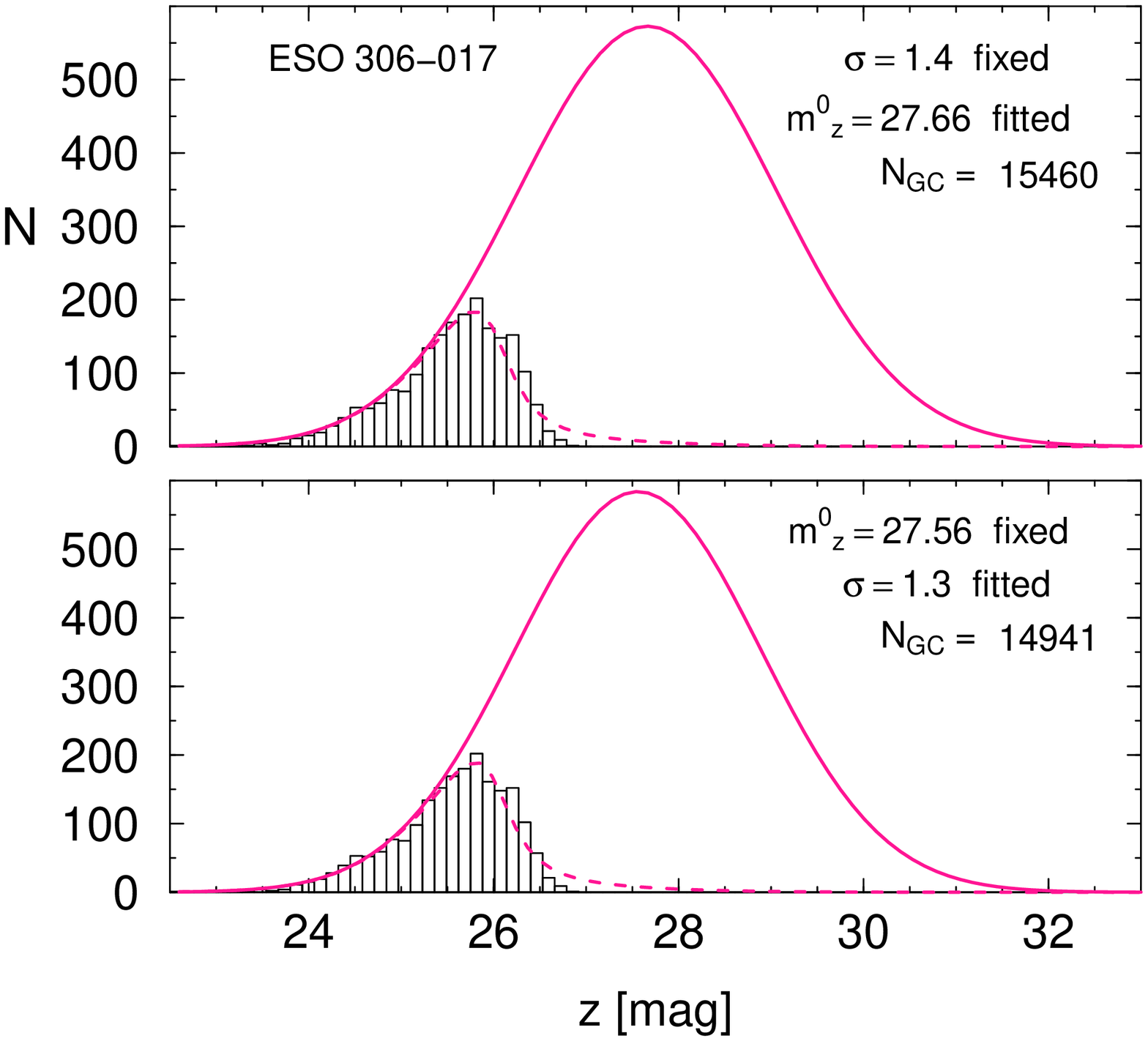}
\caption{GC luminosity functions in the $g$ (left panel) and $z$ (right panel)
    bands. The dashed lines show the best-fit Gaussian $\times$ Pritchet (completeness
    function) model, and the solid lines are the respective Gaussians. 
The numbers indicate the fixed and fitted parameters.  
{\it Top to Bottom}: NGC\,6482, NGC\,1132 and ESO\,306-017.}
\label{all_LFgausspritch}
\end{figure*}

To parameterize the GCLF, the GC magnitudes, $m$, were binned, again using an optimum bin size (Eq.\,\ref{binsize}). 
Figure\,\ref{all_LFgausspritch} show the GCLFs for NGC\,6482, NGC\,1132, and ESO\,306-017, respectively. Because we are photometrically limited, we do not detect all the GCs 
in the FOV. The GCLFs that we recover are the convolution of the intrinsic GCLFs and the incompleteness functions. To get the total number of GCs in the FOV, ${N_{\rm GC}^{\rm total}}$, we fit the product of 
a Gaussian and a Pritchet function to the GC magnitudes. The function we fit is:
 
\begin{equation}
f(m)=\frac{a_0}{\sqrt{2\pi\sigma_{\rm GCLF}^2}}e^{\frac{-(m-m^0)}{2\sigma_{\rm GCLF}^2}}*
\frac{1}{2}[1-\frac{\alpha(m-m_{\rm lim})}{\sqrt{1+\alpha^2(m-m_{\rm lim})^2}}].
\end{equation}

The values of the parameters $\alpha$ and $m_{\rm lim}$ are obtained from the incompleteness test.
We perform two fits. Firstly, we treat the mean value, $m^0$, and amplitude, $a^0$, of the Gaussian as free 
parameters, and fix the dispersion of the Gaussian to $\sigma_{\rm GCLF}=1.4$ (Jord\'an et al 2006). 
Secondly, we fit for $\sigma_{\rm GCLF}$ and $a^0$, and set ${m^0}$. 
The fits were done by $\chi^2$ minimization using the {\it Optim} task (described in Sec.\,\ref{section_starlightDIST}).
The same procedure was performed independently in g and z band, obtaining better results in z band and fixing $\sigma_{\rm GCLF}$.
Table\,\ref{gclf_parameters} shows the best fit values for $\sigma_{\rm GCLF}$ and $m^0$ recovered and their estimated ${N_{\rm GC}^{\rm total}}$, respectively. 
From Fig.\,\ref{all_LFgausspritch} we can see that the recovered values are consistent with
  the expected values for NGC\,1132 and ESO\,306-017, however for NGC\,6482 the recovered $m^0$ and $\sigma_{\rm GCLF}$ are shifted by $\sim$0.5 and 0.2, respectively. 

\begin{table*}[ht!]
\centering
\begin{tabular}{c  c  c  c  c  p{1in}}
\multicolumn{5}{c}  {GCLF fitted parameters.}                    \\
\hline \hline
  ${\rm System}$ &  ${m^0}$ (${N_{\rm GC}}^{\rm total} \pm \Delta{N}$) & ${\chi^2}_{m^0}$ & ${\sigma_{\rm GCLF}}$ (${N_{\rm GC}}^{\rm total} \pm \Delta{N}$) & ${\chi^2}_{\sigma_{GCLF}}$ \\
 \hline
\multicolumn{5}{c}  {g band}                    \\
\hline
 NGC6482 & 25.94 (787 $\pm$ 40) & 1.04 & 1.62 (1101 $\pm$ 49) & 1.23 \\
 NGC1132 & 27.52 (3298 $\pm$ 173) & 1.03 & 1.42 (3990 $\pm$ 174) & 1.36\\
 ESO\,306-017 & 28.52 (9643 $\pm$ 807) & 1.19 &  1.44 (11778 $\pm$ 597) & 1.27 \\
\hline
\multicolumn{5}{c}  {z band}                    \\
\hline
 NGC6482 & 24.84 (863 $\pm$ 36) & 1.01 & 1.61 (1130 $\pm$ 42) & 1.13  \\
 NGC1132 & 26.86 (4911 $\pm$ 252) & 0.93 & 1.3 (4161 $\pm$ 133) & 1.29 \\
 ESO\,306-017 & 27.66 (15460 $\pm$ 1224) & 1.18 &  1.33 (14941 $\pm$ 692) & 1.11 \\
\hline
\end{tabular}
\caption{{\it Col.\,1}: System; {\it Col.\,2}: ${m^0}$ and total number of GC fixing $\sigma_{\rm GCLF}=1.4$; {\it Col.\,3}: reduced chi-squared fitting ${m^0}$; {\it Col.\,4}:$\sigma_{\rm GCLF}$ and total number of GC fixing ${m^0}$; {\it Col.\,5}: reduced chi-squared fitting $\sigma_{\rm GCLF}$; {\it Col.\,6}: total number of GC using $\sigma_{\rm GCLF}=1.4$ and ${m^0}$.}
\label{gclf_parameters}
\end{table*}

\begin{table*}[ht!]
\centering
\begin{tabular}{c  c  c  c  c  c  p{1in}}
\multicolumn{6}{c}  {Specific frequency }                    \\
\hline \hline
  ${\rm System}$ &  ${N_{\rm GC,0}^{\rm total}}$ & ${M_g}^{\rm FOV}$ &  ${M_z}^{\rm FOV}$ & ${M_V}^{\rm FOV}$ & ${S_N}^{\rm FOV}$  \\
  \hline
  NGC\,6482 & 1140 $\pm$ 87 & -21.65 &  -23.20 & -21.95 & 1.89 $\pm$ 0.14 \\
 NGC\,1132 & 3613 $\pm$ 295 &  -22.39 &  -23.84 & -22.65 & 3.15 $\pm$ 0.26 \\
 ESO\,306-017 & 11577$\pm$ 1046 & -22.88 &  -24.38 & -23.16 & 6.30 $\pm$ 0.57 \\
\hline
\end{tabular}
\caption{{\it Col.\,1}: System; {\it Col.\,2}: total number of GC in the FOV using the expected ${m^0}_g$ and $\sigma_{GCLF}$; {\it Col.\,3,\,4} and {\it 5}: absolute g,z and V magnitudes in the FOV, respectively; {\it Col.\,6}: specific frequency in the FOV.}
\label{mags_S_N}
\end{table*}

\subsection{Specific Frequency}
\label{specificFrequency_label}

As noted previously, the specific frequency $S_N$ is the number of GCs
per unit galaxy luminosity, defined such that
\begin{equation}
S_N=N_{\rm GC} 10^{0.4(M_V + 15)} \,,
\end{equation} 

\noindent
where $N_{\rm GC}$ is the number of GCs, $M_V$ is the absolute magnitude of the galaxy
in $V$ band, and both quantities measured over the same area.
Although the parameters recovered from the GCLF fitting are consistent with expectations for
NGC\,1132 and ESO\,306-017, they are not for NGC\,6482.  Thus, to estimate $S_N$ we use the expected
turnover magnitudes (Col.~10 in Table\,\ref{sample_basicdata}) and fix $\sigma_{GCLF}$=1.4 to
calculate more conservative values of the total GC number, ${N_{\rm GC,0}^{\rm total}}$, for the
three FGs.  
Although we do not fit for the GCLF parameters in this case, our error estimates 
for the total GC number include a contribution arising from the uncertainties in the GCLF
parameters. The resulting values of ${N_{\rm GC,0}^{\rm total}}$ and their uncertainties are 
listed in Col.\,2 of Table\,\ref{mags_S_N} and are generally consistent with the values 
estimated previously as part of the GCLF analysis in Sec.\,\ref{sec:gclf}.

Calculation of the specific frequency also requires knowledge of the galaxy luminosity.
To estimate the absolute $V$-band magnitudes of the galaxies, we first calculate the $g$ and $z$
magnitudes from the counts in our 2-D isophotal model images, subtracting the sky values estimated
from the S\'ersic fits.  Then, we apply the transformation $V=g+0.320-0.399(g{-}z)$ given by
Kormendy et al.\ (2009).

The specific frequencies and absolute magnitudes in the $g$, $z$ and $V$ bandpasses within the FOV
are listed in Table~\ref{mags_S_N}.  
Both the GC population size and the $S_N$ increase with the galaxy luminosity for these three FGs.
There is a range of more than a factor of~3 in $S_N$, from $\lesssim2$ for NGC\,6482 to $\gtrsim6$
for \esogxy.  This spans the full range of values for
normal bright ellipticals in the Virgo cluster, as we discuss in the following section.

We note that these results are largely independent of the galaxy profile fits, which are mainly
used for estimating the sky background.  In particular, although the unusually high S\'ersic index 
$n\approx11$ for ESO\,306-017 could in part result from the limited spatial range of the fitted profile,
this would not affect the $S_N$, as we do not directly use the \sersic\ fit to calculate the
galaxy luminosity.  Sun \etal\ (2004) found from much lower resolution ground-based imaging that
the $n$ value is lower for \esogxy\ at radii beyond the ACS~FOV.
Adopting a \sersic\ profile with a lower $n$ would result in a larger sky estimate, slightly
decreasing the luminosity estimate and increasing the $S_N$ value.  However, if our
estimated profile is accurate within the FOV, the estimated sky and $S_N$ should also be accurate.

Finally, we could extrapolate the GC radial density profiles and the galaxy surface brightness fits
in order to estimate ``global'' $S_N$ values, but the extrapolated galaxy profiles are very sensitive
to the sky levels, which we cannot measure directly for our larger galaxies.  Moreover, as noted
above, there is evidence from the literature that an extrapolation of our \sersic\ fit for
\esogxy\ would not accurately describe its outer profile.  The GC number density profiles 
in Fig.~\ref{spatialdist_sersic_all} also become very uncertain at large radii.
We therefore opt to report the more reliable, directly observed, values within the FOV. 
The general tendency of GC systems to be more extended than the galaxy light means that the
global $S_N$ might be slightly higher.

\begin{figure*}[ht]
\centering
\includegraphics[width=0.45\textwidth,angle=-90]{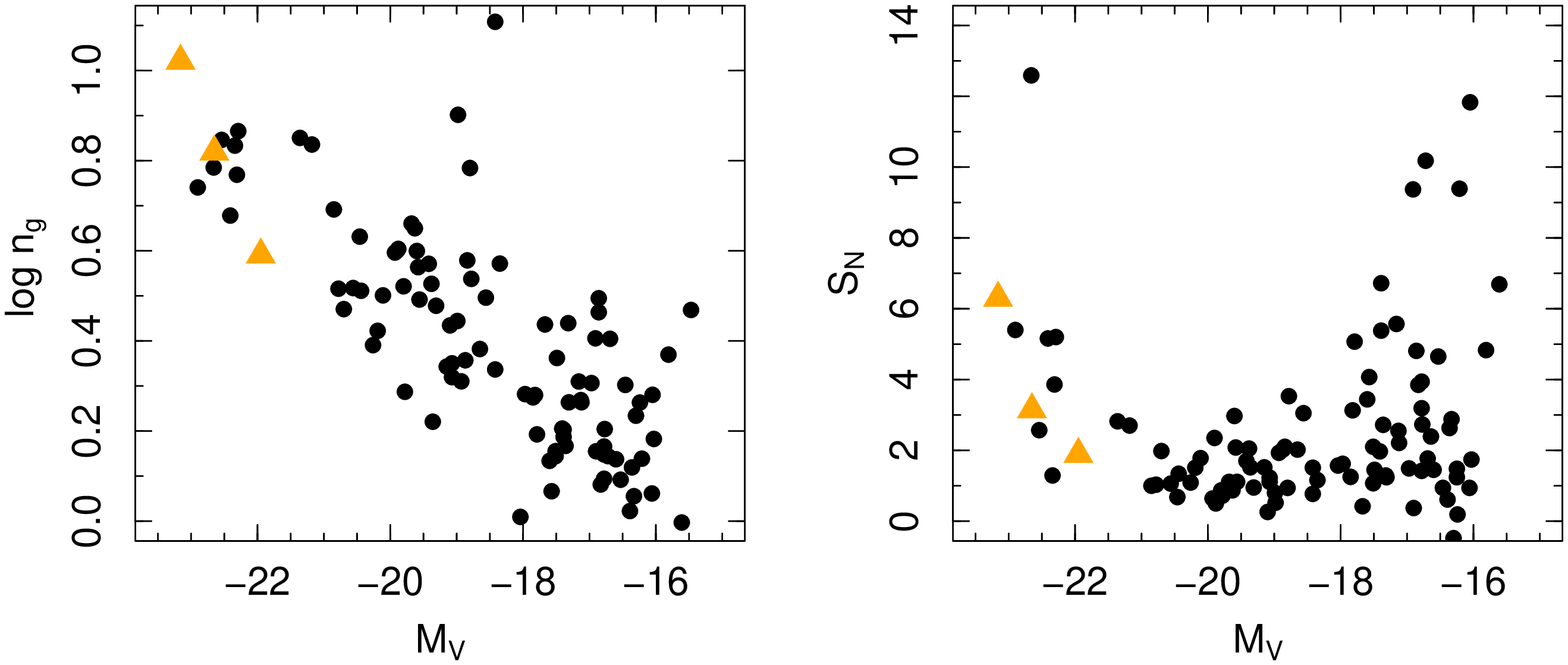}
\caption{Fossil groups in context: galaxy S\'ersic index (left) and $S_N$ are
    plotted as a function of $M_V$ for the fossil groups studied here (orange triangles)
    and the ACS Virgo Cluster Survey galaxies (black circles). The Virgo data are from
 Ferrarese et al.\ (2006 )and Peng et al.\ (2008).}
\label{sersic_Sn_M_ACS}
\end{figure*}

\section{Discussion}
\label{section_discussion}

\subsection{Comparison with Cluster Ellipticals}
\label{section_acsvcs}

One of the most basic questions to address is whether the GC systems in these FG
ellipticals differ from those of cluster ellipticals.  To compare our results with the
literature, we chose the ACS Virgo Cluster Survey (ACSVCS) sample of 100 early-type
galaxies (\cote\ et al.\ 2004), comprising the largest and most homogeneous data set on
the GC populations of cluster early-type galaxies.  Moreover, the observations were taken
with the same instrument and filters as the current study.  The ACS Fornax Cluster Survey
(\jordan\ \etal\ 2007a) targeted a sample of early-type galaxies in Fornax, but the
complete results on the GC systems remain to be published.  Cho \etal\ (2012) studied the
GC systems in a smaller sample of elliptical galaxies in low-density environments, but the
results were generally similar to those found for galaxies in Virgo.

Peng \etal\ (2006) found that the GC color distributions of the bright ACSVCS galaxies were
essentially all bimodal, with both the red and blue peaks becoming redder for more
luminous galaxies.  Likewise, we find that the color distributions of the GCs in all three
of our FG galaxies are bimodal.  Given the small size of our sample and the large scatters in
the relations between GC peak colors and galaxy luminosity, our measurements are generally
consistent with those Peng \etal\ (2006).  However, one surprising result is that the
lowest luminosity FG galaxy NGC\,6482 actually has the reddest mean GC color, 
$\langle g{-}z\rangle = 1.26$ mag, equal to that of NGC4649 (M60), the ACSVCS galaxy with
the reddest mean GC color.  We discuss this further in the following section.

The specific frequencies of the ACSVCS galaxies, and the implications for formation
efficiencies, were investigated in detail by Peng et al.\ (2008).  
Even though the FOV encloses different physical scales due to different distances of the Virgo cluster and our FG galaxies, most of the ACSVCS galaxies are small enough to encompass the entire GC system in the FOV, while for the largest galaxies, an extrapolated value to large radius was reported. 
Although $S_N$ tends to
increase with luminosity among giant ellipticals, the mean $S_N$ attains a minimum value
of $\sim\,2$ for early-type galaxies with $M_V\approx-20$, then increases again, but with
much larger scatter, at lower luminosities.
Fig.\,\ref{sersic_Sn_M_ACS} (right panel) shows that the three FGs studied here follow the trend
defined by the most luminous members of the 100 early-type ACSVCS galaxies.  
In particular, ESO\,306-017 has the highest luminosity in the combined sample, and its value
of $S_N=6.3$ is greater than that of all other 50$+$ galaxies with $M_V<-18$ except
for M87, which is the extreme high-$S_N$ outlier in this plot.  The left panel of
Fig.\,\ref{sersic_Sn_M_ACS} shows that the FGs also follow the tendency of higher
luminosity ellipticals to have larger \sersic\ indices (Ferrarese \etal\ 2006).
Thus, it appears that the FG galaxies accord well with the general $S_N$ and surface
brightness trends of normal giant ellipticals, and lack (at least in this limited sample)
the anomalously large $S_N$ values seen in some cD galaxies such as M87.

\subsection{Comparisons with X-ray Data}
\label{section_xray}

The definition of FG includes a minimum X-ray luminosity, thus, if an object is classified
as FG, it must be detected in X-rays. It has been noted that the X-ray luminosities
of the extended haloes of FGs are similar those of the intracluster media
in galaxy clusters (Mendes de Oliveira et al.\ 2009; Fasano et al.\ 2010).  For each FG
studied here, we show in Table~\ref{xray_properties} the literature values for its
X-ray luminosity, metal abundance of the intra-group gas, and M/L ratio determined
from X-ray data.

For NGC\,6482, the luminosity was obtained from Khosroshahi et al. (2004) who analyzed 
{\it Chandra} data. They also noted a central point source, possibly an AGN. 
Yoshioka et al.\ (2004) analyzed {\it ASCA} data for NGC\,1132 and classified it as an ``isolated
X-ray overluminous elliptical galaxy'' (IOLEG).  For ESO\,306-017, we use the results from Sun et
al. (2004), who analyzed {\it Chandra} and {\it XMM-Newton} data and averaged the results of
these analyses.  They also noted an X-ray {\it finger} emanating from the central X-ray peak to the
brightest companion galaxy, similar in location to the tidal feature reported in the present study
(see Sec.~\ref{section_starlightDIST}). The X-ray emission of this finger is 20\%-30\% brighter
than other regions at the same radius. 

\begin{table}[ht!]
\centering
\begin{tabular}{c  c  c  c p{1in}}
\multicolumn{4}{c}  {X-ray properties}                    \\
\hline \hline
  ${\rm System}$ & $L_{X,bol} $ & $Z/Z_{\odot}$ & M/L \\
    &  $(10^{42}{h_{71}}^{-2}{\rm erg\,s}^{-1})$ &  &  (M$_{\odot}$/L$_{\odot}$) \\
  \hline
 NGC\,6482 & 1.07  & 0.76 $\pm$ 0.28 & 71$^{\rm ~a}$ \\
 NGC\,1132 & 2.12 & 0.19 $\pm$ 0.1 & 180$^{\rm ~b}$\\
 ESO\,306-017 & 64.15 $\pm$ 0.26 & 0.44 $\pm$ 0.03 & 150$^{\rm ~a}$ \\
\hline
\end{tabular}
\caption{{\it Col.\,1}: System; {\it Col.\,2}: Bolometric X-ray luminosity; {\it Col.\,3}: metal abundance of the X-ray intra-group gas; {\it Col.\,4}: mass-to-light ratio. $^{\rm a}$~R band ;$^{\rm b}$~B band. }
\label{xray_properties}
\end{table}

Table\,\ref{xray_properties} includes the literature values of the M/L ratios. 
However, we must be cautious in comparing these values because of different assumptions
made by different authors, mainly regarding the optical luminosity and because of the large
uncertainties in mass estimates (see Khosroshahi et al. 2007 discussion). For instance, Yoshioka et al.\ (2004)
assumed a much lower contribution for the non-brightest galaxies to the total optical luminosity,
compared with Khosroshahi et al.\ (2006) and Sun et al.\ (2004), a fact that increases their M/L estimate. 
Thus, systematic uncertainties can make it difficult to determine
whether or not there is any trend in GC properties with M/L. 

While three objects are an insufficient sample for drawing general conclusions, we do see hints of
interesting trends when comparing the properties of the X-ray gas and GC systems
(Table\,\ref{kmm_parameters}).  The results are consistent with positive correlations between X-ray
luminosity, GC number, and $S_N$, similar to the trends found for BCGs (e.g., Blakeslee 1999; West
et al. 1995).  There may also be a trend with optical luminosity.  Intriguingly, the mean color of
the GC population ($\langle g{-}z\rangle =$ 1.26, 1.11 and 1.20 for NGC\,6482, NGC\,1132 and
ESO\,306-017, respectively) becomes redder as the metal abundance of the intra-group gas increases.
The apparent correlations of the properties of the GC systems with X-ray intra-group gas abundance
and galaxy luminosity suggest that these components might likely formed at similar epochs.  

One might expect the systems with the highest X-ray luminosities and masses to possess the most
metal-rich GC systems, but this is apparently not the case for this small sample of FGs. The mean
color of the GC system does not correlate with X-ray luminosity, but interestingly, it does with X-ray
metallicity. This could be an indication that the metal enrichment is not set mainly by the mass
but by the concentration of the halo (Atlee \& Martini 2012). The dark matter concentration, $c$,
is measured as the ratio of the virial radius to the radius of a sphere enclosing 0.2 the virial
mass (von Benda-Beckmann et al. 2008). Higher $c$ values are associated with an earlier formation
epoch (Khosroshahi et al. 2007; von Benda-Beckmann et al. 2008).
Numerical simulations indicate that FGs have higher values of $c$ than galaxy groups with similar
masses but not classified as FGs ($c=6.4$ for FGs and $c=5.5$ for normal groups).  The $c$ values reported for
NGC\,6482, NGC\,1132 and ESO\,306-017 are 60, 38 and 8.5 (see Fig\,9 from Khosroshahi et al. 2007); 
all of these are high, with the value for NGC\,6482 being quite extreme.  However, the dark matter
concentration on these scales is notoriously difficult to constrain observationally.

High concentrations can help fuel larger rates of star-formation, producing a higher proportion of
massive stellar systems such as GCs (Larsen \& Richtler 2000; Peng \etal\ 2008), as well as making
self-enrichment more efficient.
Thus, the possible link between dark matter halo concentration, X-ray gas metallicity, and
GC mean color may be understood in the context of early and violent dissipative assembly, whereby a
first generation of GCs contaminate the surrounding media from which the bulk of the stars formed.
Eventually the feedback from star formation (and possibly AGN) heats the gas and suppresses
subsequent star formation (see Johansson et al.\ 2012).  Afterwards, the galaxy continues growing by
non-dissipative minor mergers (De Lucia et al. 2006; Naab et al. 2009).  Of course, a larger sample
of FGs is needed to confirm these correlations before drawing more detailed conclusions on the star
formation histories of FG galaxies.

%%%%%%%%%%%%%%%%%%%%%%%%%%%%%%%%%%%%%%%%%%%%%%

\section{Conclusions}
\label{section_conclusions}

The main results of this study are: 

\begin{enumerate}
\item{We detected 369, 1410 and 1918 GCs down to magnitude $z$=24.7, 26.2 and 26.2 in NGC\,6482, NGC\,1132 and ESO\,306-017, respectively; after completeness correction and assuming a Gaussian GCLF with expected ${M^0}_g=-7.2$ and $\sigma_{GCLF}=1.4$, the number of GCs in each FOV are: 1140$\pm$87, 3613$\pm$295 and 11577$\pm$1046, respectively.\\
}
\item{The GC color distributions for all three FGs are better described by a bimodal, rather than unimodal  Gaussian model. For the heteroscedastic cases, NGC\,1132 and ESO\,306-017, the dispersion of the red population is wider than the blue. The mean $g{-}z$ colors are: 1.26, 1.11 and 1.20 for NGC\,6482, NGC\,1132 and ESO\,306-017, respectively.  The mean color for NGC\,6482 is unusually red; it is interesting that this is also the FG with the highest X-ray gas metallicity.\\
}
\item{For all three FGs, the spatial distribution of the starlight is more concentrated than that of the GCs, and the blue GCs are more extended than the red ones, similar to the case for other ellipticals.\\ 
}
\item{The derived values of $S_N$ are: $1.9\pm0.1$, $3.1\pm0.3$ and $6.3\pm0.6$ for NGC\,6482, NGC\,1132 and ESO\,306-017, respectively. These span the full range for normal ellipticals in the Virgo cluster.  The results are consistent with both the total number of GCs and $S_N$ increasing with the optical luminosity of the galaxy and the X-ray luminosity from the intra-group gas. \\
}
\item{ From analysis of the surface brightness distributions, we found evidence of recent
  interactions, particularly in ESO\,306-017, which shows a tidal feature coincident with the {\it
    finger} reported previously in X-rays, and NGC\,1132, which has shell-like structure.  
While NGC\,6482 is well-described by a standard \sersic\ profile, the two brighter galaxies are better fitted by core-\sersic\ models. 
All three galaxies contain central dust features.
These observations are consistent with numerical simulations indicating
that signs of recent merging should be fairly common in first-ranked FG
galaxies (D\'iaz-Gim\'enez et al.\ 2008), and they suggest that the
paradigm of FGs as relaxed, undisturbed systems needs to be reconsidered.  \\
}
\end{enumerate}

Larger samples of GC systems in FGs are needed in order to make more definite conclusions. 
However, overall we conclude that the GC properties (colors, spatial distributions,
specific frequencies) in FG central galaxies are
generally similar to those seen in other giant ellipticals, which mainly reside in
clusters.  Although the environments differ, this study suggests that the GC systems formed under
very similar conditions.  
These results might therefore be taken as a confirmation that the same
basic formation processes are responsible for the buildup of massive early-type
galaxies in all environments.

\begin{acknowledgements}
 We thank the anonymous referee for his/her helpful comments and suggestions that helped to improve the clarity of the manuscript. AJ acknowledges support from Ministry of Economy ICM Nucleus P07-021-F, Anillo ACT-086 and BASAL CATA PFB-06. K.A.A-M acknowledges the support of ESO through a studenship and CONACyT (Mexico). 
      Support for program \#10558 was provided by NASA through a grant from the Space Telescope Science Institute, which is operated by the Association of Universities for Research in Astronomy, Inc., under NASA contract NAS 5-26555. 
      
\end{acknowledgements}


\begin{thebibliography}{}
%A
\bibitem[Atlee \& Martini(2012)]{2012arXiv1201.2957A} Atlee, D.~W., \& Martini, P.\ 2012, arXiv:1201.2957 
%B
\bibitem[Bertin \& Arnouts(1996)]{1996A&AS..117..393B} Bertin, E., \& Arnouts, S.\ 1996, \aaps, 117, 393 
\bibitem[Blakeslee et al.(1997)]{1997AJ....114..482B} Blakeslee, J.~P., Tonry, J.~L., \& Metzger, M.~R.\ 1997, \aj, 114, 482 
\bibitem[Blakeslee(1999)]{1999AJ....118.1506B} Blakeslee, J.~P.\ 1999, \aj, 118, 1506 
\bibitem[Blakeslee et al.(2003)]{2003ASPC..295..257B} Blakeslee, J.~P., Anderson, K.~R., Meurer, G.~R.\ et al.\ 2003, Astronomical Data Analysis Software and Systems XII, 295, 257 
\bibitem[Blakeslee et al.(2012)]{2012ApJ...746...88B} Blakeslee, J.~P., Cho, H., Peng, E.~W., et al.\ 2012, \apj, 746, 88
\bibitem[Brodie \& Huchra(1991)]{1991ApJ...379..157B} Brodie, J.~P., \& Huchra, J.~P.\ 1991, \apj, 379, 157 
\bibitem[Brodie \& Strader(2006)]{2006ARA&A..44..193B} Brodie, J.~P., \& Strader, J.\ 2006, \araa, 44, 193 
\bibitem[Byun et al.(1996)]{1996AJ....111.1889B} Byun, Y.-I., Grillmair, 
C.~J., Faber, S.~M., et al.\ 1996, \aj, 111, 1889 
%C
\bibitem[Chies-Santos et 
al.(2012)]{2012A&A...539A..54C} Chies-Santos, A.~L., Larsen, S.~S., Cantiello, M., et al.\ 2012, \aap, 539, A54 
\bibitem[Cho et al.(2012)]{2012MNRAS.422.3591C} Cho, J., Sharples, R.~M., 
Blakeslee, J.~P., et al.\ 2012, \mnras, 422, 3591 
\bibitem[Cohen et al.(1998)]{1998ApJ...496..808C} Cohen, J.~G., Blakeslee, 
J.~P., \& Ryzhov, A.\ 1998, \apj, 496, 808 
\bibitem[C{\^o}t{\'e} et al.(1998)]{1998ApJ...501..554C} C{\^o}t{\'e}, P., Marzke, R.~O., \& West, M.~J.\ 1998, \apj, 501, 554 
J.~P., \& C{\^o}t{\'e}, P.\ 2003, \apj, 592, 866
\bibitem[C{\^o}t{\'e} et al.(2004)]{2004ApJS..153..223C} C{\^o}t{\'e}, P., Blakeslee, J.~P., Ferrarese, L., et al.\ 2004, \apjs, 153, 223 
\bibitem[C{\^o}t{\'e} et al.(2007)]{2007ApJ...671.1456C} C{\^o}t{\'e}, P., Ferrarese, L., Jord{\'a}n, A., et al.\ 2007, \apj, 671, 1456 
\bibitem[Cui et al.(2011)]{2011MNRAS.416.2997C} Cui, W., Springel, V., 
Yang, X., De Lucia, G., \& Borgani, S.\ 2011, \mnras, 416, 2997 
\bibitem[Cypriano et al.(2006)]{2006AJ....132..514C} Cypriano, E.~S., 
Mendes de Oliveira, C.~L., \& Sodr{\'e}, L., Jr.\ 2006, \aj, 132, 514 
%D
\bibitem[Dariush et al.(2007)]{2007MNRAS.382..433D} Dariush, A., 
Khosroshahi, H.~G., Ponman, T.~J., et al.\ 2007, \mnras, 382, 433 
\bibitem[De Lucia et al.(2006)]{2006MNRAS.366..499D} De Lucia, G., Springel, V., White, S.~D.~M., Croton, D., \& Kauffmann, G.\ 2006, \mnras, 366, 499
\bibitem[D'Onghia et al.(2005)]{2005ApJ...630L.109D} D'Onghia, E., 
Sommer-Larsen, J., Romeo, A.~D., et al.\ 2005, \apjl, 630, L109 
\bibitem[D{\'{\i}}az-Gim{\'e}nez et 
al.(2008)]{2008A&A...490..965D} D{\'{\i}}az-Gim{\'e}nez, E., Muriel, H., \& Mendes de Oliveira, C.\ 2008, \aap, 490, 965 
\bibitem[D{\'{\i}}az-Gim{\'e}nez et 
al.(2011)]{2011A&A...527A.129D} D{\'{\i}}az-Gim{\'e}nez, E., Zandivarez, A., Proctor, R., Mendes de Oliveira, C., \& Abramo, L.~R.\ 2011, \aap, 527, A129 
%E
\bibitem[Eigenthaler \& Zeilinger(2012)]{2012arXiv1202.4470E} Eigenthaler, P., \& Zeilinger, W.~W.\ 2012, arXiv:1202.4470 
%F
\bibitem[Fall \& Rees(1985)]{1985ApJ...298...18F} Fall, S.~M., \& Rees, M.~J.\ 1985, \apj, 298, 1
\bibitem[Fasano et al.(2010)]{2010MNRAS.404.1490F} Fasano, G., Bettoni, D., Ascaso, B., et al.\ 2010, \mnras, 404, 1490 
\bibitem[Ferrarese et al.(2006)]{2006ApJS..164..334F} Ferrarese, L., 
C{\^o}t{\'e}, P., Jord{\'a}n, A., et al.\ 2006, \apjs, 164, 334 
\bibitem[Fleming et al.(1995)]{1995AJ....109.1044F} Fleming, D.~E.~B., 
Harris, W.~E., Pritchet, C.~J., \& Hanes, D.~A.\ 1995, \aj, 109, 1044 
\bibitem[Forbes et al.(1996)]{1996ApJ...467..126F} Forbes, D.~A., Franx, M., Illingworth, G.~D., \& Carollo, C.~M.\ 1996, \apj, 467, 126 
\bibitem[Forbes et al.(1997)]{1997AJ....113..887F} Forbes, D.~A., Brodie, J.~P., \& Huchra, J.\ 1997, \aj, 113, 887 
%G
\bibitem[Gebhardt 
\& Kissler-Patig(1999)]{1999AJ....118.1526G} Gebhardt, K., \& Kissler-Patig, M.\ 1999, \aj, 118, 1526 
\bibitem[Geisler et al.(1996)]{1996AJ....111.1529G} Geisler, D., Lee, M.~G., \& Kim, E.\ 1996, \aj, 111, 1529 
\bibitem[Graham 
\& Driver(2005)]{2005PASA...22..118G} Graham, A.~W., \& Driver, S.~P.\ 2005, \pasa, 22, 118 
\bibitem[Graham et al.(1996)]{1996ApJ...465..534G} Graham, A., Lauer, T.~R., Colless, M., \& Postman, M.\ 1996, \apj, 465, 534 
%H
\bibitem[Harris \& van den Bergh(1981)]{1981AJ.....86.1627H} Harris, W.~E., \& van den Bergh, S.\ 1981, \aj, 86, 1627 
\bibitem[Harris(1991)]{1991AJ....102.1348H} Harris, W.~E.\ 1991, \aj, 102, 1348 
\bibitem[Harris(1991)]{1991ARA&A..29..543H} Harris, W.~E.\ 1991, \araa, 29, 543 
\bibitem[Harris \& Harris(2002)]{2002AJ....123.3108H} Harris, W.~E., \& Harris, G.~L.~H.\ 2002, \aj, 123, 3108 
%I
\bibitem[Izenman(1991)]{1991AJ....102.1348H} Izenman, A.~J.\ 1991, Am. Stat. Assoc., 86, 205 
%J
\bibitem[Jedrzejewski(1987)]{1987MNRAS.226..747J} Jedrzejewski, R.~I.\ 1987, \mnras, 226, 747 
\bibitem[Johansson et al.(2012)]{2012arXiv1202.3441J} Johansson, P.~H., 
Naab, T., \& Ostriker, J.~P.\ 2012, arXiv:1202.3441
\bibitem[Jones et al.(2003)]{2003MNRAS.343..627J} Jones, L.~R., Ponman, T.~J., Horton, A., et al.\ 2003, \mnras, 343, 627 
\bibitem[Jord{\'a}n et al.(2004)]{2004AJ....127...24J} Jord{\'a}n, A., 
C{\^o}t{\'e}, P., West, M.~J., et al.\ 2004, \aj, 127, 24 
%
\bibitem[Jord{\'a}n et al.(2005)]{2005ApJ...634.1002J} Jord{\'a}n, A., 
C{\^o}t{\'e}, P., Blakeslee, J.~P., et al.\ 2005, \apj, 634, 1002 
%
\bibitem[Jord{\'a}n et al.(2007)]{2007ApJS..171..101J} Jord{\'a}n, A., 
McLaughlin, D.~E., C{\^o}t{\'e}, P., et al.\ 2007, \apjs, 171, 101 
%
\bibitem[Jord{\'a}n et al.(2007)]{2007ApJS..169..213J} Jord{\'a}n, A., 
Blakeslee, J.~P., C{\^o}t{\'e}, P., Ferrarese, L., Infante, L., Mei, S., 
Merritt, D., Peng, E.~W., Tonry, J.~L., West, M.~J.\
  2007, \apjs, 169, 213
\bibitem[Jord{\'a}n et al.(2006)]{2006ApJ...651L..25J} Jord{\'a}n, A., 
McLaughlin, D.~E., C{\^o}t{\'e}, P., et al.\ 2006, \apjl, 651, L25 
%K
\bibitem[Khosroshahi et al.(2004)]{2004MNRAS.349.1240K} Khosroshahi, H.~G., Jones, L.~R., \& Ponman, T.~J.\ 2004, \mnras, 349, 1240 
\bibitem[Khosroshahi et al.(2006)]{2006MNRAS.372L..68K} Khosroshahi, H.~G., 
Ponman, T.~J., \& Jones, L.~R.\ 2006, \mnras, 372, L68
\bibitem[Khosroshahi et al.(2007)]{2007MNRAS.377..595K} Khosroshahi, H.~G., 
Ponman, T.~J., \& Jones, L.~R.\ 2007, \mnras, 377, 595 
\bibitem[Kormendy et al.(2009)]{2009ApJS..182..216K} Kormendy, J., Fisher, 
D.~B., Cornell, M.~E., \& Bender, R.\ 2009, \apjs, 182, 216 
%L
\bibitem[La Barbera et al.(2012)]{2012MNRAS.422.3010L} La Barbera, F., 
Paolillo, M., De Filippis, E., \& de Carvalho, R.~R.\ 2012, \mnras, 422, 3010 
\bibitem[Larsen 
\& Richtler(2000)]{2000A&A...354..836L} Larsen, S.~S., \& Richtler, T.\ 2000, \aap, 354, 836 
%M
\bibitem[Madrid(2011)]{2011ApJ...737L..13M} Madrid, J.~P.\ 2011, \apjl, 737, L13 
\bibitem[McLaughlin 
\& Fall(2008)]{2008ApJ...679.1272M} McLaughlin, D.~E., \& Fall, S.~M.\ 2008, \apj, 679, 1272 
\bibitem[Mendes de Oliveira \& Carrasco(2007)]{2007ApJ...670L..93M} Mendes de Oliveira, C.~L., \&
  Carrasco, E.~R.\ 2007, \apjl, 670, L93  
\bibitem[Mendes de Oliveira et al.(2009)]{2009AJ....138..502M} Mendes de Oliveira, C.~L., Cypriano,
  E.~S., Dupke, R.~A., \& Sodr{\'e}, L., Jr.\ 2009, \aj, 138, 502  
\bibitem[M{\'e}ndez-Abreu et 
al.(2012)]{2012A&A...537A..25M} M{\'e}ndez-Abreu, J., Aguerri, J.~A.~L., Barrena, R., et al.\ 2012, \aap, 537, A25 
\bibitem[Muratov 
\& Gnedin(2010)]{2010ApJ...718.1266M} Muratov, A.~L., \& Gnedin, O.~Y.\ 2010, \apj, 718, 1266 
%N
\bibitem[Naab et al.(2009)]{2009ApJ...699L.178N} Naab, T., Johansson, 
P.~H., \& Ostriker, J.~P.\ 2009, \apjl, 699, L178 
\bibitem[Nair et al.(2011)]{2011ApJ...734L..31N} Nair, P., van den Bergh, 
S., \& Abraham, R.~G.\ 2011, \apjl, 734, L31
\bibitem[Nelder \& Mead(1965)]{nelder-mead1965} Nelder, J. A. \& Mead, R.\ 1965, The Computer Journal, 308, 313 \\
%O
%P
\bibitem[Peng et al.(2006)]{2006ApJ...639...95P} Peng, E.~W., Jord{\'a}n, A., C{\^o}t{\'e}, P., et al.\ 2006, \apj, 639, 95 
\bibitem[Peng et al.(2008)]{2008ApJ...681..197P} Peng, E.~W., Jord{\'a}n, A., C{\^o}t{\'e}, P., et al.\ 2008, \apj, 681, 197 
\bibitem[Ponman et al.(1994)]{1994Natur.369..462P} Ponman, T.~J., Allan, D.~J., Jones, L.~R., et al.\ 1994, \nat, 369, 462 
\bibitem[Postman \& Lauer(1995)]{1995ApJ...440...28P} Postman, M., \& Lauer, T.~R.\ 1995, \apj, 440, 28 \\
\bibitem[Press \& Schechter(1974)]{1974ApJ...187..425P} Press, W.~H., \& Schechter, P.\ 1974, \apj, 187, 425 
\bibitem[Proctor et al.(2011)]{2011MNRAS.418.2054P} Proctor, R.~N., de 
Oliveira, C.~M., Dupke, R., et al.\ 2011, \mnras, 418, 2054
\bibitem[Puzia et al.(2005)]{2005A&A...439..997P} Puzia, T.~H., Kissler-Patig, M., Thomas, D., et al.\ 2005, \aap, 439, 997 
%R
\bibitem[Rhode \& Zepf(2004)]{2004AJ....127..302R} Rhode, K.~L., \& Zepf, S.~E.\ 2004, \aj, 127, 302
%S
\bibitem[Sales et al.(2012)]{2012MNRAS.423.1544S} Sales, L.~V., Navarro, 
J.~F., Theuns, T., et al.\ 2012, \mnras, 423, 1544 
\bibitem[Sersic(1968)]{1968adga.book.....S} S\'ersic, J.~L.\ 1968, Atlas de Galaxias Australes (Cordoba: Observatorio Astronomico)
\bibitem[Schlegel et al.(1998)]{1998ApJ...500..525S} Schlegel, D.~J., 
Finkbeiner, D.~P., \& Davis, M.\ 1998, \apj, 500, 525 
\bibitem[Schombert \& Smith(2012)]{2012arXiv1203.2578S} Schombert, J., \& Smith, A.~K.\ 2012, arXiv:1203.2578 
\bibitem[Sirianni et al.(2005)]{2005PASP..117.1049S} Sirianni, M., Jee, 
M.~J., Ben{\'{\i}}tez, N., et al.\ 2005, \pasp, 117, 1049 
\bibitem[de Souza et al.(2004)]{2004ApJS..153..411D} de Souza, R.~E., 
Gadotti, D.~A., \& dos Anjos, S.\ 2004, \apjs, 153, 411 
\bibitem[Stetson(1987)]{1987PASP...99..191S} Stetson, P.~B.\ 1987, \pasp, 99, 191 
\bibitem[Sun et al.(2004)]{2004ApJ...612..805S} Sun, M., Forman, W., Vikhlinin, A., et al.\ 2004, \apj, 612, 805 
%T
\bibitem[Tody(1986)]{1986SPIE..627..733T} Tody, D.\ 1986, \procspie, 627, 
733 
\bibitem[Tovmassian(2010)]{2010RMxAA..46...61T} Tovmassian, H.\ 2010, \rmxaa, 46, 61
\bibitem[Trujillo et al.(2004)]{2004AJ....127.1917T} Trujillo, I., Erwin, 
P., Asensio Ramos, A., \& Graham, A.~W.\ 2004, \aj, 127, 1917 
%V
\bibitem[van den Bergh(1975)]{1975ARA&A..13..217V} van den Bergh, S.\ 1975, \araa, 13, 217 
\bibitem[Vikhlinin et al.(1999)]{1999ApJ...520L...1V} Vikhlinin, A., McNamara, B.~R., Hornstrup, A., et al.\ 1999, \apjl, 520, L1 
\bibitem[Villegas et al.(2010)]{2010ApJ...717..603V} Villegas, D., 
Jord{\'a}n, A., Peng, E.~W., et al.\ 2010, \apj, 717, 603
\bibitem[von Benda-Beckmann et al.(2008)]{2008MNRAS.386.2345V} von 
Benda-Beckmann, A.~M., D'Onghia, E., Gottl{\"o}ber, S., et al.\ 2008, 
\mnras, 386, 2345 
%W 
\bibitem[West et al.(1995)]{1995ApJ...453L..77W} West, M.~J., Cote, P., 
Jones, C., Forman, W., \& Marzke, R.~O.\ 1995, \apjl, 453, L77 
\bibitem[West et al.(2004)]{2004Natur.427...31W} West, M.~J., C{\^o}t{\'e}, 
P., Marzke, R.~O., \& Jord{\'a}n, A.\ 2004, \nat, 427, 31 
%Y
\bibitem[Yoon et al.(2006)]{2006Sci...311.1129Y} Yoon, S.-J., Yi, S.~K., \& Lee, Y.-W.\ 2006, Science, 311, 1129 
\bibitem[Yoon et al.(2011)]{2011ApJ...743..150Y} Yoon, S.-J., Lee, S.-Y., 
Blakeslee, J.~P., et al.\ 2011, \apj, 743, 150
\bibitem[Yoshioka et al.(2004)]{2004AdSpR..34.2525Y} Yoshioka, T., 
Furuzawa, A., Takahashi, S., et al.\ 2004, Advances in Space Research, 34, 
2525 
%Z
\bibitem[Zepf \& Ashman(1993)]{1993MNRAS.264..611Z} Zepf, S.~E., \& Ashman, K.~M.\ 1993, \mnras, 264, 611 
\bibitem[Zinn(1985)]{1985ApJ...293..424Z} Zinn, R.\ 1985, \apj, 293, 424 
\end{thebibliography}
\end{document}